\documentclass[a4paper,11pt]{article}
\pdfoutput=1

\usepackage{jcappub}
\usepackage[T1]{fontenc} 

\usepackage{amsmath,amssymb}
\usepackage{bbold}
\usepackage{dsfont}
\usepackage{dcolumn}
\usepackage{tikz,hyperref}
\usepackage[utf8]{inputenc}
\usepackage{verbatim}
\usepackage{mathtools}
\usepackage{animate,media9}
\usepackage{slashed}
\usepackage{braket}
\usepackage{subfigure}
\usepackage{slashed}
\usepackage{hyperref}
\usepackage{multirow}

\usepackage{graphicx}
\usepackage{bm}

\definecolor{lime}{HTML}{A6CE39}
\DeclareRobustCommand{\orcidicon}{%
        \begin{tikzpicture}
        \draw[lime, fill=lime] (0,0)
        circle [radius=0.16]
        node[white] {{\fontfamily{qag}\selectfont \tiny ID}};
        \draw[white, fill=white] (-0.0625,0.095)
        circle [radius=0.007];
        \end{tikzpicture}
        \hspace{-2mm}
}

\foreach \x in {A, ..., Z}{%
        \expandafter\xdef\csname orcid\x\endcsname{\noexpand\href{https://orcid.org/\csname orcidauthor\x\endcsname}{\noexpand\orcidicon}}
}

\definecolor{lime}{HTML}{A6CE39}
\DeclareRobustCommand{\orcidicon}{%
        \begin{tikzpicture}
        \draw[lime, fill=lime] (0,0)
        circle [radius=0.16]
        node[white] {{\fontfamily{qag}\selectfont \tiny ID}};
        \draw[white, fill=white] (-0.0625,0.095)
        circle [radius=0.007];
        \end{tikzpicture}
        \hspace{-2mm}
}

\foreach \x in {A, ..., Z}{%
        \expandafter\xdef\csname orcid\x\endcsname{\noexpand\href{https://orcid.org/\csname orcidauthor\x\endcsname}{\noexpand\orcidicon}}
}

\title{A Parametric Model for Self-Interacting Dark Matter Halos}

\author[a]{Daneng Yang,\footnote{Corresponding author.}\orcidA{}}
\author[b,c]{Ethan O.~Nadler,\orcidB{}}
\author[a]{Hai-Bo Yu,\orcidC{}}
\author[d,e]{Yi-Ming Zhong\orcidD{}}

\emailAdd{danengy@ucr.edu}
\emailAdd{enadler@carnegiescience.edu}
\emailAdd{haiboyu@ucr.edu}
\emailAdd{yimzhong@cityu.edu.hk}

\affiliation[a]{Department of Physics and Astronomy, University of California, Riverside, CA 92521, USA}
\affiliation[b]{Carnegie Observatories, 813 Santa Barbara Street, Pasadena, CA 91101, USA}
\affiliation[c]{Department of Physics $\&$ Astronomy, University of Southern California, Los Angeles, CA, 90007, USA}
\affiliation[d]{Department of Physics, City University of Hong Kong, Tat Chee Avenue, Kowloon, Hong Kong SAR, China}
\affiliation[e]{Kavli Institute for Cosmological Physics, University of Chicago, Chicago, IL 60637, USA}

\abstract{

We propose a parametric model for studying self-interacting dark matter (SIDM) halos. The model uses an analytical density profile, calibrated using a controlled N-body SIDM simulation that covers the entire gravothermal evolution, including core-forming and -collapsing phases. By normalizing the calibrated density profile, we obtain a universal description for SIDM halos at any evolution phase. The model allows us to infer properties of SIDM halos based on their cold dark matter (CDM) counterparts. As a basic application, we only require two characteristic parameters of an isolated CDM halo at $z=0$. We then extend the model to incorporate effects induced by halo mass changes, such as major mergers or tidal stripping, making it applicable to both isolated halos and subhalos. The parametric model is tested and validated using cosmological zoom-in SIDM simulations available in the literature.

}

\begin{document}

\maketitle

\date{\today}

\flushbottom

\tableofcontents

In self-interacting dark matter (SIDM), dark matter particles can interact with each other, in addition to their interactions through gravity; see refs.~\cite{Tulin170502358,Adhikari220710638} for reviews. The self-interactions can thermalize the inner region of dark matter halos and modify its structure~\cite{Spergel9909386}. An SIDM halo could experience gravothermal core-forming and -collapsing phases sequentially, and its inner density profile varies with time dynamically. Compared to collisionless cold dark matter (CDM), SIDM predicts more diverse dark matter distributions in {\it both} isolated halos and subhalos, see, e.g.,~\cite{Correa:2022dey,Yang:2022mxl}. In particular, the phenomenon of gravothermal collapse~\cite{Balberg0110561,Balberg:2002ue,Koda11013097} could be probed using observations of the Milky Way satellite galaxies~\cite{Nishikawa:2019lsc,Sameie:2019zfo,Zavala:2019sjk,Kaplinghat:2019svz,Kahlhoefer:2019oyt,Correa:2020qam,Turner:2020vlf, Slone:2021nqd,Silverman:2022bhs,Correa:2022dey,Yang:2022mxl} and strong gravitational lensing events~\cite{Minor:2020hic,Yang:2021kdf,Gilman:2021sdr,Gilman:2022ida,Nadler:2023nrd}. The collapsed central halo could also provide a seed for supermassive black holes~\cite{Pollack:2014rja,Choquette:2018lvq,Feng:2020kxv,Feng210811967,Xiao:2021ftk,Meshveliani:2022rih}.  
 
Cosmological N-body simulations are essential for studying the formation and evolution of SIDM halos in realistic environments, see, e.g.,~\cite{Vogelsberger:2012ku,Rocha:2012jg,Peter12083026,Zavala:2012us,Vogelsberger:2015gpr,Fitts:2018ycl,Robles:2019mfq,Banerjee:2019bjp,Sameie:2021ang,Nadler:2020ulu,Ebisu:2021bjh}. These simulations are often computationally expensive as they need a high resolution in order to resolve the central halo, where the dark matter collision rate is the highest. In some cases, controlled high-resolution N-body simulations can be used to study SIDM predictions, see, e.g.,~\cite{Elbert:2016dbb,Creasey:2017qxc,Sameie:2018chj,Huo:2019yhk,Yang:2020iya,Zeng:2021ldo,Burger:2022cjo,Ray:2022ydr}. A conducting fluid model is also broadly used~\cite{Balberg:2002ue,Koda11013097,Essig:2018pzq,Nishikawa:2019lsc,Feng:2020kxv,Yang:2022zkd,Yang:2023stn, Zhong:2023yzk}, as it can resolve the central region of a collapsed halo, but the model is based on an idealized setup. Another complication is that dark matter self-scattering is generally angular- and velocity-dependent~\cite{Feng09110422,Feng09053039,Tulin13023898,Kahlhoefer:2013dca,Agrawal11611004611,Robertson:2016qef,Kahlhoefer:2017umn,Kummer:2017bhr,Colquhoun:2020adl,Yang220503392,Girmohanta:2022dog}, and it is challenging to directly use these simulation tools in the search for entire particle parameter space. 

Ref.~\cite{Kaplinghat150803339} introduced a semi-analytical model for inferring SIDM density profiles from their CDM counterparts. It assumes that the dark matter distribution in the inner halo follows an isothermal distribution, and determines the distribution parameters through a matching procedure. The model has been tested and validated in both controlled and cosmological SIDM simulations with and without baryons~\cite{Elbert:2016dbb,Sameie:2018chj,Robertson:2020pxj,Jiang:2022aqw}. It has been used to analyze the rotation curves of a large sample of spiral galaxies~\cite{Kamada:2016euw,Ren180805695,2020MNRAS.495...58S,Zentner:2022xux}. However, the basic version of this model is only valid for isolated halos in the core-forming phase, but see \cite{Yang:2023stn}. 

In this work, we propose a parametric model that can ``transfer'' CDM halos into their SIDM counterparts for both core-forming and -collapsing phases, and it works for isolated halos {\it and} subhalos. Our model is based on two main observations. For velocity- and angular-dependent dark matter self-scattering, there is a constant effective cross section~\cite{Yang220503392,Yang:2022zkd,Outmezguine:2022bhq} that can dynamically incorporate the variation of the halo's mass and concentration over its evolution history. Furthermore, gravothermal evolution of an SIDM halo is largely self-similar~\cite{Balberg:2002ue,Balberg0110561,Koda11013097,Pollack:2014rja,Essig:2018pzq,Outmezguine:2022bhq,Zhong:2023yzk}. We will normalize halo evolution time with its collapse timescale; see also~\cite{Outmezguine:2022bhq,Zhong:2023yzk}. The rescaled evolution trajectories of characteristic halo parameters become universal, as they do not have an explicit dependence on the cross section. 

We will introduce an analytical density profile and calibrate it using a controlled N-body SIDM simulation from ref.~\cite{Yang220503392} for the entire evolution history. Then, we normalize the calibrated density profile and obtain a universal description for SIDM halos. We consider two approaches in applying the parametric model. In the basic application, we take density parameters of an isolated CDM halo at $z=0$ as input and reconstruct the evolution history of its SIDM counterpart. We extend the model to incorporate effects induced by halo mass changes, such as major mergers or tidal stripping, making it applicable to both isolated halos and subhalos. We will validate the parametric model using the cosmological zoom-in SIDM simulation of a Milky Way analog in ref.~\cite{Yang:2022mxl}. Since our model is accurate and computationally inexpensive, it can be used to efficiently scan large parameter space of particle physics models of SIDM. We will provide an example for such an exercise. 

The rest of the paper is organized as follows. In Sec.~\ref{sec:sim}, we discuss the simulation data used in this work. In Sec.~\ref{sec:cNFW}, we introduce the universal density profile and calibrate it with the controlled N-body SIDM simulation~\ref{sec:cNFW}. In Sec.~\ref{sec:application}, we discuss applications of the parametric model based on basic and integral approaches. We validate the parametric model with isolated halos and subhalos from the cosmological zoom-in simulation, in Secs.~\ref{sec:isolated} and ~\ref{sec:subs}, respectively. We present an example of using the parametric model to constrain particle physics parameters of SIDM in Sec.~\ref{sec:scan} and conclude in Sec.~\ref{sec:discussion}.

\section{Simulation data}
\label{sec:sim}

In this study, we use two types of simulation datasets from the literature. We first take the results from an isolated N-body SIDM simulation in ref.~\cite{Yang220503392} to calibrate a universal density profile over the course of gravothermal evolution, an essential ingredient of our parametric model. Then, we apply the model to a large population of isolated halos and subhalos from a cosmological zoom-in SIDM simulation of a Milky Way analog from ref.~\cite{Yang:2022mxl}, and show the parametric model can successfully predict their properties. Both simulations were performed using the~\textsc{Gadget2} program~\citep{Springel0505010}, with an SIDM module that adopts similar techniques described in refs.~\cite{Robertson:2016xjh,Rocha:2012jg}.

For calibration, we take simulated halos from ref.~\cite{Yang220503392} that have an NFW initial condition with the scale density $\rho_s = 2.74 \times 10^8~\rm M_{\odot}/kpc^3$ and scale density $r_s = 0.141~\rm kpc$. These halos are collectively named as ``BM2'' in ref.~\cite{Yang220503392}, and we will follow the same naming convention for convenience. The simulated halos contain four million particles, assuming various SIDM scenarios, such as Rutherford- and M{\o}ller-like scatterings, both having novel angular- and velocity-dependence, as well as isotropic scatterings with a constant cross section. Ref.~\cite{Yang220503392} proposed and tested a constant effective cross section to model the halo evolution for velocity- and angular-dependent SIDM interactions; see also refs.~\cite{Yang:2022zkd,Outmezguine:2022bhq}. For a differential cross section of $d \sigma/d \cos\theta$, the effective cross section is evaluated as
\begin{eqnarray}
\label{eq:eff}
\sigma_{\rm eff} &=& \frac{2 \int d v d \cos\theta \frac{d \sigma}{d \cos\theta} \sin^2\theta v^5 f_{\rm MB}(v,\nu_{\rm eff}) }{\int d v d \cos\theta \sin^2\theta v^5 f_{\rm MB}(v,\nu_{\rm eff}) } \\ \nonumber
&=& \frac{1}{512 \nu_{\rm eff}^8 } \int d v d\cos\theta \frac{d \sigma}{d\cos\theta} v^7 \sin^2\theta \exp \left[-\frac{v^2}{4\nu_{\rm eff}^2}\right],
\end{eqnarray}
where we have assumed that dark matter particles follow a Maxwell-Boltzmann velocity distribution
\begin{eqnarray}
f_{\rm MB} (v,\nu_{\rm eff}) \propto v^2 \exp \left[-\frac{v^2}{4\nu_{\rm eff}^2}\right], 
\end{eqnarray}
with $\nu_{\rm eff} =0.64 V_{\rm max, NFW} \approx 1.05 r_{\rm eff} \sqrt{G\rho_{\rm eff}}$ being a characteristic velocity dispersion of dark matter particles in the halo, 
$v$ is the relative velocity between the two incoming particles, and $\theta$ the polar angle that takes values in $[0,\pi]$. 
We have used the relations $\rho_{\rm eff} = (V_{\rm max}/(1.648 r_{\rm eff}))^2/G$ and $r_{\rm eff} = R_{\rm max}/2.1626$, where $V_{\rm max}$ is the maximal circular velocity of a halo and $R_{\rm max}$ its corresponding radius. For an NFW halo, $r_{\rm eff}$ and $\rho_{\rm eff}$ reduce to $r_s$ and $\rho_s$, respectively. In cosmological simulations, $V_{\rm max}$ and $R_{\rm max}$ are more useful for specifying a halo, as they do not rely on a specific form of density profiles.
The kernel $\sin^2\theta v^5$ in eq.~(\ref{eq:eff}) is motived by the evaluation of thermal conductivity in kinetic theory of a fluid in the short-mean-free-path regime. 
Although for the bulk of SIDM evolution, the inner halo is not in that regime, the kernel does provide an accurate weighting factor for velocity- and angular-dependent dark matter self-interactions as demonstrated using simulations~\cite{Yang220503392,Yang:2022zkd,Outmezguine:2022bhq}.

To be concrete, we will consider a Rutherford scattering benchmark in ref.~\cite{Yang220503392}, which assumes the BM2 initial NFW condition and the following differential cross section
\begin{eqnarray}
\label{eq:xsr}
\frac{d\sigma}{d \cos\theta} = \frac{\sigma_{0}w^4}{2\left[w^2+{v^{2}}\sin^2(\theta/2)\right]^2 },
\end{eqnarray}
where $\sigma_0/m=2.4\times10^4~{\rm cm^2/g}$ and $w=1~{\rm km/s}$. The corresponding effective cross section is $\sigma_{\rm eff}/m=7.1~{\rm cm^2/g}$. 
Ref.~\cite{Yang220503392} performed N-body simulations with the differential cross section in eq.~(\ref{eq:xsr}) and a constant cross section of $\sigma/m=10~{\rm cm^2/g}$.\footnote{The simulations assume all particles participate in scattering. See the relevant discussion in ref.~\cite{Yang220503392} for a self-consistent interpretation in terms of quantum statistics.} It further rescaled the simulation result of $\sigma/m=10~{\rm cm^2/g}$ and obtained the density profile at a given snapshot for $\sigma_{\rm eff}/m=7.1~{\rm cm^2/g}$. In this work, we will calibrate our parametric model using the density profile of $\sigma_{\rm eff}/m=7.1~{\rm cm^2/g}$ from the rescaling, as well as the simulated one based on the differential cross section in eq.~(\ref{eq:xsr}), and show they agree with each other well. Then we take the calibrated model and apply it to halos from cosmological simulations, as we discuss next.

The cosmological zoom-in simulation in ref.~\cite{Yang:2022mxl} contains a Milky Way analog and the differential cross section is given in eq.~(\ref{eq:xsr}), with $\sigma_0/m=147.1~\rm cm^2/g$ and $w=24.33~\rm km/s$. We consider simulated halos above a mass threshold of $10^8~{\rm M_{\odot}}/h$ and each halo contains more than $3500$ simulation particles. Applying this selection condition, we obtain $125$ and $115$ subhalos for CDM and SIDM, respectively. For isolated halos at $z=0$, we further require them to reside between $0.3\textup{--}3~{\rm Mpc}$ the center of the main halo so that these candidates do not suffer from numerical contamination. There are $647$ CDM and $626$ SIDM isolated halos. Among them, some are slingshot and splashback halos, which have experienced close encounters with other halos at low redshifts. In addition, $32$ subhalos and $58$ isolated halos have their collapse timescale less than $10~{\rm Gyr}$~\cite{Yang:2022mxl}.

\section{A universal density profile}
\label{sec:cNFW}

The underlying mechanism of our parametric model is that the evolution of SIDM halos exhibits universality~\cite{Outmezguine:2022bhq,Zhong:2023yzk}, i.e., after proper normalization, the evolution of the density profile does not explicitly depend on the initial condition {\it and} the self-scattering cross section, unless its central region is deeply collapsed. In this section, we first introduce an analytical density profile and determine its parameters by fitting to the simulated halo for a given snapshot. We then propose accurate fitting functions to smoothly model the evolution of the parameters. These functions are formulated without explicit dependence on the initial halo condition and the cross section, and hence they represent a universal solution to gravothermal evolution of dark matter halos. 

In SIDM, a halo first develops a shallow density core, and then the core size shrinks and central density increases. We introduce the following function to model the density profile of an isolated halo over the course of gravothermal evolution,
\begin{eqnarray}
\label{eq:cnfw}
\rho_{\rm SIDM}(r) = \frac{\rho_s}{\frac{\left(r^{\beta}+r_c^{\beta} \right)^{1/{\beta}}}{r_s} \left(1 + \frac{r}{r_s} \right)^2},  
\end{eqnarray}
where $\rho_s$ and $r_s$ are scale density and radius, respectively, and $r_c$ is the core size. The transition between inner core and outer NFW profile is controlled by $\beta$. We found that increasing $\beta$ from $1$ to $4$, the fit to the simulated halo can be improved significantly, and its remains almost unchanged for $\beta\geq 4$. We fix $\beta=4$, as in ref.~\cite{Robertson:2016qef}, where a similar trick was applied for halos with a Hernquist density profile. 
In the limit $r_c\to 0$, eq.~(\ref{eq:cnfw}) smoothly reduces to an NFW profile. The density profile in eq.~(\ref{eq:cnfw}) only applies to isolated halos. We will discuss subhalos in Sec.~\ref{sec:subs}. 
In Appendix~\ref{sec:appA}, we will discuss an alternative cored density profile from refs.~\cite{2016MNRAS.459.2573R,2016MNRAS.462.3628R}, which works as well as the one in eq.~(\ref{eq:cnfw}). 

\begin{figure*}[htbp]
  \centering
  \includegraphics[height=4.8cm]{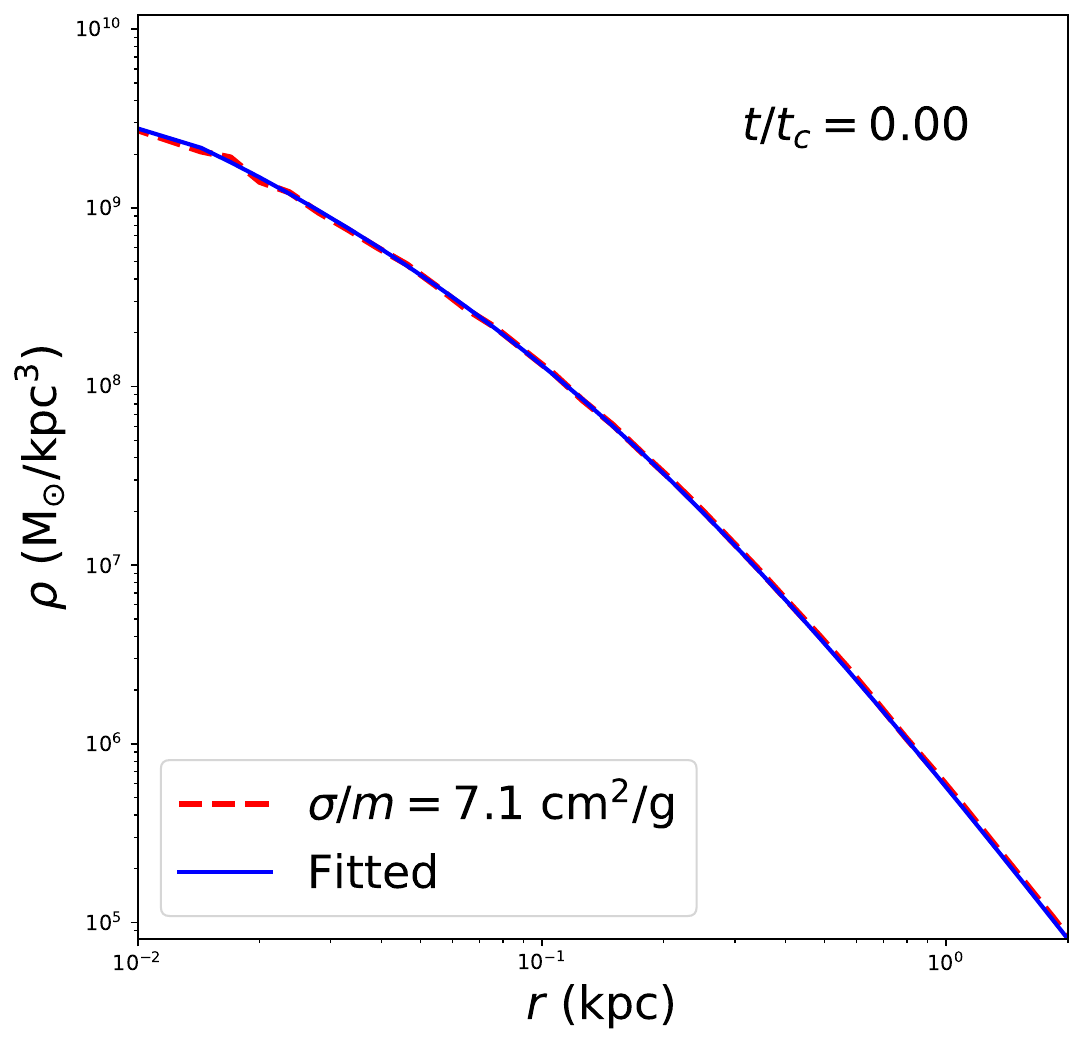}
  \includegraphics[height=4.8cm]{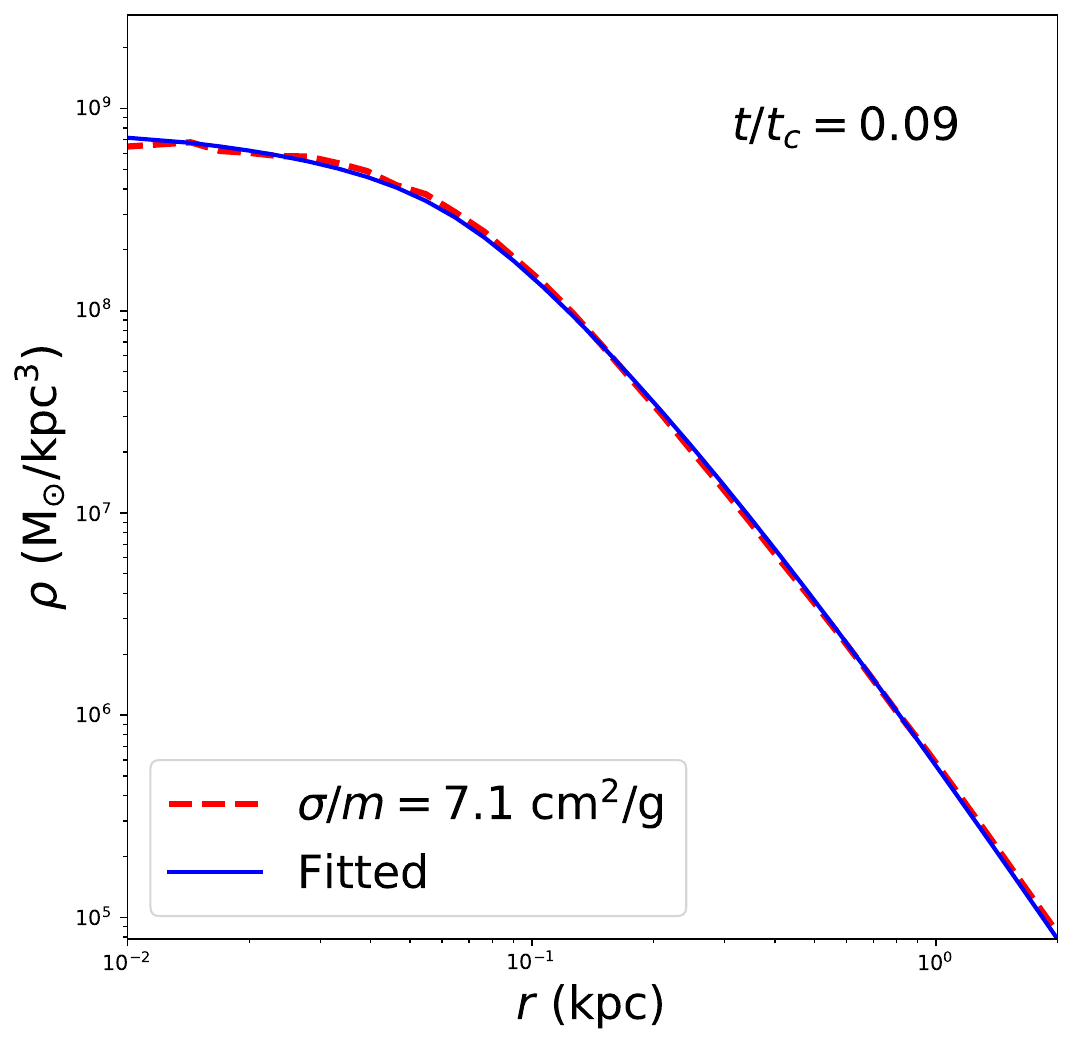}
  \includegraphics[height=4.8cm]{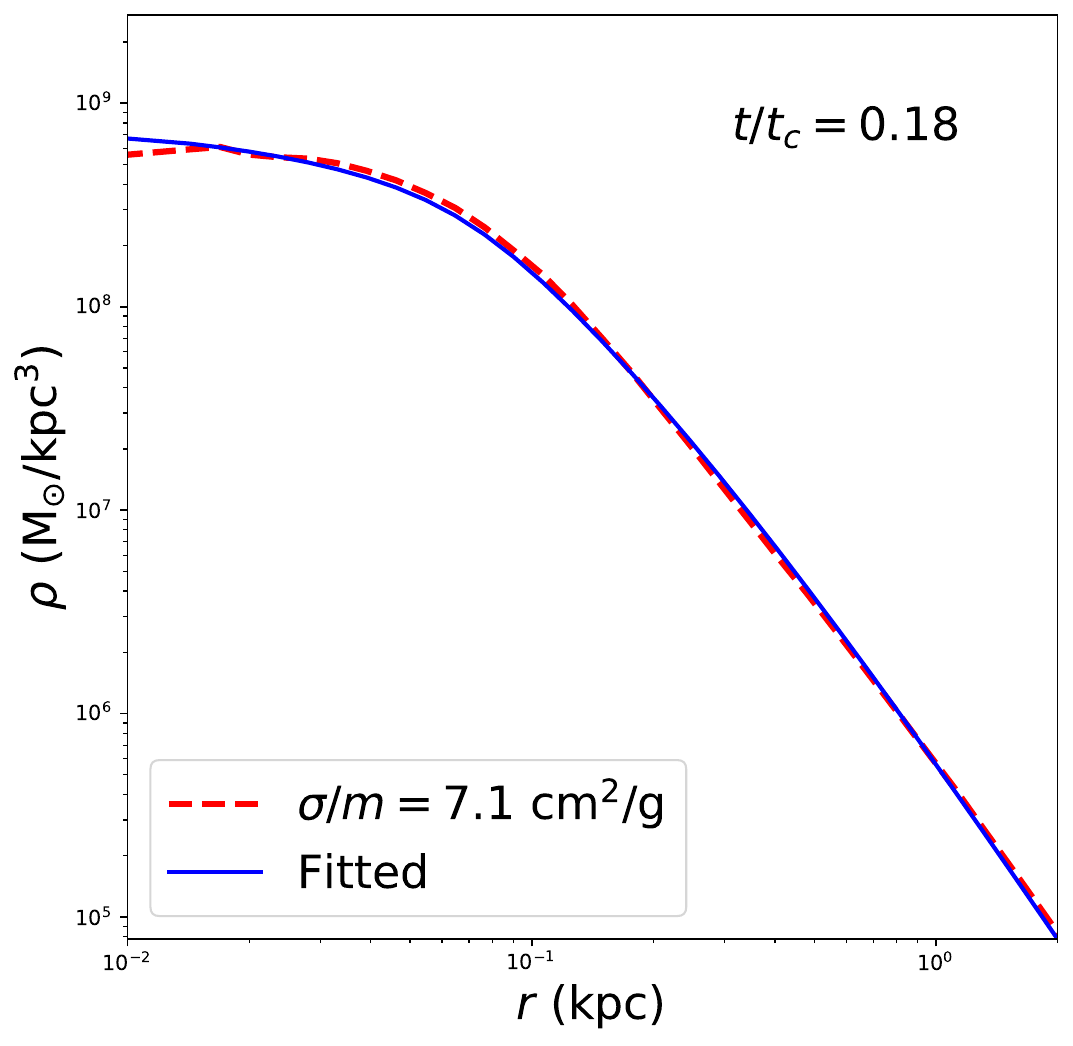}\\
  \includegraphics[height=4.8cm]{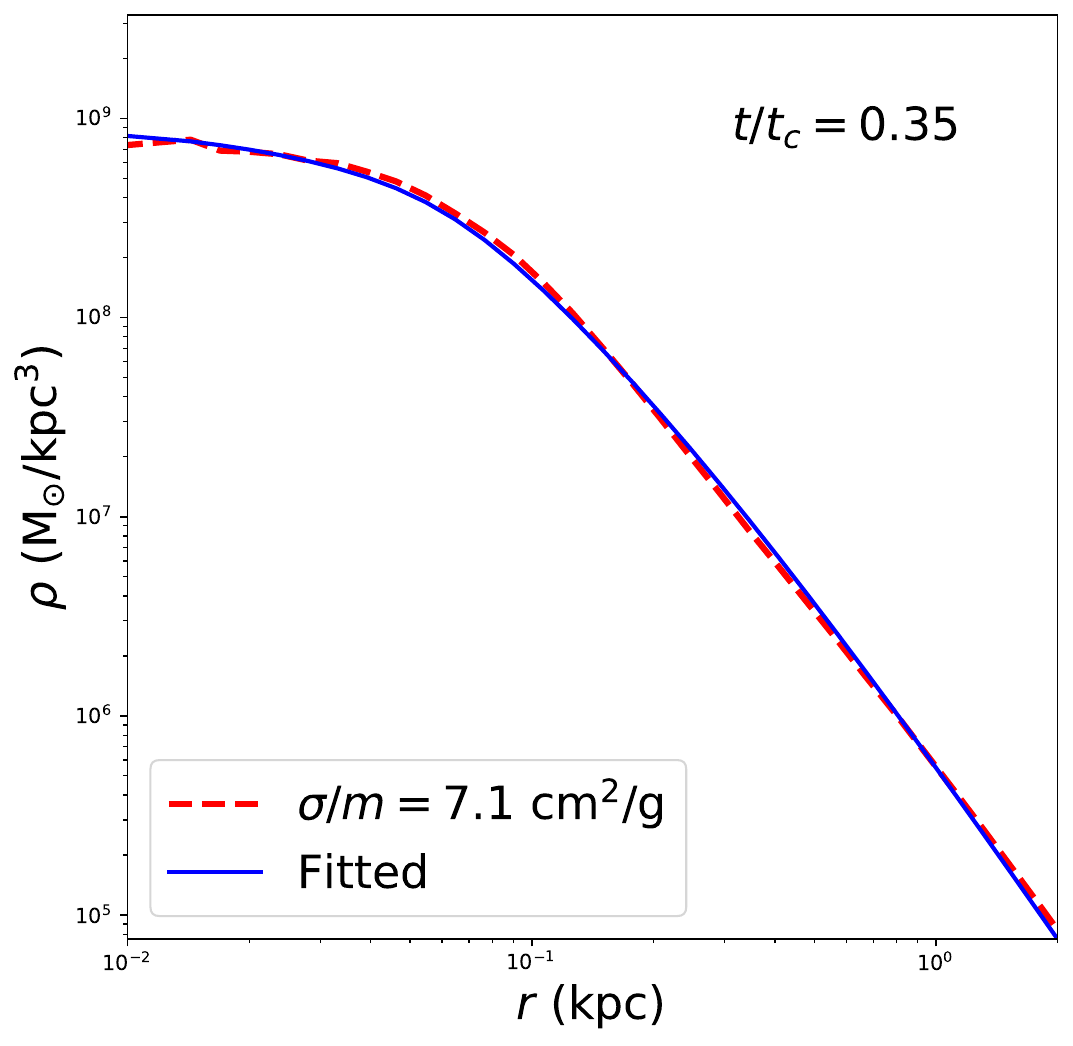}
  \includegraphics[height=4.8cm]{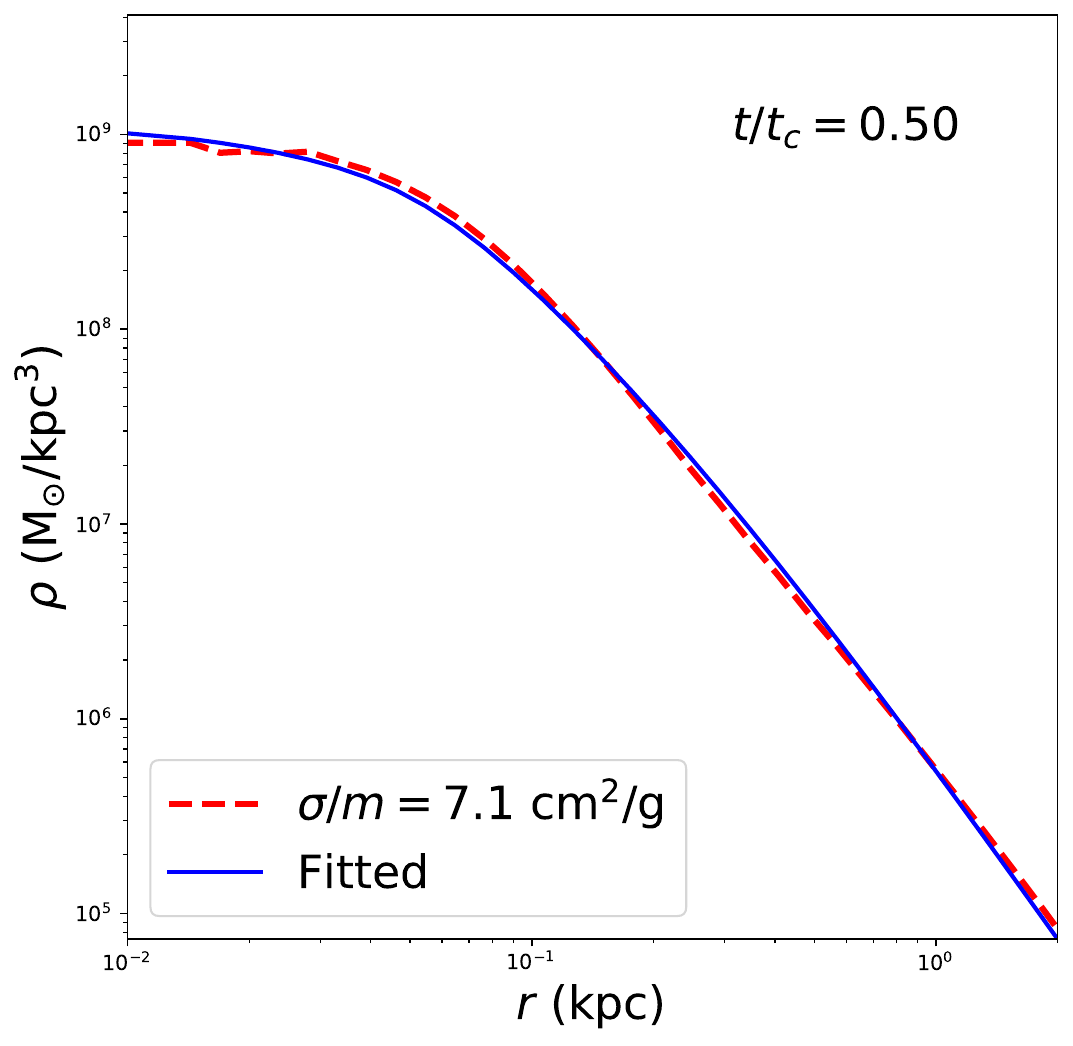}
  \includegraphics[height=4.8cm]{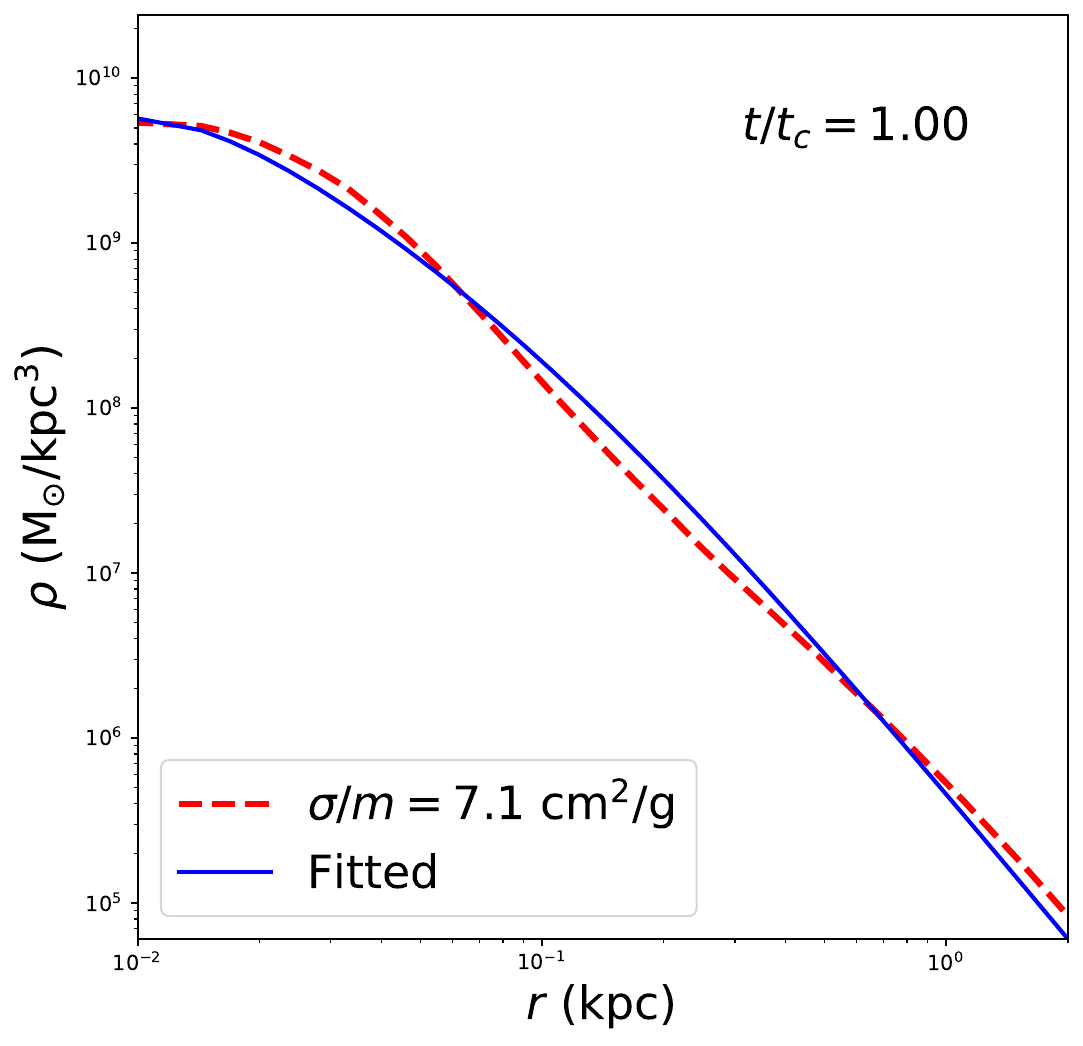}
  \caption{\label{fig:fitrho} Density profiles of the simulated BM2 halo at different times for an effective constant cross section of $\sigma_{\rm eff}/m=7.1~{\rm cm^2/g}$ (dashed-red); data from ref.~\cite{Yang220503392}. Each density profile is fitted using the density profile in eq.~(\ref{eq:cnfw}) (solid-blue). }
\end{figure*}

To develop a universal solution for parameterizing SIDM halo evolution, we need to use a dimensionless timescale and eliminate explicit dependence on the cross section. This can be achieved by normalizing the evolution time with the collapse timescale $t_c$, which is approximately a fixed multiple of the relaxation time. We estimate $t_c$ as~\cite{Balberg:2002ue,Pollack:2014rja,Essig:2018pzq} 
\begin{eqnarray}
\label{eq:tc0}
t_{\rm c}  &=& \frac{150}{C} \frac{1}{(\sigma_{\rm eff}/m) \rho_{\rm eff} r_{\rm eff}} \frac{1}{\sqrt{4\pi G \rho_{\rm eff}}},
\end{eqnarray}
where $C$ is constant that can be calibrated with N-body simulations~\cite{Koda11013097,Essig:2018pzq,Nishikawa:2019lsc,Yang:2022zkd}, and we fix $C=0.75$. For the BM2 halo we consider, $t_c\approx28.7~{\rm Gyr}$ for $\sigma_{\rm eff}=7.1~\rm cm^2/g$~\cite{Yang220503392}.

Figure~\ref{fig:fitrho} shows the simulated density profile of the BM2 halo at $t/t_{c} =0,\,0.09,\,0.18,\,0.25,$ $0.50,$ and $1$ ($\sigma_{\rm eff}/m=7.1~{\rm cm^2/g}$, dashed-red), as well as the one fitted using the density profile in eq.~(\ref{eq:cnfw}) (solid-blue). Our fit is performed over the entire range, upper to the virial radius, and the agreement is very good. Even at $t/t_c=1$, the agreement is within $10\%$. While introducing additional parameters could further improve the fit, we prioritize simplicity over sub-$10\%$ accuracy in our approach.

We further fit the density profile of the simulated BM2 halo at successive time intervals of $0.2~{\rm Gyr}$, and obtain the evolution trajectories for $r_s$, $\rho_s$ and $r_c$ in eq.~(\ref{eq:cnfw}) over the dimensionless time variable $\tau\equiv t/t_c$. They can be expressed as the following functions 
\begin{eqnarray}
\label{eq:m0}
\frac{\rho_s}{\rho_{s,0}} &=& 2.033 + 0.7381 \tau + 7.264 \tau^5 -12.73 \tau^7  + 9.915 \tau^9 + (1-2.033) (\ln 0.001)^{-1} \ln \left( \tau + 0.001 \right), \nonumber \\
\frac{r_s}{r_{s,0}} &=& 0.7178 - 0.1026 \tau +  0.2474 \tau^2 -0.4079 \tau^3 + (1-0.7178) (\ln 0.001)^{-1} \ln \left( \tau + 0.001 \right), \nonumber \\
\frac{r_c}{r_{s,0}} &=& 2.555 \sqrt{\tau} -3.632 \tau + 2.131 \tau^2 -1.415 \tau^3 + 0.4683 \tau^4,  
\end{eqnarray}
where the subscript $``0"$ denotes the corresponding value of the initial NFW profile. To enforce $\rho_s/\rho_{s,0}=1$ and $r_s/r_{s,0}=1$ at $\tau=0$, we have chosen a term of $\alpha + (1-\alpha) (\ln 0.001)^{-1} \ln \left( \tau + 0.001 \right)$ and $\alpha$ is the fitting parameter. The numerical factor $0.001$ inside the logarithm is introduced to avoid singularity at $\tau=0$. Although $t_c$ may vary slightly with a different calibration of $C$, the functional forms in eq.~(\ref{eq:m0}), including the coefficients, are robust to this variation. For the choice of terms in the fitting functions, we did it through trial and error with the assumption that the three halo parameters can be expressed in terms of polynomials of the normalized time, aside from the logarithmic term. We had included more terms with different powers of time, and then dropped those with small coefficients before performing the final fit. In this way, we achieve a balance between accuracy and simplicity of the fitting functions.

\begin{figure*}[tp]
  \centering
  \includegraphics[height=4.8cm]{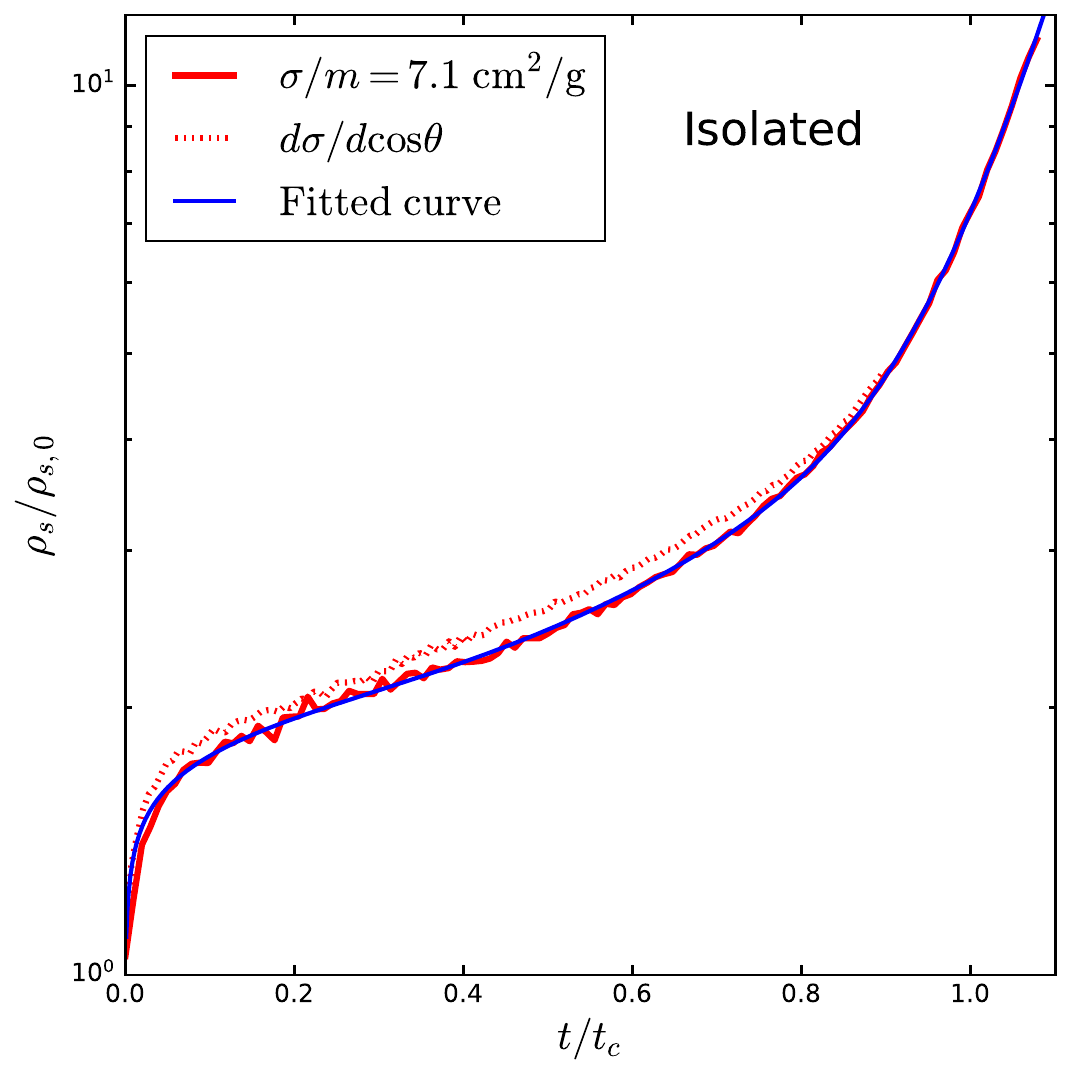}
  \includegraphics[height=4.8cm]{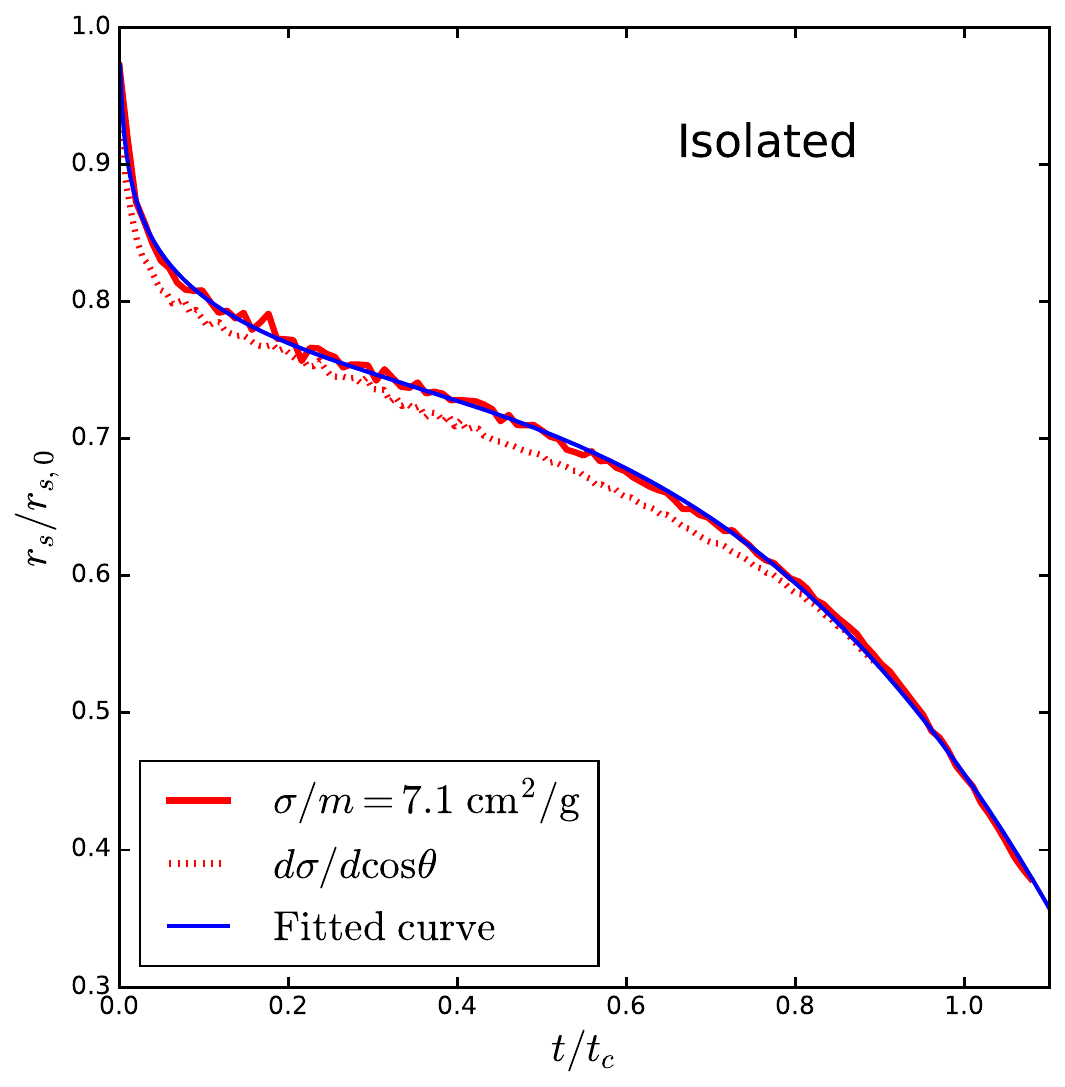}
  \includegraphics[height=4.8cm]{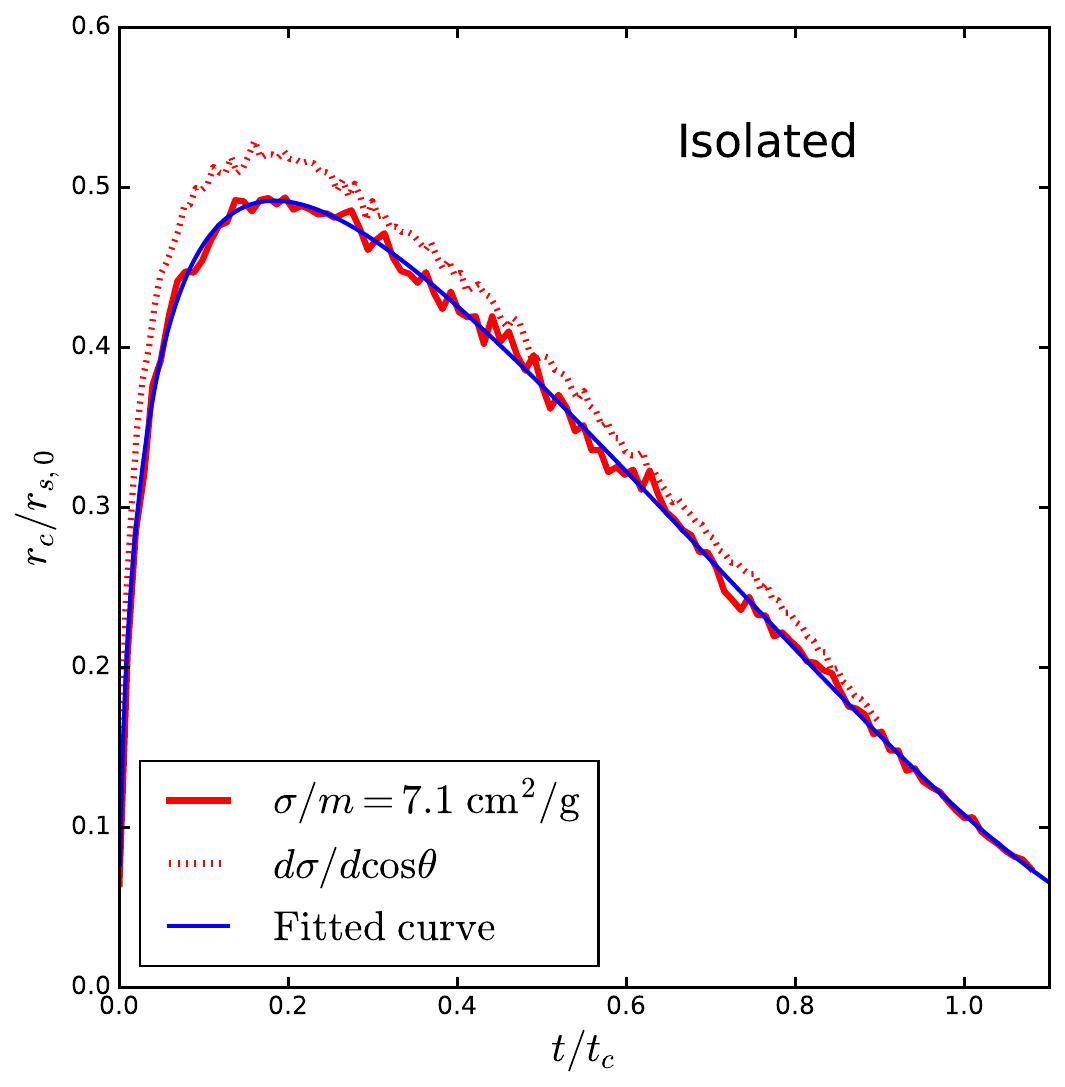}
  \caption{\label{fig:bm2a} Evolution of normalized density parameters $\rho_s/\rho_{s,0}$ (left), $r_s/r_{s,0}$ (middle), and $r_c/r_{s,0}$ (right) from the calibrated functions in eq.~(\ref{eq:m0}) (solid-blue) and the N-body SIDM simulation with the effective constant cross section $\sigma_{\rm eff}/m=7.1~{\rm cm^2/g}$~\cite{Yang220503392} (solid-red). For comparison, the SIDM simulation~\cite{Yang220503392} based on the differential scattering cross section $d\sigma/d\cos\theta$ is also shown (dotted-red). 
 }
\end{figure*}

In figure~\ref{fig:bm2a}, we present the evolution of $\rho_s/\rho_{s,0}$ (left), $r_s/r_{s,0}$ (middle), and $r_c/r_{c,0}$ (right). Our fitting functions in eq.~(\ref{eq:m0}) (solid-blue) well describe the simulation results ($\sigma_{\rm eff}/m = 7.1~\rm cm^2/g$, solid-red). We further show the results from the N-body simulation~\cite{Yang220503392} with the differential cross section ($d\sigma/d\cos\theta$, dotted-red), which agree with those from the simulation with the effective cross section $\sigma_{\rm eff}/m = 7.1~\rm cm^2/g$ (solid-red), within $\sim5\%$. The agreement further confirms that the effective constant cross section can accurately capture the halo evolution, setting the base for constructing a parametric model that can be applied to SIDM models with velocity- and angular-dependent dark matter self-interactions, as we will discuss later. Since ref.~\cite{Yang220503392} has more extended snapshots for the BM2 halo using $\sigma_{\rm eff}/m$ ($t\approx30~{\rm Gyr}$) than the one using the differential cross section $d\sigma/d\cos\theta$ ($t\approx25~{\rm Gyr}$), we will present our main results based on the parametric model calibrated with the former.

\begin{figure*}[htbp]
  \centering
  \includegraphics[height=4.8cm]{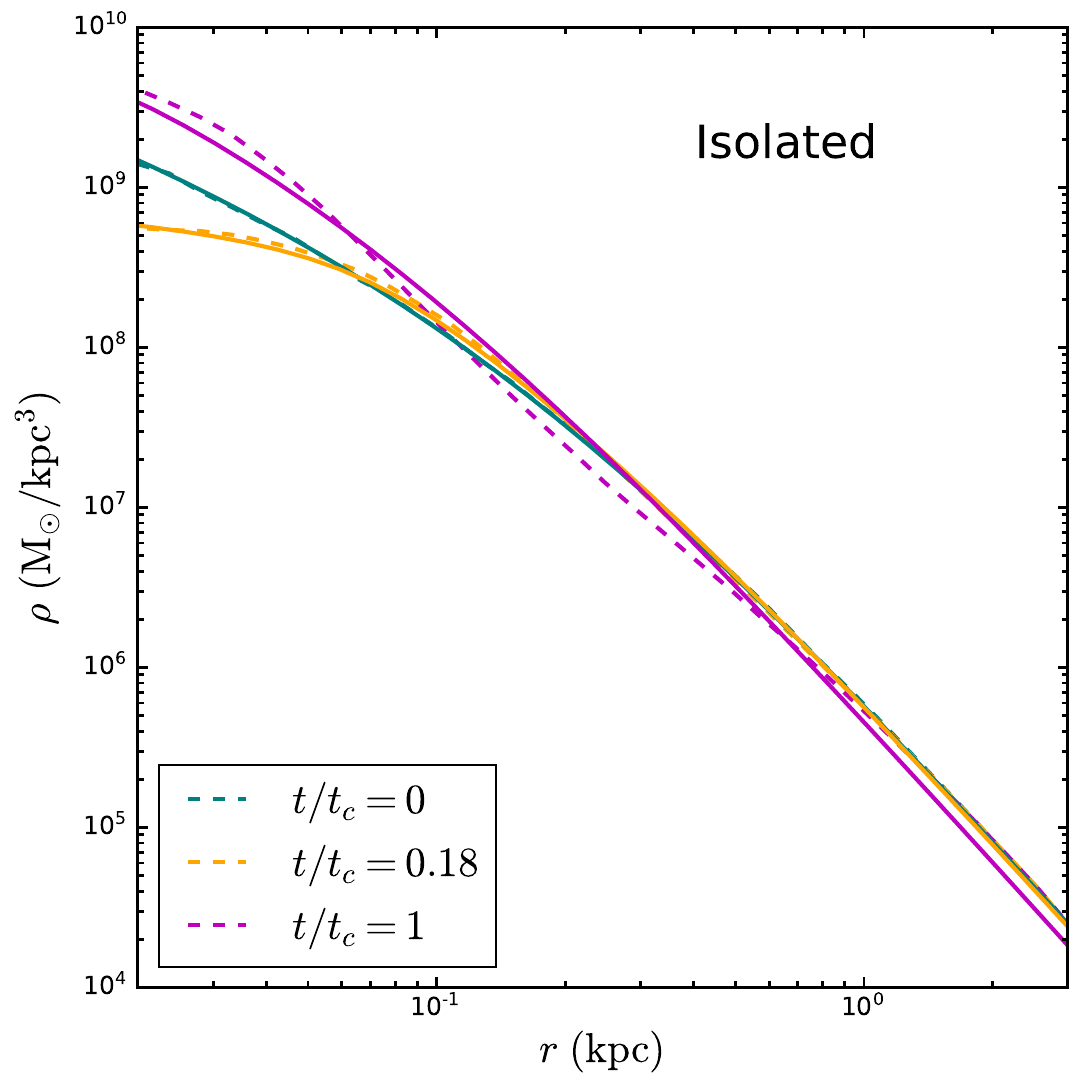}
  \includegraphics[height=4.8cm]{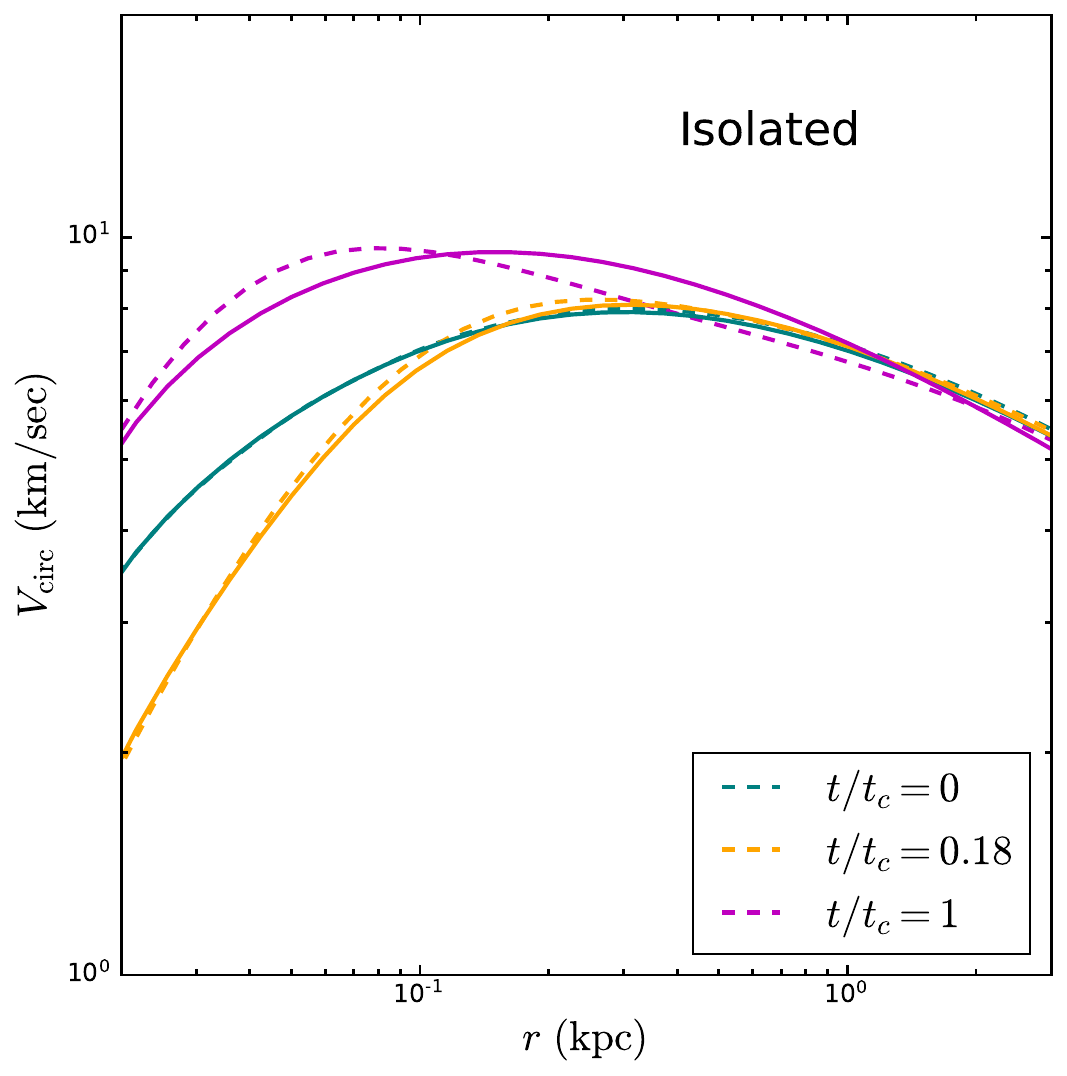}
  \includegraphics[height=4.8cm]{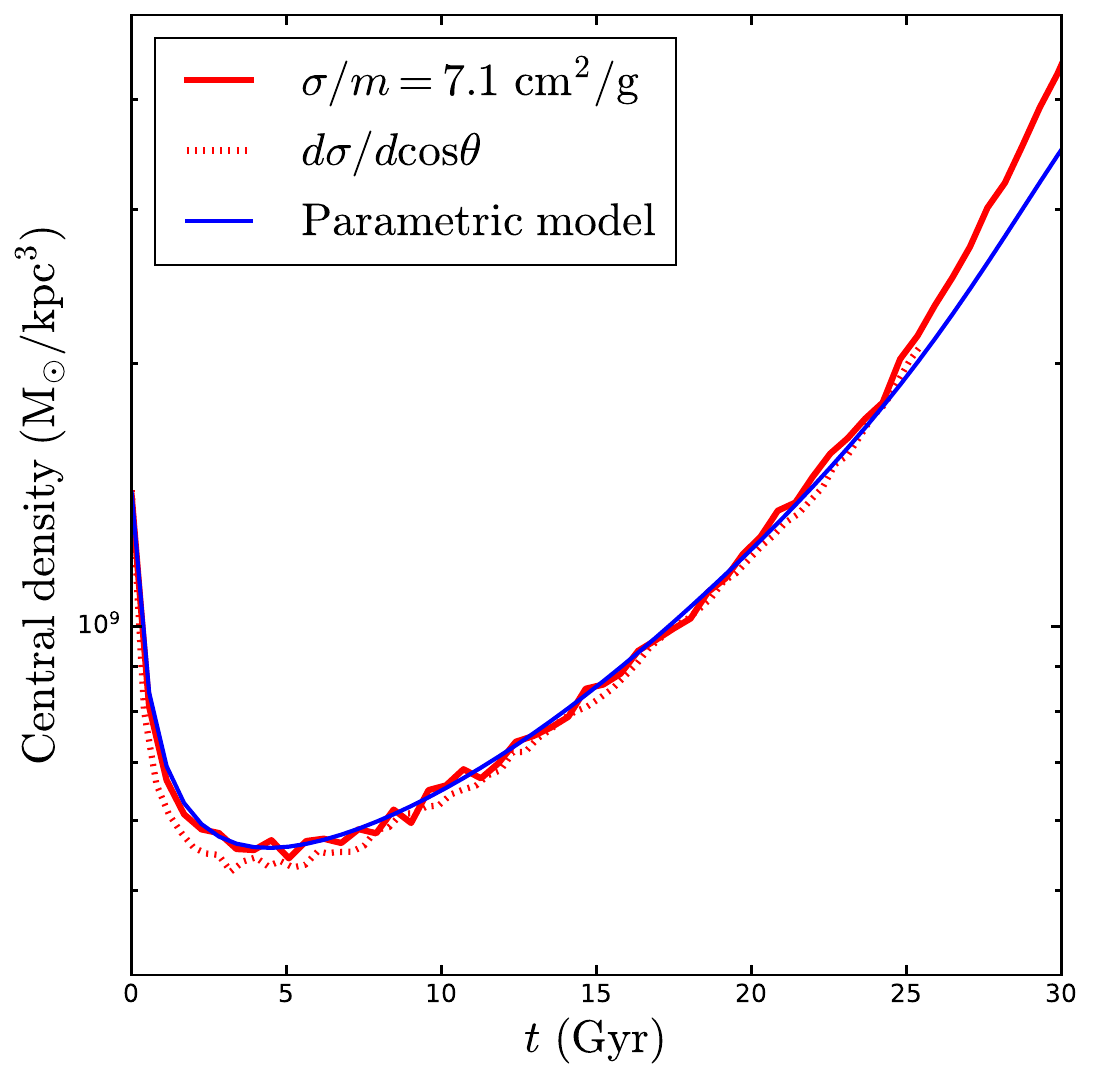}
  \caption{\label{fig:bm2b} Density (left) and circular velocity (middle) profiles from eq.~(\ref{eq:m0}) (solid) and the SIDM simulation with the effective cross section $\sigma_{\rm eff}/m = 7.1~\rm {cm^2/g}$~\cite{Yang220503392} (dashed) at $t/t_c=0,~0.18$ and $1$. The evolution curves of the central dark matter density (right) from eq.~(\ref{eq:m0}) (solid-blue) and the simulations~\cite{Yang220503392} with $\sigma_{\rm eff}/m = 7.1~\rm cm^2/g$ (solid-red) and $d\sigma/d\cos\theta$ (dashed-red) are also shown.
 }
\end{figure*}

In figure~\ref{fig:bm2b}, we show the density (left) and circular velocity (middle) profiles obtained from the N-body simulation (dashed) and our model in eq.~(\ref{eq:m0}) (solid) at three representative times, $t/t_c=0$, $0.18$, and $1$. The overall agreement is good, with only a minor discrepancy at $t/t_c\approx 1$, which is expected from the fit shown in figure~\ref{fig:fitrho}. Figure~\ref{fig:bm2b} (right) further shows the evolution of central densities from the N-body simulation with the effective cross section ($\sigma/m = 7.1~\rm cm^2/g$; solid-red) and the one with $d\sigma/d\cos\theta$ (dotted-red), as well as that obtained using eq.~(\ref{eq:m0}) (solid-blue). We see that they agree with each other very well up to $t/t_c=25/28.7\approx 0.87$, beyond which eq.~(\ref{eq:m0}) underestimates the central density by $\sim 15\%$ at $t/t_c=1$. 

We can use eq.~(\ref{eq:m0}) and obtain $V_{\rm max}$ and $R_{\rm max}$ from the reconstructed halo rotation curve at each snapshot. For the BM2 halo with $\sigma/m=7.1~{\rm cm^2/g}$, we find that the evolution trajectories of $V_{\rm max}$ and $R_{\rm max}$ can be well fitted by 
\begin{eqnarray}
\label{eq:m1}
\nonumber
\frac{V_{\rm max}}{V_{\rm max,0}} &=& 1+ 0.1777 \tau -4.399 \tau^3 + 16.66 \tau^4 - 18.87 \tau^5 + 9.077 \tau^7 - 2.436 \tau^9  \\ 
\frac{R_{\rm max}}{R_{\rm max,0}} &=& 1 + 0.007623 \tau - 0.7200 \tau^2 + 0.3376 \tau^3 -0.1375 \tau^4,  
\end{eqnarray}
where $\tau=t/t_c$ and the subscript $``0"$ denotes the corresponding value of the initial NFW profile. 

\section{Transforming CDM halos into SIDM halos} 
\label{sec:application}

To apply the functions in eq. (\ref{eq:m0}) for halos in cosmological simulations, we need to determine $\rho_{s,0}$, $r_{s,0}$, $t_c$, and evolution time after halo formation. In this section, we first introduce a basic approach for the parametric model that simply utilizes the corresponding CDM halo parameters at $z=0$ as input to make predictions for a given cross section specified in eq.~(\ref{eq:xsr}). We then present a more synthesized approach that makes use of the CDM halo properties over the evolution history, yielding more detailed and reliable SIDM predictions. To demonstrate the accuracy and effectiveness, we will take the CDM halos from the cosmological simulation in ref.~\cite{Yang:2022mxl}, use the parametric model to obtain their corresponding SIDM halos, which will be further compared to the SIDM halos directly from the N-body simulation.  

To facilitate discussion, we rank the simulated halos in ref.~\cite{Yang:2022mxl} by their masses in descending order. We identify the corresponding CDM counterpart for each simulated SIDM halo by examining the evolution history. We label these pairs as ``Cosmo-\#.'' For example, Cosmo-501 represents the halo ranked 501st from the top in terms of mass.

\subsection{The basic approach}
\label{sec:basic}

\begin{figure*}[htbp]
  \centering
  \includegraphics[height=4.8cm]{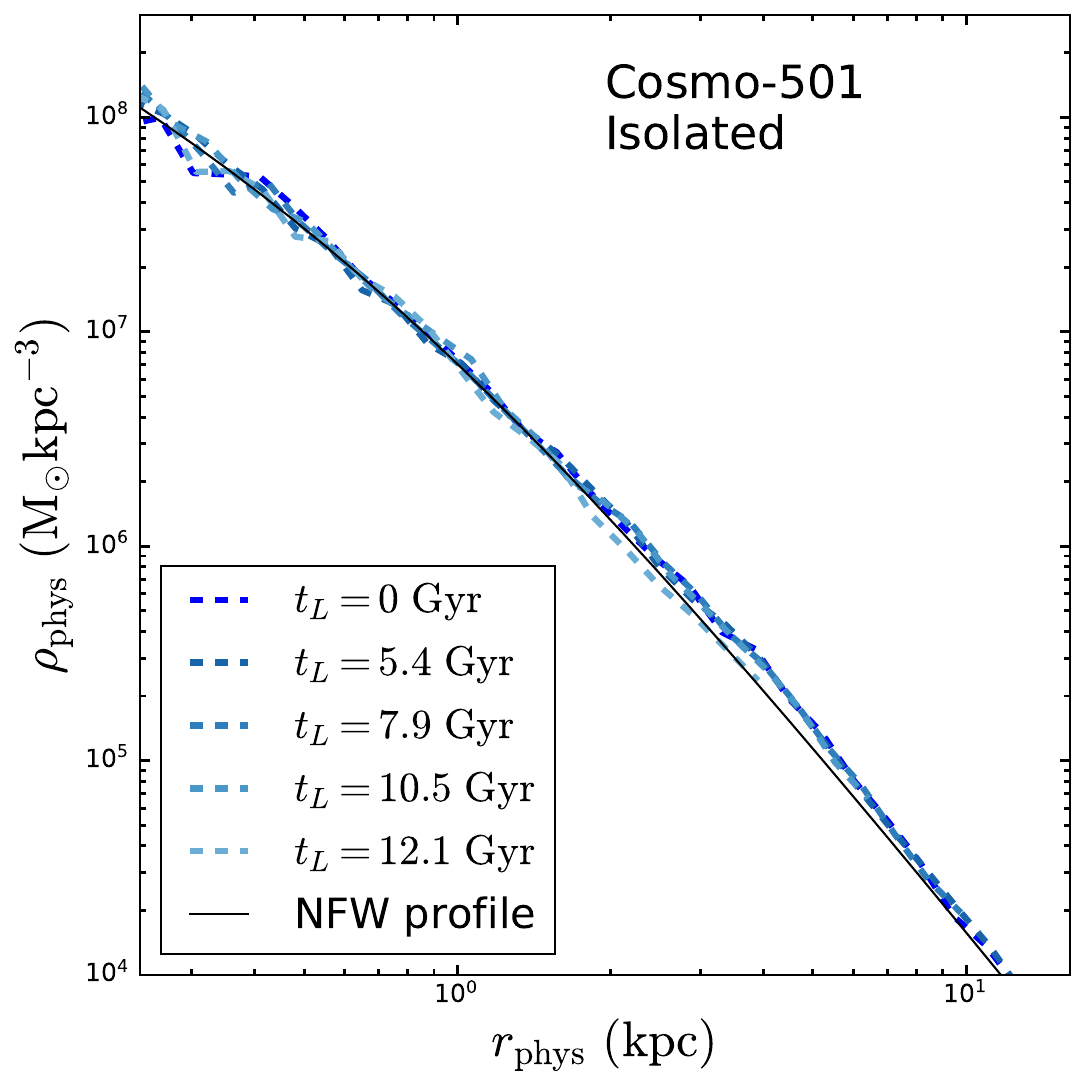}
  \includegraphics[height=4.8cm]{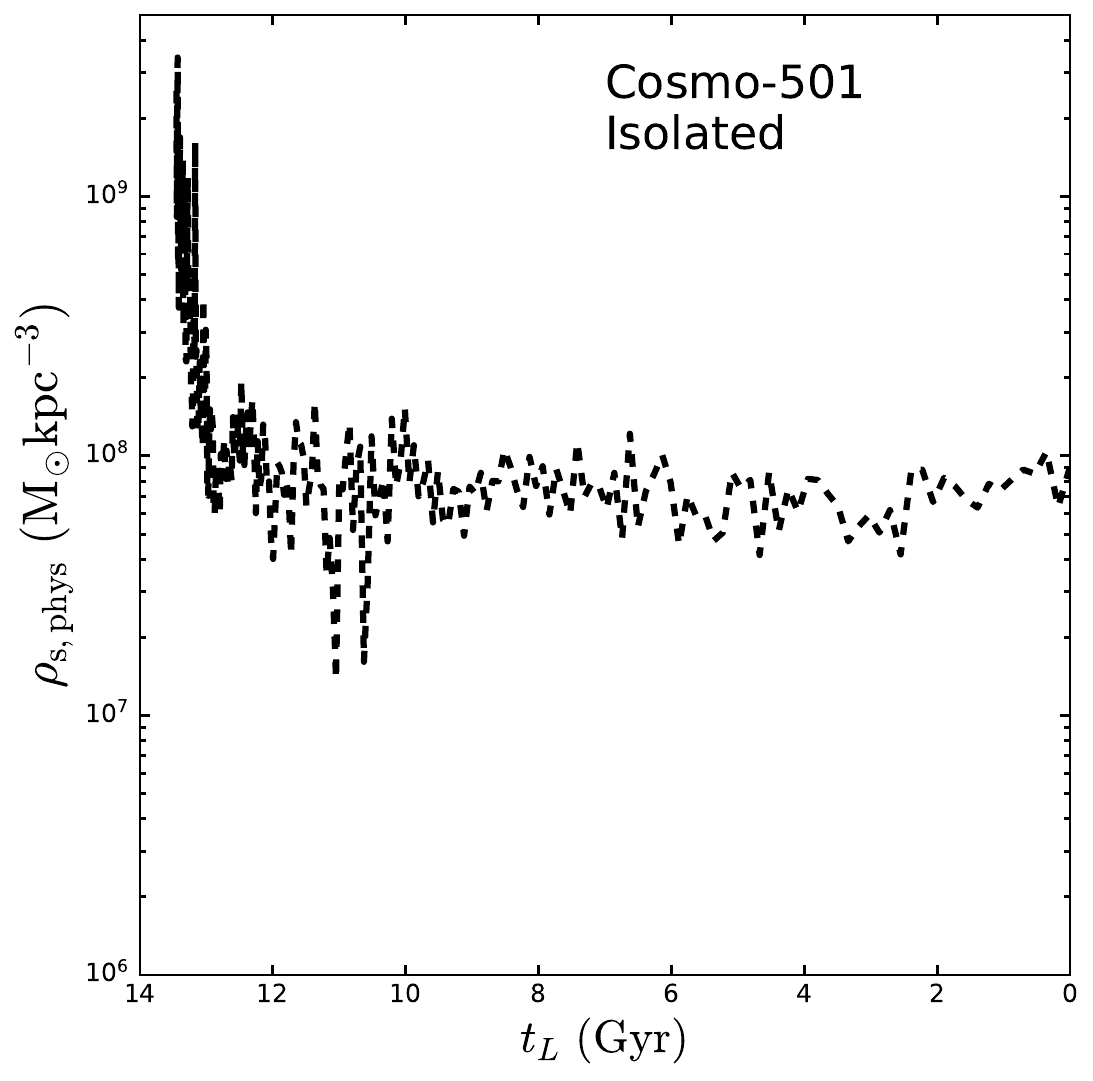} 
  \includegraphics[height=4.8cm]{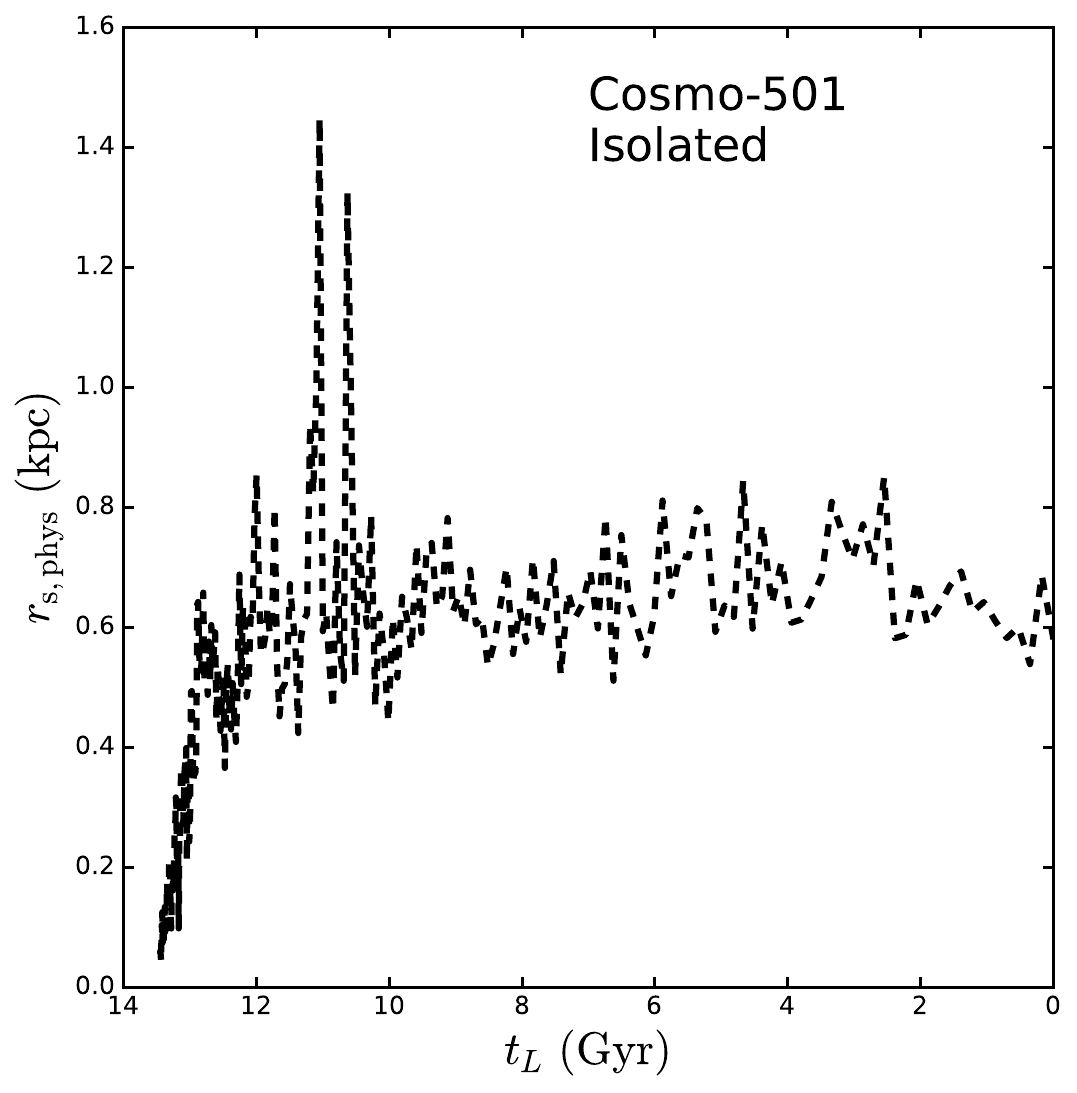} 
  \caption{\label{fig:test}
  Density profiles (left) of an isolated CDM halo at different lookback times, from the cosmological zoom-in simulation~\cite{Yang:2022mxl}. For $t_L\lesssim0.2~{\rm Gyr}$, the halo density can be described by a single NFW profile. Evolution of the NFW parameters $\rho_s$ (middle) and $r_s$ (right) constructed from the simulation~\cite{Yang:2022mxl}. The subscript ``phys'' denotes that the relevant quantities are evaluated in physical coordinates.  
}
\end{figure*}

We propose to use the NFW halo parameters $r_{s} = R_{\rm max}/2.1626$ and $\rho_{s} = (V_{\rm max}/(1.648 r_{s}))^2/G$ evaluated for simulated CDM halos at $z=0$ as input for the functions in eq. (\ref{eq:m0}). To justify this approach, we show in figure~\ref{fig:test} the evolution history of the density profile (left), scale density (middle) and radius (right) for the Cosmo-501 CDM halo in physical coordinates. For this halo, its inner NFW profile was established $10~{\rm Gyr}$ ago and has remained largely unchanged since then. In particular, $\rho_{s}$ and $r_s$ are approximately constants after first $3~{\rm Gyr}$. 

In general, isolated halos undergo exponential mass growth at high redshifts, during which their inner density profiles are established. At later stages, minor mergers primarily contribute mass to the outer regions, resulting in moderate mass increases, and the inner density profile remains largely unchanged. Thus we can use the NFW halo parameters at $z=0$ for the functions in eq. (\ref{eq:m0}) to make SIDM predictions, if we know the halo formation time $t_f$ at which the $\rho_{s}$ and $r_s$ become stable.

To estimate $t_f$, we use the following relation~\cite{Correa:2014xma},
\begin{eqnarray}
\label{eq:zf}
z_f = -0.0064 \left(\log_{10}\left(\frac{M_{\rm vir, 0}}{{10^{10}~\rm M_{\odot}}}\right) \right)^2 -0.1043 \log_{10} \left(\frac{M_{\rm vir, 0}}{{10^{10}~ \rm M_{\odot}}}\right) + 1.4807
\end{eqnarray}
where $z_f$ is the redshift of halo formation and $M_{\rm vir, 0}$ is the halo mass at present. The formation redshift is determined through the condition that at $z_f$ the halo mass is $M_{\rm vir, 0}/q$, where  $q\approx4.137 z^{-0.9476}_f$~\cite{Correa:2014xma}. The relation in eq.~(\ref{eq:zf}) is validated for halos in the mass range $10^8\textup{--}10^{15}~\rm M_{\odot}$~\cite{Correa:2014xma}, and it is applicable for our study as the halo masses of interest in the cosmological simulation from ref.~\cite{Yang:2022mxl} are $10^8\textup{--}10^{12}~{\rm M_\odot}$. 

The corresponding lookback time is computed as 
$t_{L}(z_f) = \int_{0}^{z_f} d z /(H\times(1+z))$, where $H$ is the Hubble rate as a function of $z$. For the cosmology with $h=0.7$, $\Omega_{m,0}=0.286$, and $\Omega_{\Lambda_0}=0.714$, as in ref.~\cite{2013ApJS..208...19H},
the lookback time can be evaluated as
\begin{eqnarray}
\label{eq:lookback}
t_{L}(z) = 13.647 - 11.020 \ln\left(\frac{1.5800}{(1 + z)^{1.5}} + \sqrt{1 + \frac{2.4965}{(1 + z)^3}}\right) \text{ Gyr}.
\end{eqnarray}
For convenience, we define $t_f = 13.647~{\rm Gyr} - t_{L}(z_f)$. 
For the Cosmo-501 isolated halo, $M_{\rm vir, 0}=1.81\times 10^9\rm ~M_{\odot}$, we have $z_f\approx 1.55$ and $t_{L}(z_f)\approx 9.48~{\rm Gyr}$ estimated using eqs. (\ref{eq:zf}) and (\ref{eq:lookback}), respectively. This is well consistent with the evolution history shown in figure~\ref{fig:test}.

In short, we take the following steps for applying the parametric model with the basic approach. For an isolated CDM halo, we take its NFW scale parameters at $z=0$, i.e., $\rho_{s,0}$ and $r_{s,0}$ in eq.~(\ref{eq:m0}), and calculate the virial mass and sequentially the formation time $t_L(z_f)$ using eqs.~(\ref{eq:zf}) and ~(\ref{eq:lookback}), which could also be obtained from halo catalogs if available. Then, we evaluate the effective cross section eq.~(\ref{eq:eff}) and collapse time eq.~(\ref{eq:tc0}) for a given particle physics model of SIDM, see eq.~(\ref{eq:xsr}). Lastly, we use the functions in eq. (\ref{eq:m0}) and obtain $\rho_s$, $r_s$ and $r_c$ as a function of time $t$, and reconstruct the density profile at $t$ based on eq.~(\ref{eq:cnfw}). In practice, we also incorporate a Gaussian scatter of $0.16$ dex in $\log_{10} (t_L(z_f)/{\rm Gyr})$~\cite{Giocoli:2011hz}, when applying to a population of halos. If $t_L(z_f)/t_c>1$, we truncate the halo's evolution at $t_c$ to avoid extrapolation. This truncation only affects the value of the central density of halos that would be deeply collapsed.

\subsection{The integral approach}
\label{sec:integral}

The basic approach is straightforward, but it has limitations when applied to halos that have undergone significant mass accretion or loss at relatively late times. Isolated halos, for instance, can experience mass increases due to minor mergers, or mass loss due to slingshot and splashback events, which cause spikes and valleys in $V_{\rm max}$ and $R_{\rm max}$. Furthermore, subhalos evolve in the tidal field and lose mass. Here, we propose an extended approach that can effectively ``integrate'' the SIDM effects over the period of mass growth or reduction.

We assume the mass accretion history in SIDM is similar to that in CDM. This is a reasonable assumption for most cases, as the self-interactions mainly modify the inner halo structure; we will discuss later about subhalos where the assumption is violated.  Given the evolution of $V_{\rm max,CDM}$ and $R_{\rm max,CDM}$, we propose the following integral functions
\begin{eqnarray}
\label{eq:int}
\nonumber 
V_{\rm max}(t)    &=&  V_{\rm max, CDM}(t_f) + \int_{t_f}^{t} d t'  \frac{d V_{\rm max,CDM}(t')}{d t'} +  \int_{t_f}^{t} \frac{dt'}{t_c(t')} \frac{d V_{\rm max, Model} (\tau') }{d \tau'} \\  
R_{\rm max}(t)    &=&  R_{\rm max, CDM}(t_f) + \int_{t_f}^{t} d t'  \frac{d R_{\rm max,CDM}(t')}{d t'} + \int_{t_f}^{t} \frac{dt'}{t_c(t')} \frac{d R_{\rm max, Model} (\tau')}{d \tau'}, 
\end{eqnarray}
where $\tau'=t'/t_c(t')$. The first integral on the right-hand side accommodates for the change due to the evolution events, such as mergers and tidal mass loss, while the second integral captures the SIDM effects by accounting for variations of the effective cross section and the collapse timescale, resulting from the change in the halo properties. The $d V_{\rm max, Model} (\tau) /d \tau$ and $d R_{\rm max, Model} (\tau) /d \tau$ terms are derivatives of the corresponding functions in eq.~(\ref{eq:m1}), i.e.,
\begin{eqnarray}
\label{eq:del2}
\nonumber 
\frac{1}{V_{\rm max, CDM}(t)} \frac{d V_{\rm max, Model} (\tau) }{d \tau} &=& 0.1777 - 13.20 \tau^2 + 66.62 \tau^3 - 94.34 \tau^4 + 63.54 \tau^6 - 21.93 \tau^8  \\ \nonumber
\frac{1}{R_{\rm max, CDM}(t)} \frac{d R_{\rm max, Model} (\tau) }{d \tau} &=& 0.007623 - 1.440 \tau + 1.013 \tau^2 - 0.5502 \tau^3,
\end{eqnarray}
where the $V_{\rm max, 0}$ and $R_{\rm max, 0}$ in eq.~(\ref{eq:m1}) have been replaced by $V_{\rm max, CDM}(t)$ and $R_{\rm max, CDM}(t)$, respectively. 

At each moment, the collapse time $t_c(t')$ is computed using $V_{\rm max,CDM}(t')$ and $R_{\rm max,CDM}(t')$. Once obtaining $V_{\rm max}(t)$, $R_{\rm max}(t)$ and $\tau=(t_L(z_f) - t_L(t))/t_c(t)$, we further use eq. (\ref{eq:m1}) to solve for $V_{\rm max,0}$ and $R_{\rm max,0}$, which in turn give rise to $\rho_{s,0}$ and $r_{s,0}$ assuming an NFW density profile. In this way, we ``construct'' an isolated CDM halo at a given moment $t$, to which we can apply the functions in eq.~(\ref{eq:m0}) for obtaining $\rho_s$, $r_s$ and $r_c$. Note that in the integral approach, $\rho_{s,0}$ and $r_{s,0}$ in eq.~(\ref{eq:m0}) are not necessary their corresponding values found in the simulation at $z=0$. With the effects of mass changes incorporated, we can apply the model to the halo earlier than the formation time used in the basic approach. For the integral approach, we take it to be half of $13.647~{\rm Gyr} - t_{L}(z_f)$, see eq.~(\ref{eq:lookback}), although the difference is small. In practice, we discretize the lookback time for a simulated halo into $1000$ intervals with equal spacing and obtain its $V_{\rm max}$ and $R_{\rm max}$ at all times incrementally. Then, we replace the integrals in eq.~(\ref{eq:int}) with discrete summation. 

The proposed integrals in eq.~(\ref{eq:int}) can be understood as follows. For a CDM halo, at each small time interval, the change in the mass due to accretion or tidal stripping, is reflected in the change of $\rho_{s}$ and $r_s$. The effective cross section eq.~(\ref{eq:eff}) with the velocity dispersion $\nu_{\rm eff}$, as well as the collapse time $t_c$, being computed using the obtained $\rho_{s}$ and $r_s$ values, effectively incorporates the SIDM effects during this time interval. Thus, the integral approach in eq.~(\ref{eq:int}) gives rise to the net change in $V_{\rm max}$ and $R_{\rm max}$ during the time interval. If the evolution history of CDM halos is known, the integral approach is easy to implement and it has broader applications than the basic approach.  

We also note that the integral approach becomes the basic one in the limit that $V_{\rm max, CDM}$ and $R_{\rm max, CDM}$ do not change after the halo formation time $t_f$. From eq.~(\ref{eq:int}), we see that in the limit the first integral vanishes, and the second integral simply reduces to two boundary terms at $t$ and $t_f$ as $t_c$ is a constant. The latter term carries a minus sign and it cancels the first term in the right-hand side of eq.~(\ref{eq:int}).

\section{Modeling the evolution of SIDM isolated halos} 
\label{sec:isolated}

We first consider a few representative halos from the cosmological simulation~\cite{Yang:2022mxl} that have evolved into late stages; see table~\ref{tab2} for their characteristic properties. Among them, Cosmo-26 and Cosmo-32 are core-forming, and both have $t_L(z_f)/t_c\ll 0.2$. All others are deeply collapsed at $z=0$, with $t_L(z_f)/t_c\gtrsim1$, and they will be truncated at $t_L(z_f)/t_c=1$. Cosmo-796 and Cosmo-1128 have relatively complex accretion histories. The former had a late merger and the latter experienced a splashback event (i.e., it passed through a larger halo before $z=0$).

\begin{table}[bthp]
\begin{center}
\begin{footnotesize}
\begin{tabular}{l|ccccc|cc}
\hline
\hline
Cosmo-ID & $M_{\rm vir,0} $  & $t_L(z_f) $ & $t_L(z_f)/t_c$ & $V_{\rm max,SIDM,0}$ & $R_{\rm max,SIDM,0}$
                                  & $V_{\rm max,Model,0}$         & $R_{\rm max,Model,0}$ \\
Isolated                &   $\rm (10^8\ M_{\odot})$    & $\rm (Gyr)$ & - & $\rm (km/s)$ & $\rm (kpc)$  & $\rm (km/s)$ & $\rm (kpc)$ \\
\hline
26   &  $186$   & 9.23 & 0.035 & 45.4 & 9.11 & 45.2 & 9.27 \\
32   &  $146$   & 9.26 & 0.018 & 40.7 & 10.4 & 40.4 & 12.6 \\
501  &  $6.47$  & 9.58 & 1.06 & 21.9 & 0.63 & 22.8 & 0.61 \\
796  &  $3.92$  & 9.62 & 1.00 & 21.1 & 0.56 & 21.7 & 0.61 \\
800  &  $4.03$  & 9.62 & 1.17 & 21.7 & 0.40 & 21.7 & 0.57 \\
1128 &  $2.64$  & 9.65 & 1.27 & 19.6 & 0.60 & 20.9 & 0.53 \\
\hline
\hline
\end{tabular}
\caption{\label{tab2} Characteristic properties of isolated halos at $z=0$, selected from the cosmological zoom-in SIDM simulation~\cite{Yang:2022mxl} for testing the parametric model. 
The virial mass, formation time, $t_L(z_f)/t_c$, $V_{\rm max}$ and $R_{\rm max}$ are reported for the halos. For comparison, their $V_{\rm max}$ and $R_{\rm max}$ values predicted in the parametric model eq.~(\ref{eq:m1}) are also listed in the last two columns.  
}
\end{footnotesize}
\end{center}
\end{table}

\begin{figure*}[htbp]
  \centering
  \includegraphics[height=4.8cm]{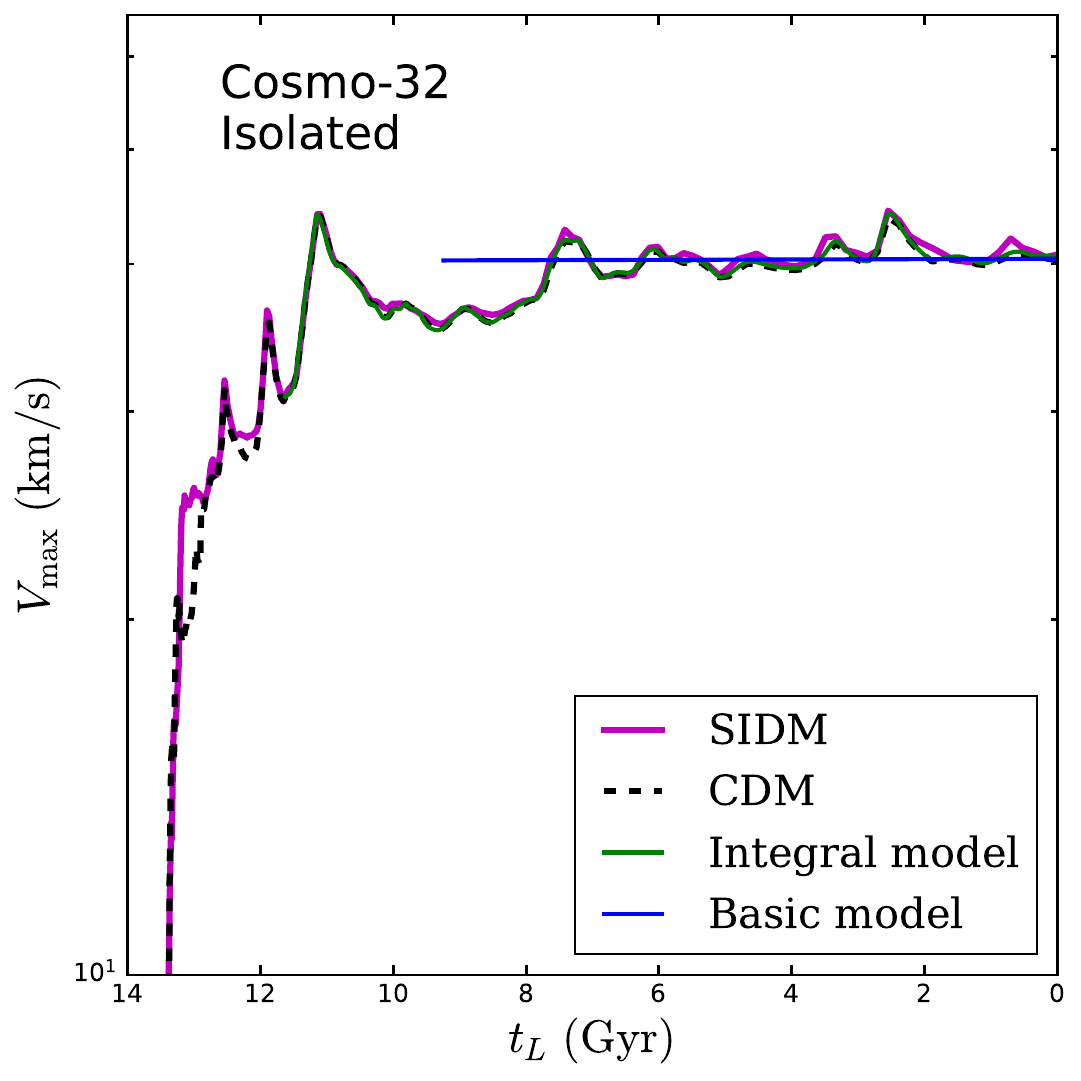}
  \includegraphics[height=4.8cm]{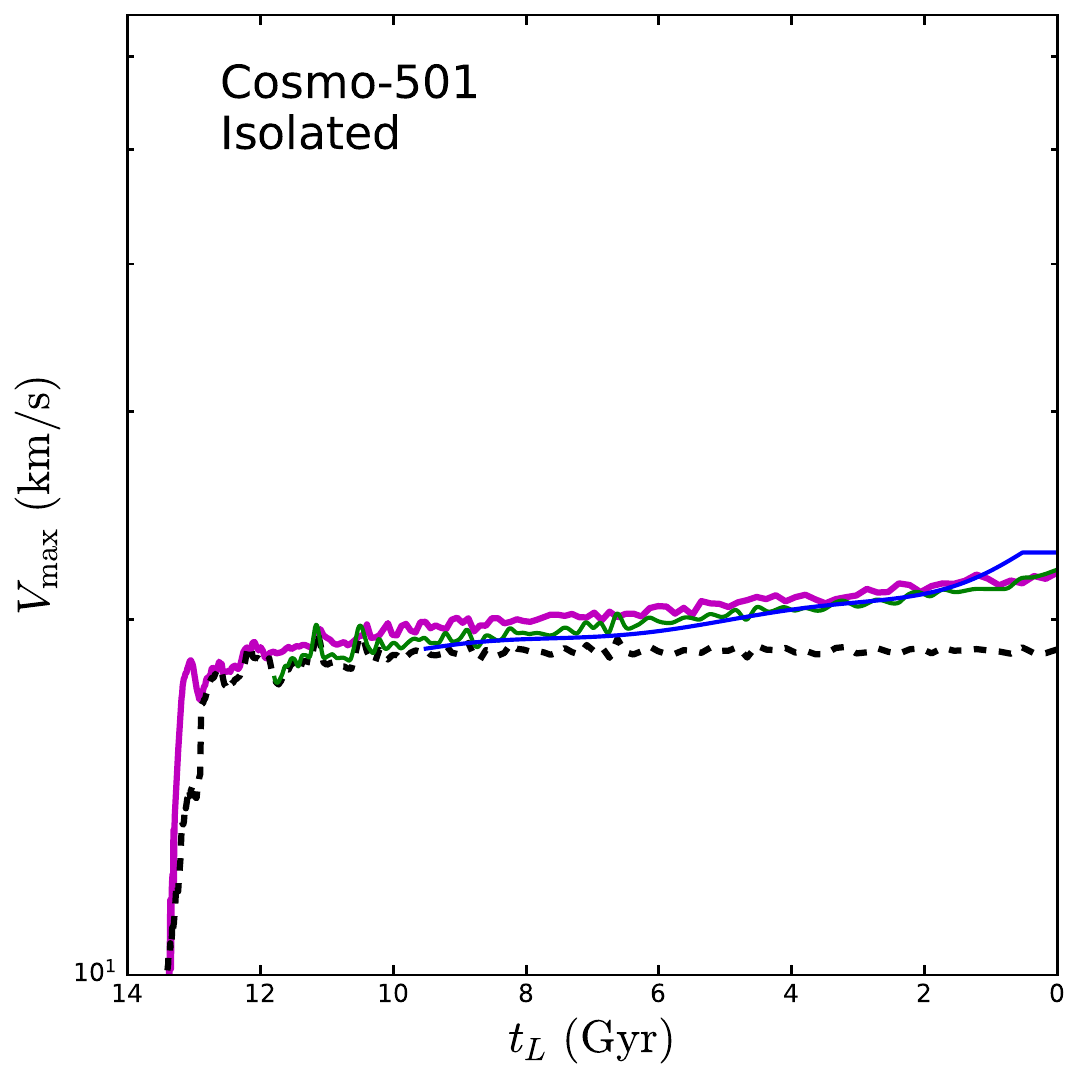}
  \includegraphics[height=4.8cm]{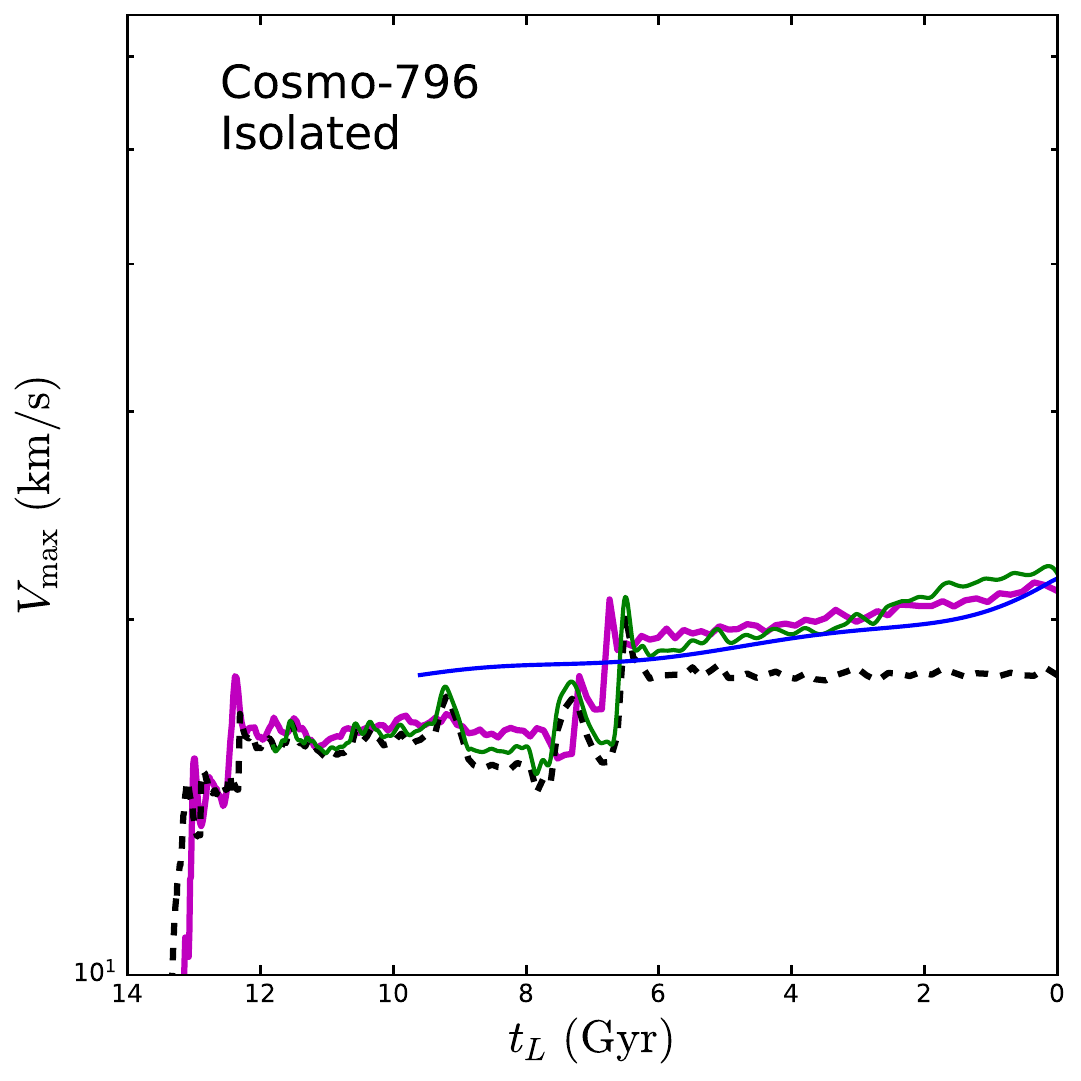} \\
  \includegraphics[height=4.8cm]{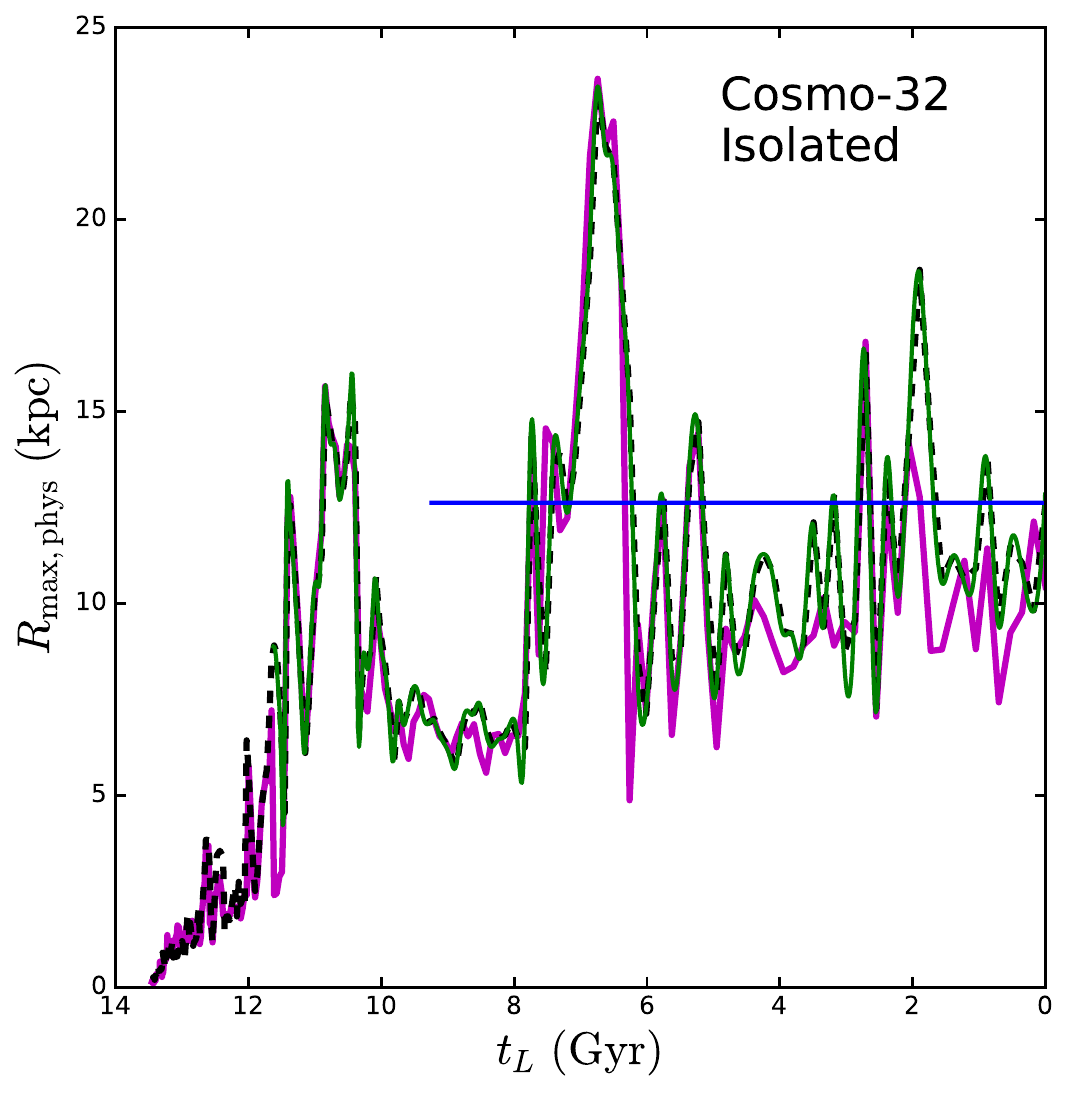}
  \includegraphics[height=4.8cm]{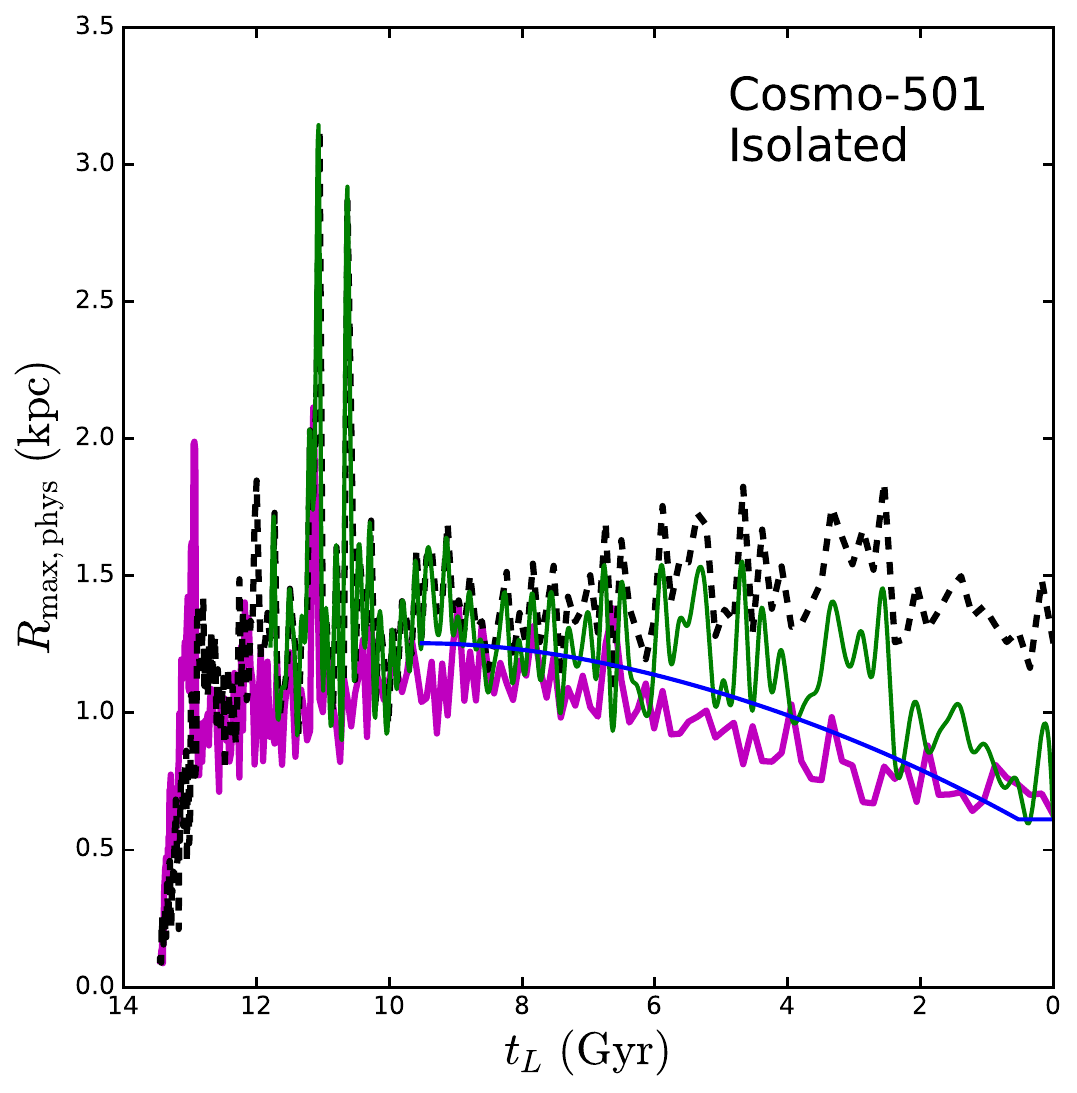}
  \includegraphics[height=4.8cm]{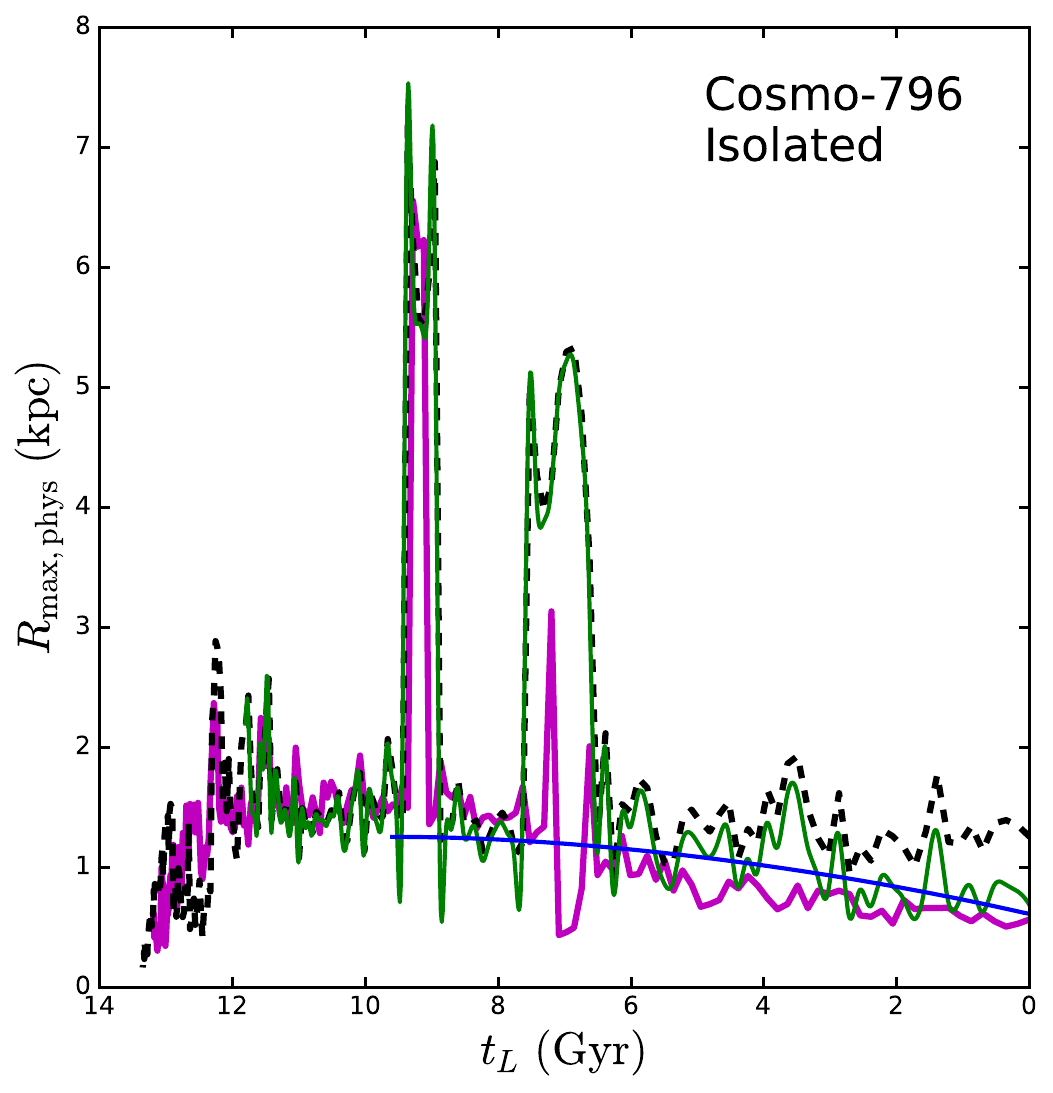}
  \caption{\label{fig:isoBM1} Evolution of $V_{\rm max}$ (top) and $R_{\rm max}$ (bottom) for three isolated halos from the parametric model with the basic (solid-blue) and integral (solid-green) approaches, as well as the N-body simulation~\cite{Yang:2022mxl} (solid-magenta). Cosmo-32 is in the core-forming phase at $z=0$, while the other two are deeply collapsed. Cosmo-796 has a late major merger at $t_L\sim7~{\rm Gyr}$. The subscript ``phys'' denotes that the relevant quantities are evaluated in physical coordinates. }
\end{figure*}

\begin{figure*}[htbp]
  \centering
  \includegraphics[height=4.8cm]{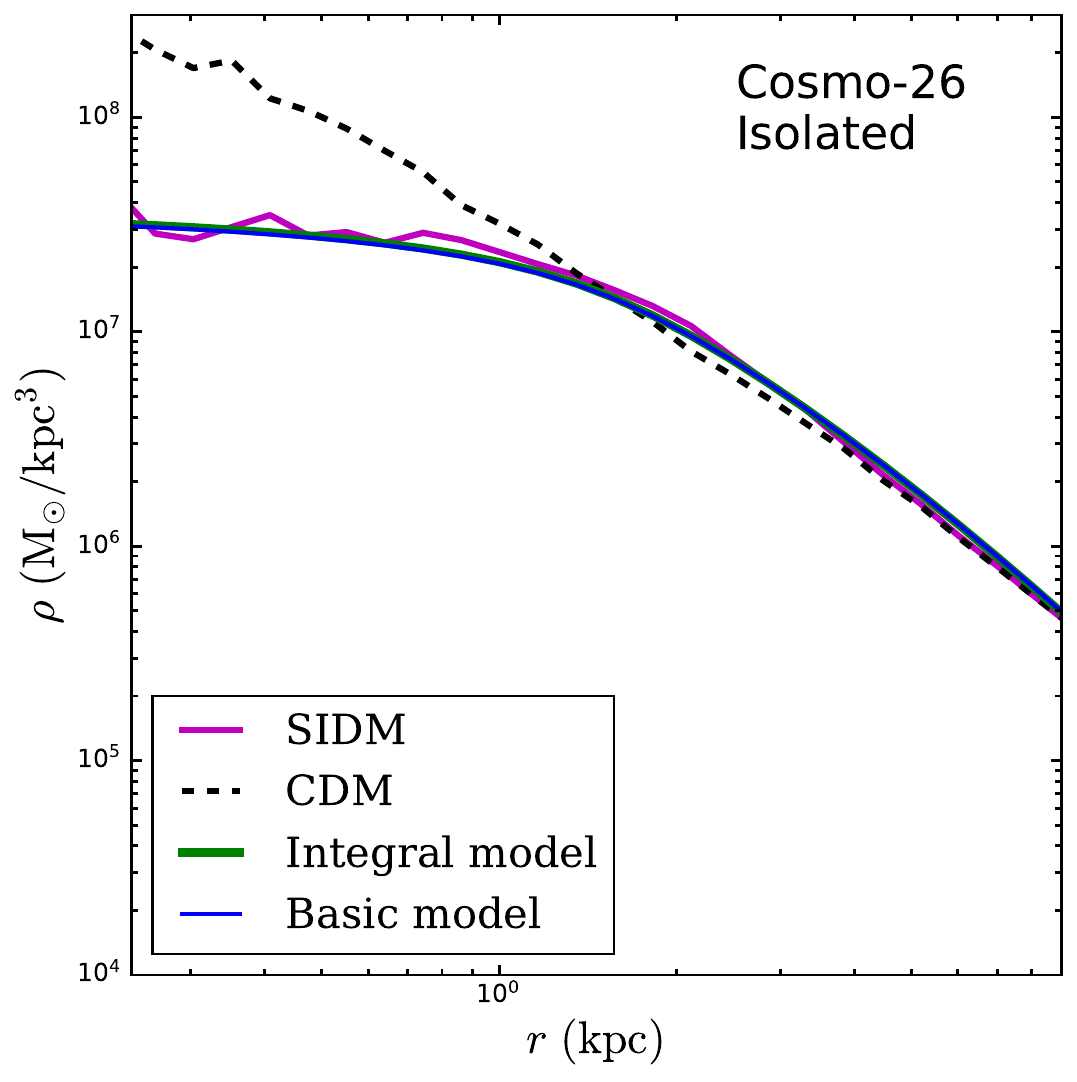}
    \includegraphics[height=4.8cm]{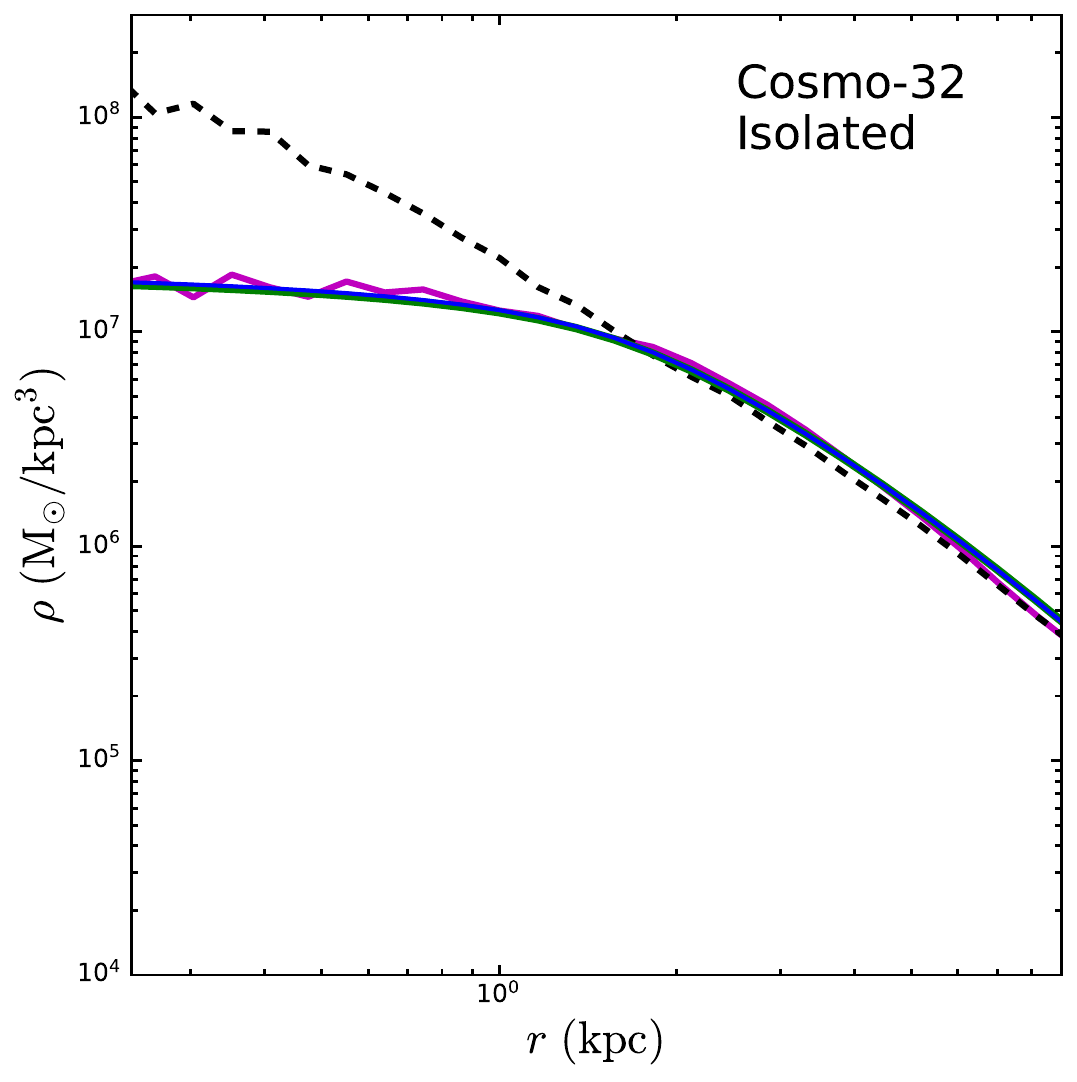}
  \includegraphics[height=4.8cm]{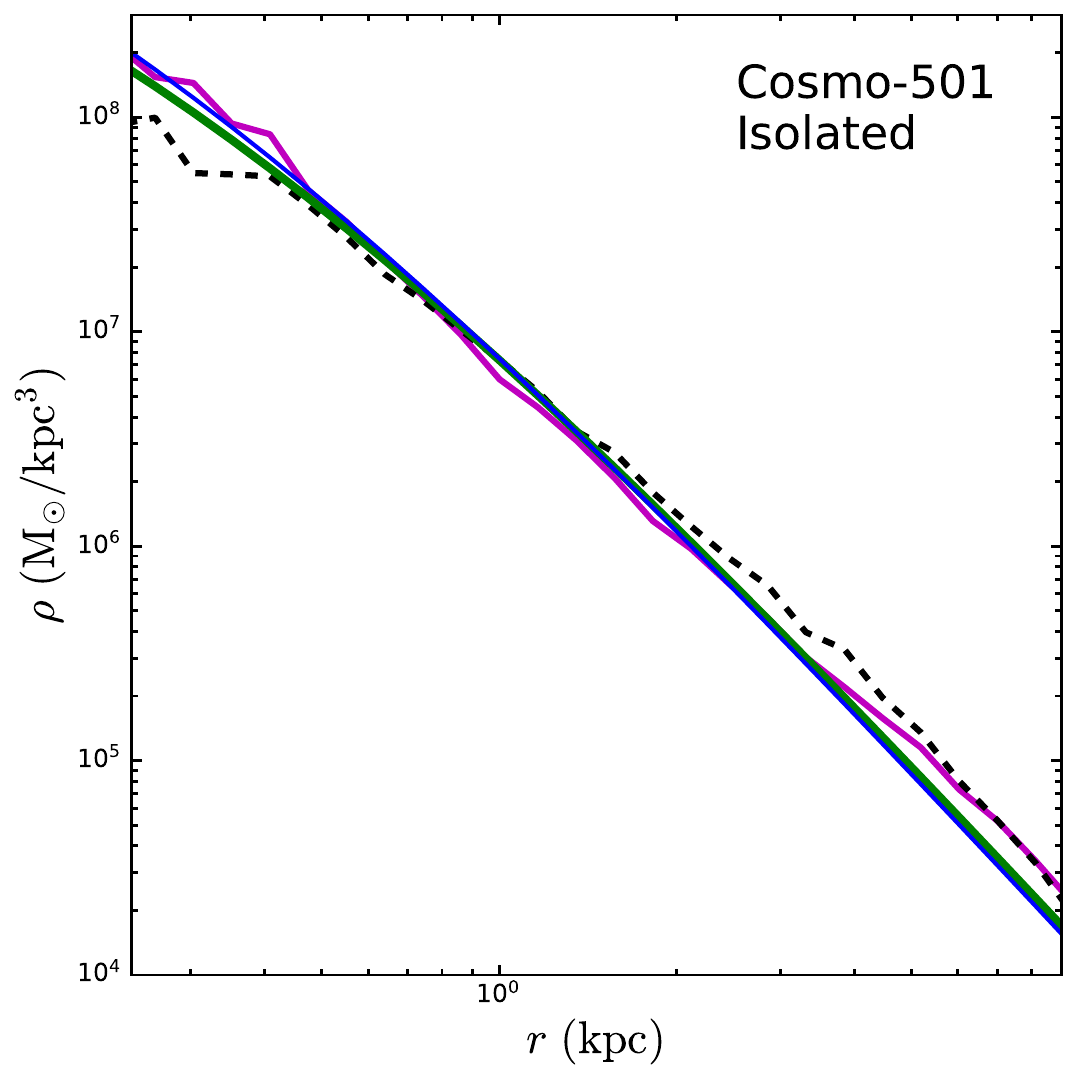}\\
    \includegraphics[height=4.8cm]{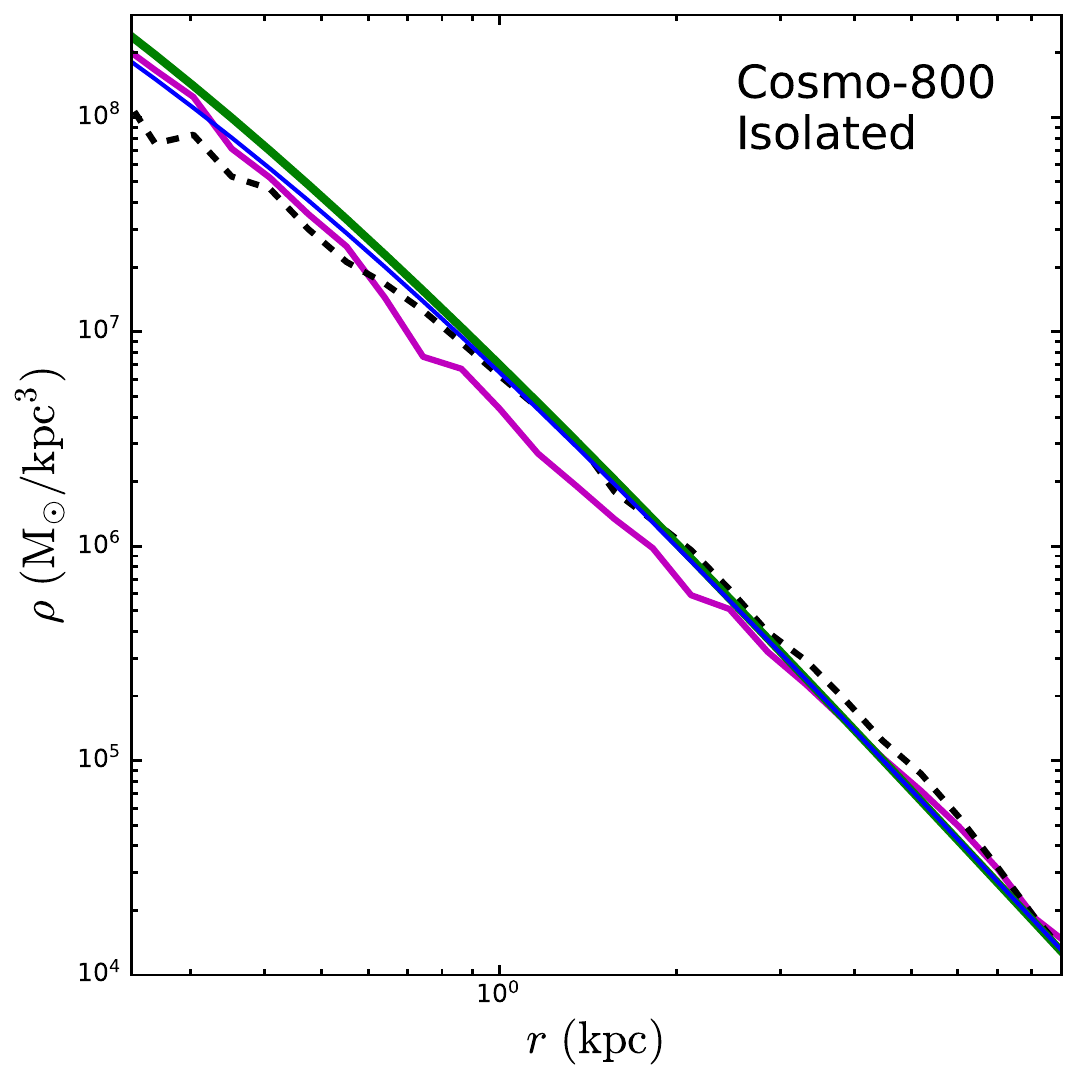}
  \includegraphics[height=4.8cm]{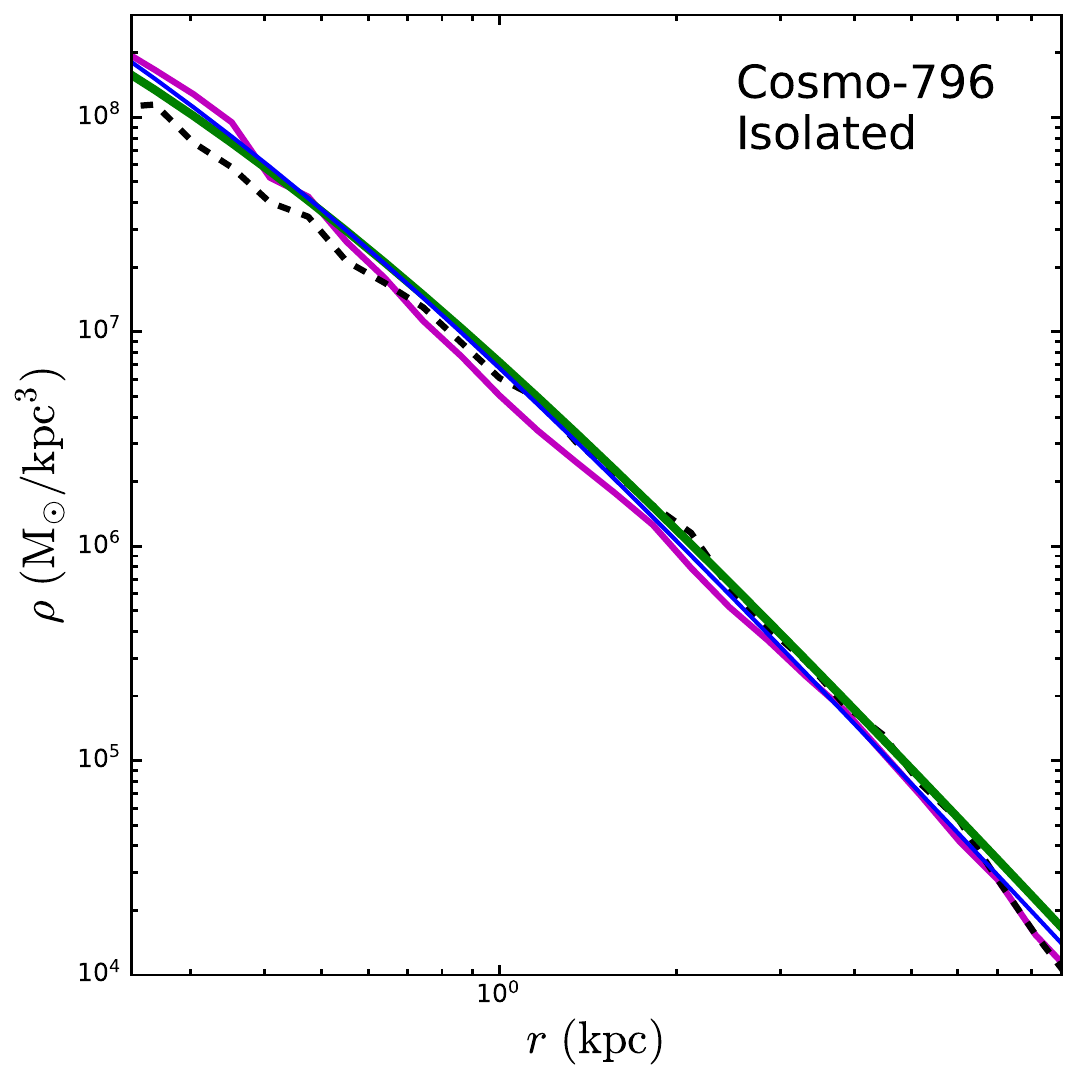}
  \includegraphics[height=4.8cm]{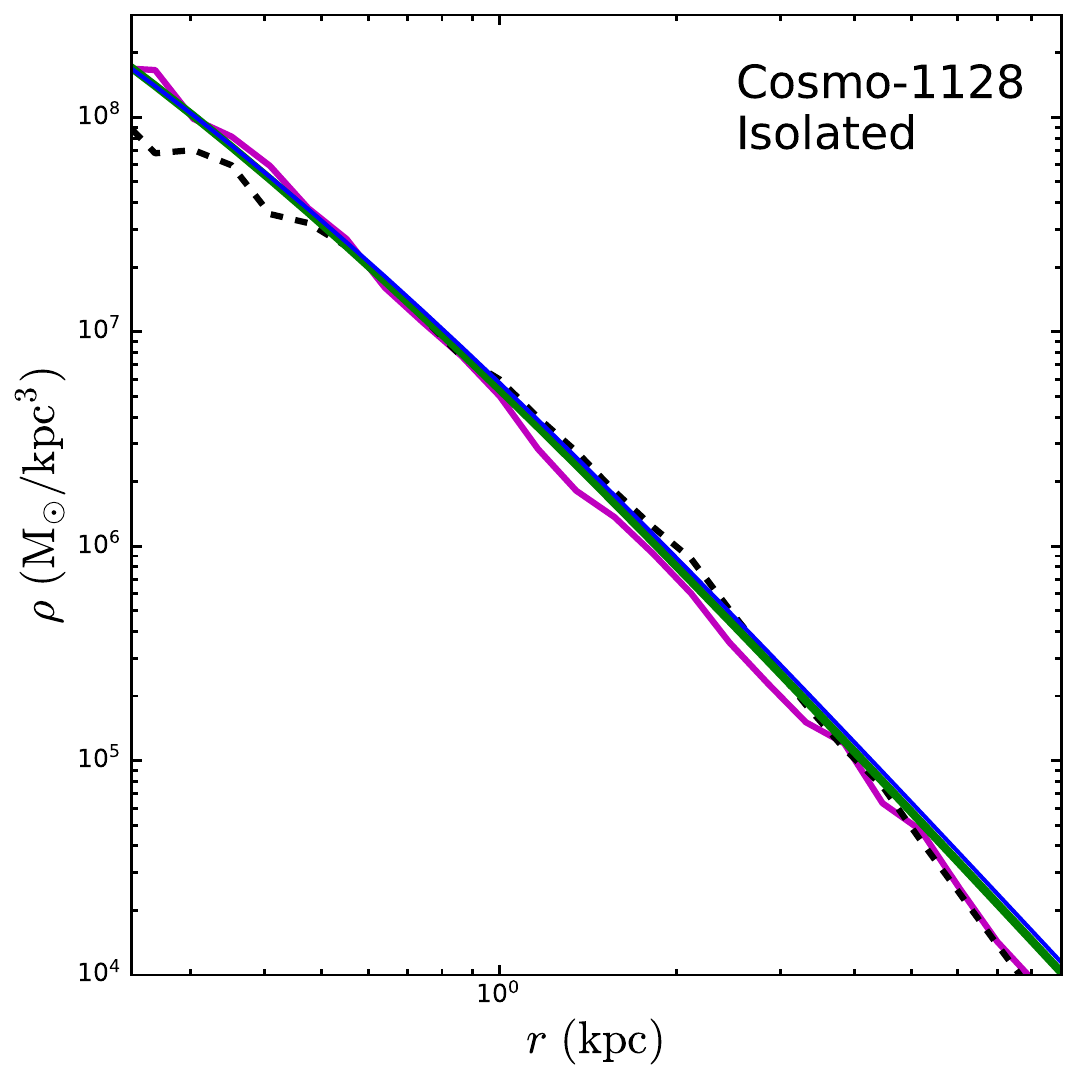}
  \caption{\label{fig:isoBM2}
Density profiles of the isolated halos in table~\ref{tab2} from the parametric model with the basic (solid-blue) and integral (solid-green) approaches, and the SIDM simulation~\cite{Yang:2022mxl} (solid-magenta), as well as the CDM simulation (dashed-black) for comparison. 
}
\end{figure*}

In figure~\ref{fig:isoBM1}, we show the $V_{\rm max}$ (top) and $R_{\rm max}$ (bottom) evolution of three halos: Cosmo-32, Cosmo-501, and Cosmo-796. Both basic (solid-blue) and integral (solid-green) approaches agree with the SIDM simulation (solid-magenta) within $\sim10\%$. Compared to their CDM  counterparts (dashed-black), Cosmo-501 and Cosmo-796 have larger $V_{\rm max}$ and smaller $R_{\rm max}$ at later stages due to core collapse. Although Cosmo-796 has a significant merger at $t_L\sim7~{\rm Gyr}$, both approaches make similar predictions, and they agree with the simulation. For Cosmo-32, the evolution is almost identical in SIDM and CDM. This halo is in the core-forming phase over the entire evolution history, as it has a higher mass and a smaller effective cross section accordingly. As a good approximation, dark matter self-interactions do not change $V_{\rm max}$ and $R_{\rm max}$ ($z=0$) for core-forming isolated SIDM halos~\cite{Yang:2022mxl}.

In figure~\ref{fig:isoBM2}, we show density profiles of all isolated halos at $z=0$ listed in table~\ref{tab2}. We again see that both basic (solid-blue) and integral (solid-green) approaches agree well with the SIDM simulation (solid-magenta). Cosmo-26 and Cosmo-32 are in the core-forming phase at $z=0$ and they have a cored density profile. The four others are deeply collapsed, and their central densities are higher compared to CDM (dashed-black).

We have tested both approaches with a sample of $647$ isolated CDM halos resolved in the simulation~\cite{Yang:2022mxl}. As discussed previously, the integral approach requires well-resolved evolution history of a CDM halo, but numerical noise could be large for low-mass halos in the sample that are close to the resolution limit. In addition, the basic and integral approaches agree well for the representative examples shown in figures~\ref{fig:isoBM1} and \ref{fig:isoBM2}. Thus, we show the comparison between N-body simulation and basic approach for the whole sample of simulated halos. 

\begin{figure*}[htbp]
  \centering
  \includegraphics[height=7.cm]{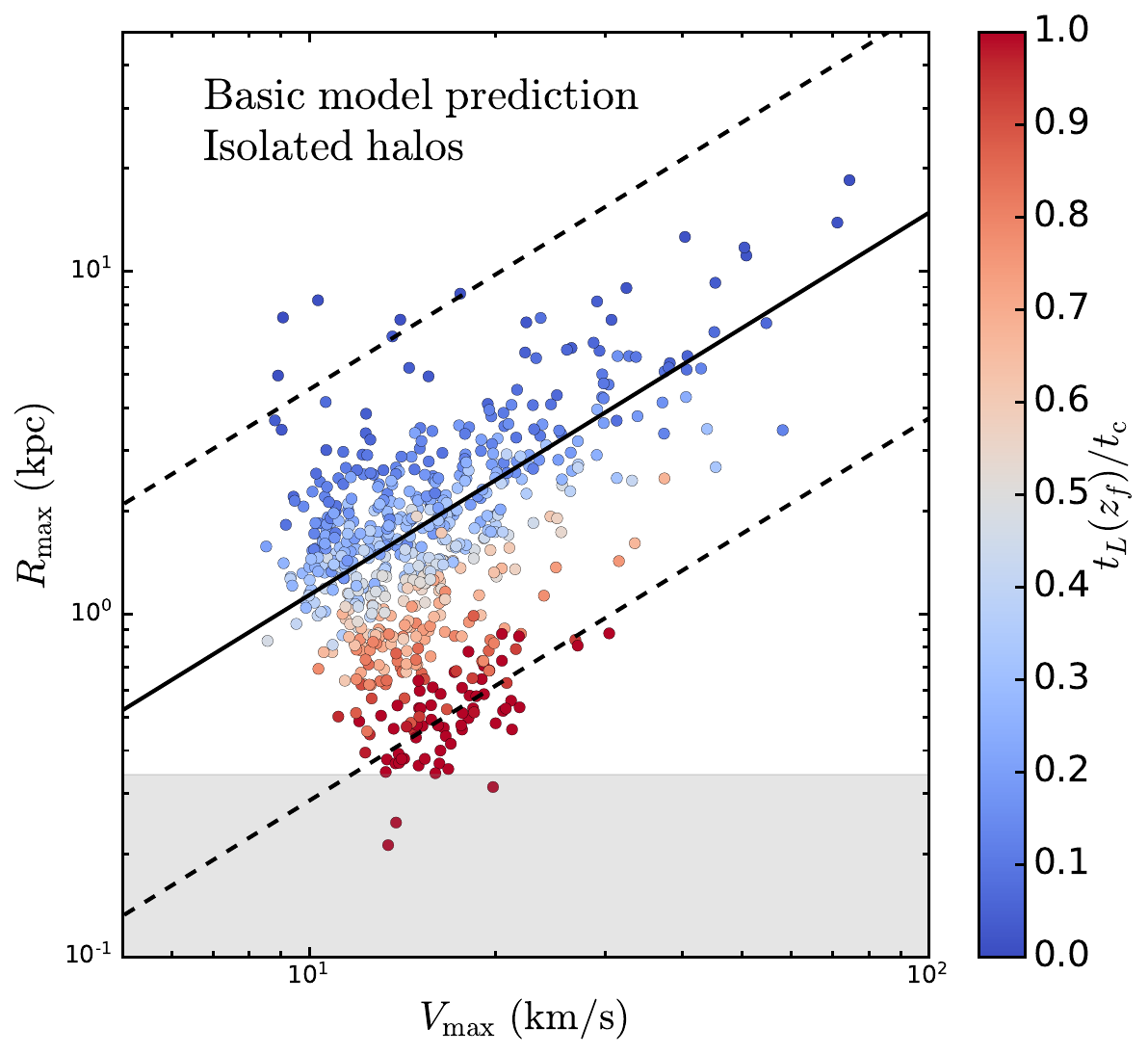}
  \includegraphics[height=7.cm]{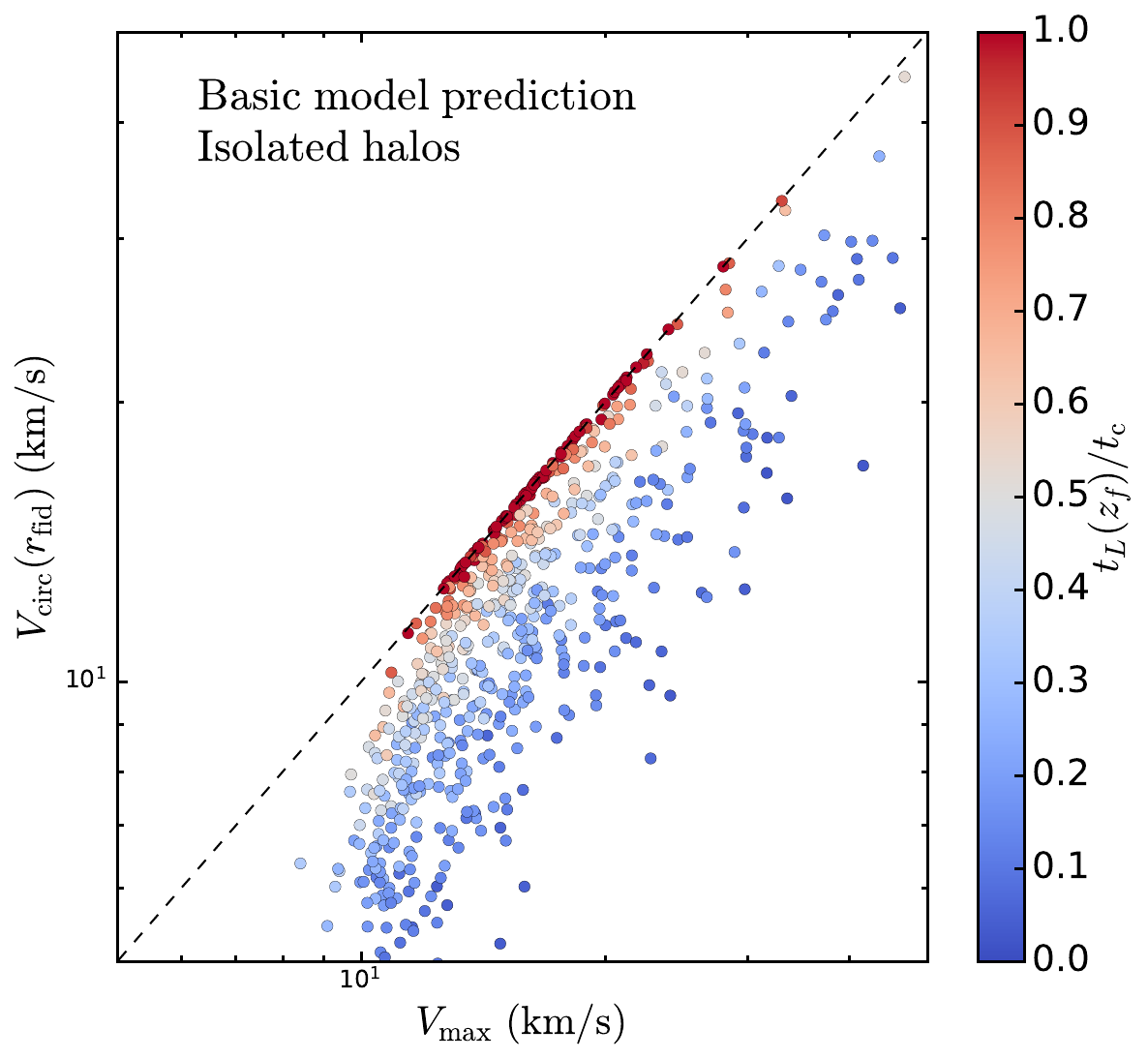}\\
  \includegraphics[height=7.cm]{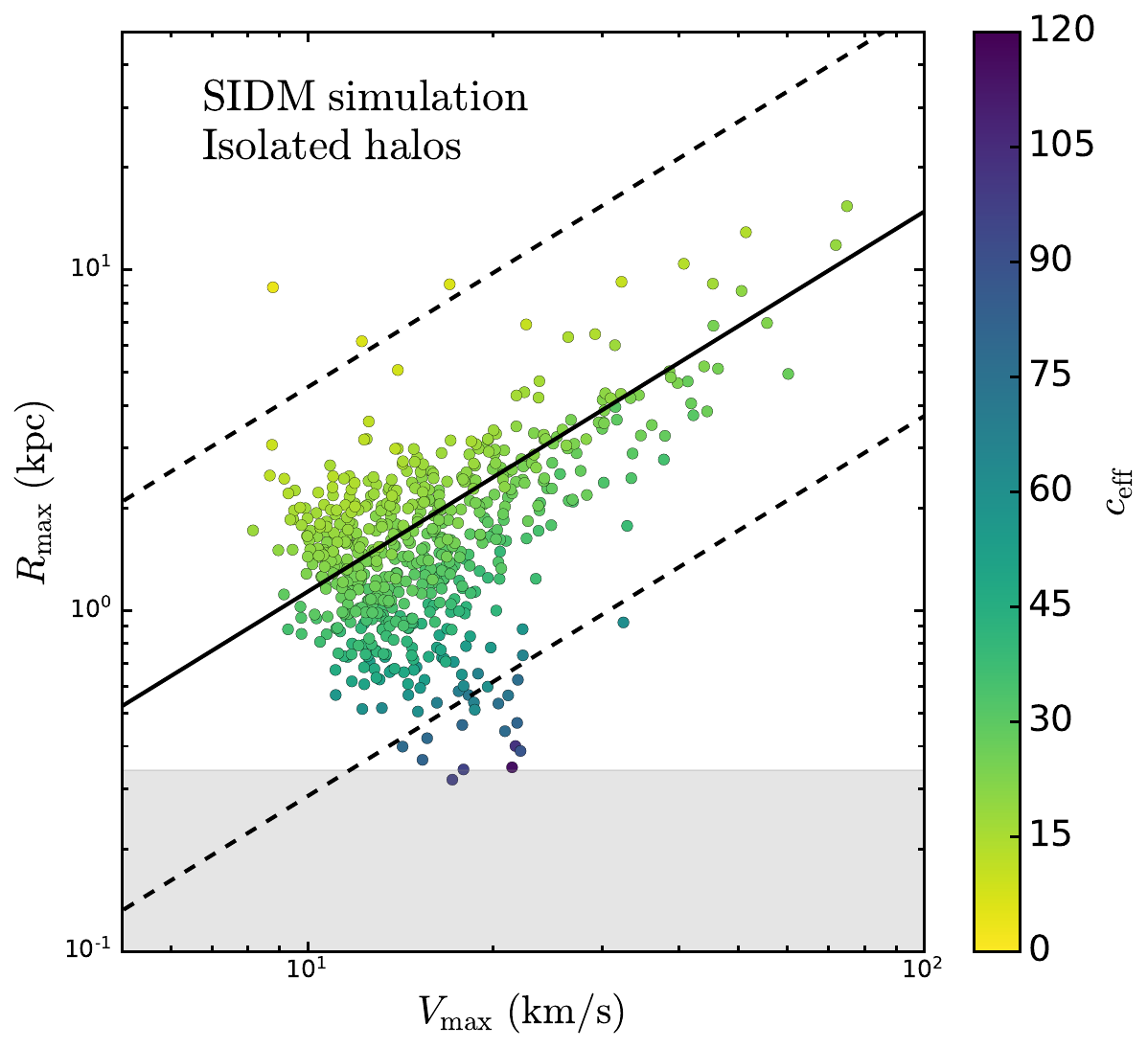}
  \includegraphics[height=7.cm]{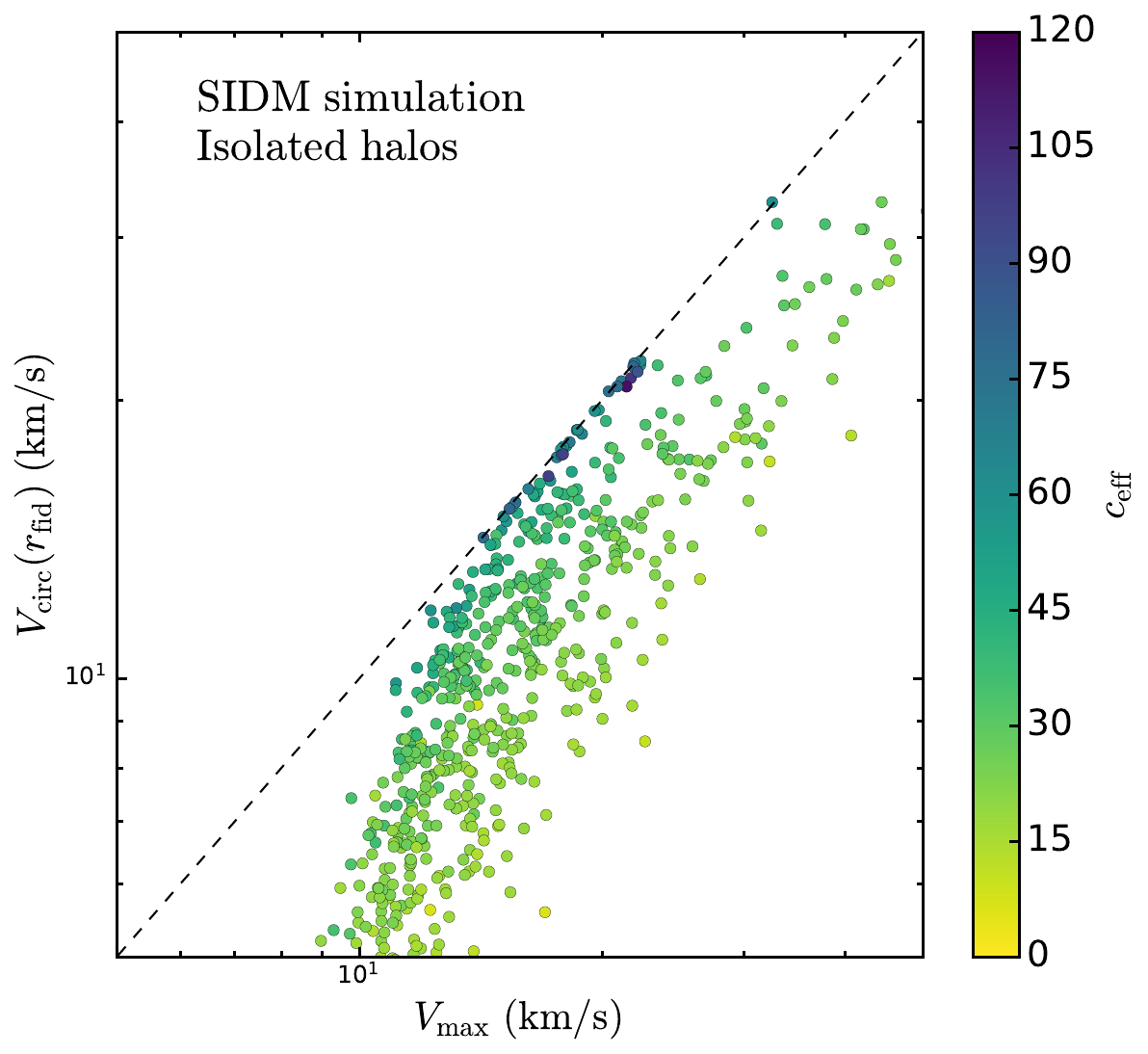}
  \caption{\label{fig:comp0} The $R_{\rm max}\textup{--}V_{\rm max}$ (left) and $V_{\rm circ}(r_{\rm fid})\textup{--}V_{\rm max}$ (right) distributions of isolated SIDM halos at $z=0$ predicted using the parametric model with the basic approach (top), where each halo is color coded according to $t_L(z_f)/t_c$. For comparison, the corresponding distributions from the N-body SIDM simulation are also shown (bottom), and the color coding is based on the effective concentration $c_{\rm eff}=R_{\rm vir}/(R_{\rm max}/2.1626)$; taken from~\cite{Yang:2022mxl}. As in~\cite{Yang:2022mxl}, the median CDM isolated halo $V_{\rm max}\textup{--}R_{\rm max}$ relation (solid-black) together with a $\pm0.6$ dex band (dashed-black), and resolution limit (gray-shaded) are shown in the left panel; the $1\textup{:}1$ relation (dashed-black) in the right panel. }
\end{figure*}

Figure~\ref{fig:comp0} (top) shows $R_{\rm max}\textup{--}V_{\rm max}$ and $V_{\rm circ}(r_{\rm fid})\textup{--}V_{\rm max}$ distributions predicted using the basic approach, where $r_{\rm fid} = 2 V_{\rm max}/(70~{\rm km/s})\ {\rm kpc}$ and each halo is color coded according to $t_L(z_f)/t_c$. We see that the predicted distributions agree well with those directly from the N-body SIDM simulation~\cite{Yang:2022mxl} as shown in figure~\ref{fig:comp0} (bottom), where the simulated SIDM halo is colored based on its effective concentration as in~\cite{Yang:2022mxl}. In particular, our parametric model successfully reproduces halos that are deeply collapsed. These halos have the lowest $R_{\rm max}$ and highest $V_{\rm circ}(r_{\rm fid})$ values for given $V_{\rm max}$. 

\begin{figure}[htbp]
  \centering
  \includegraphics[height=4.8cm]{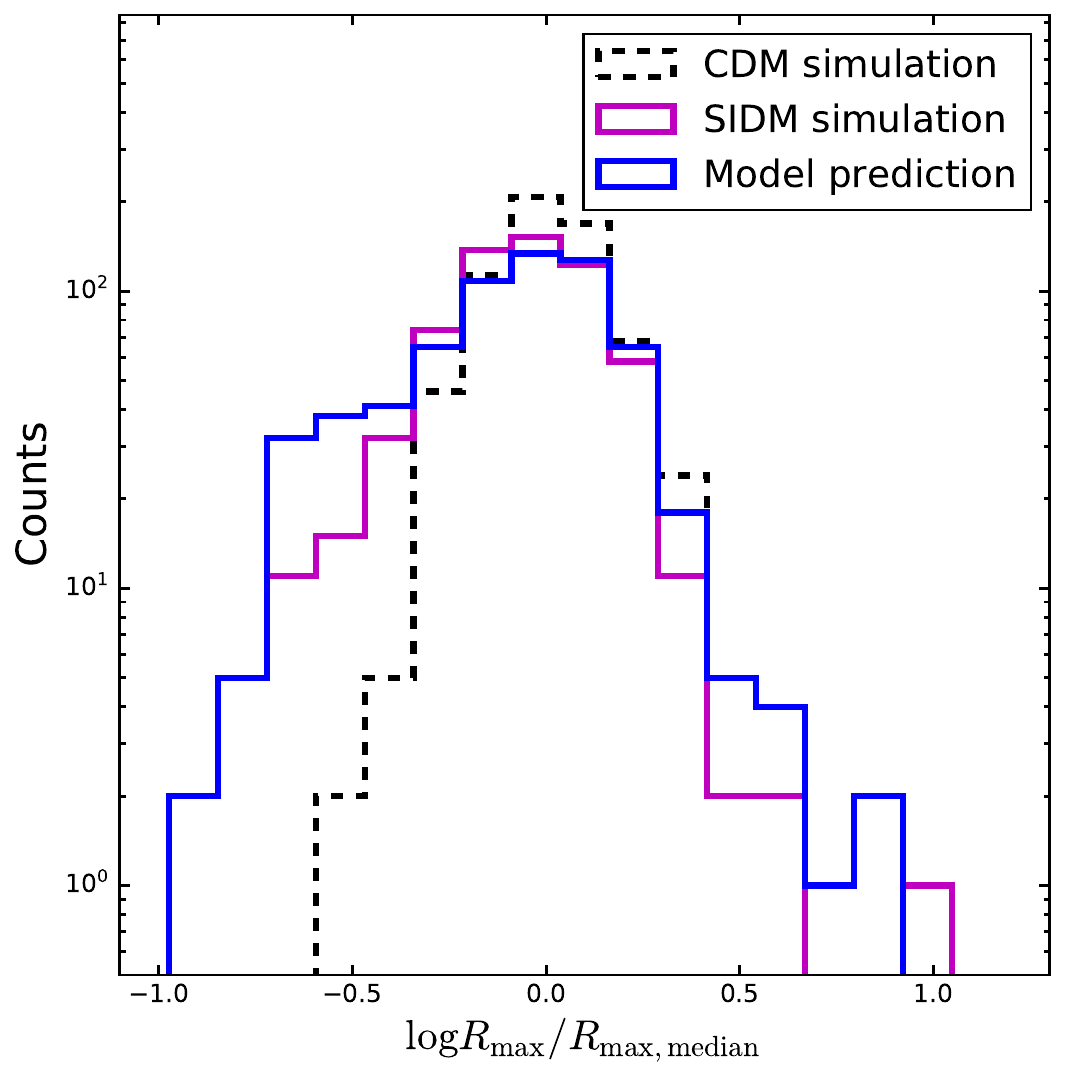}
  \includegraphics[height=4.8cm]{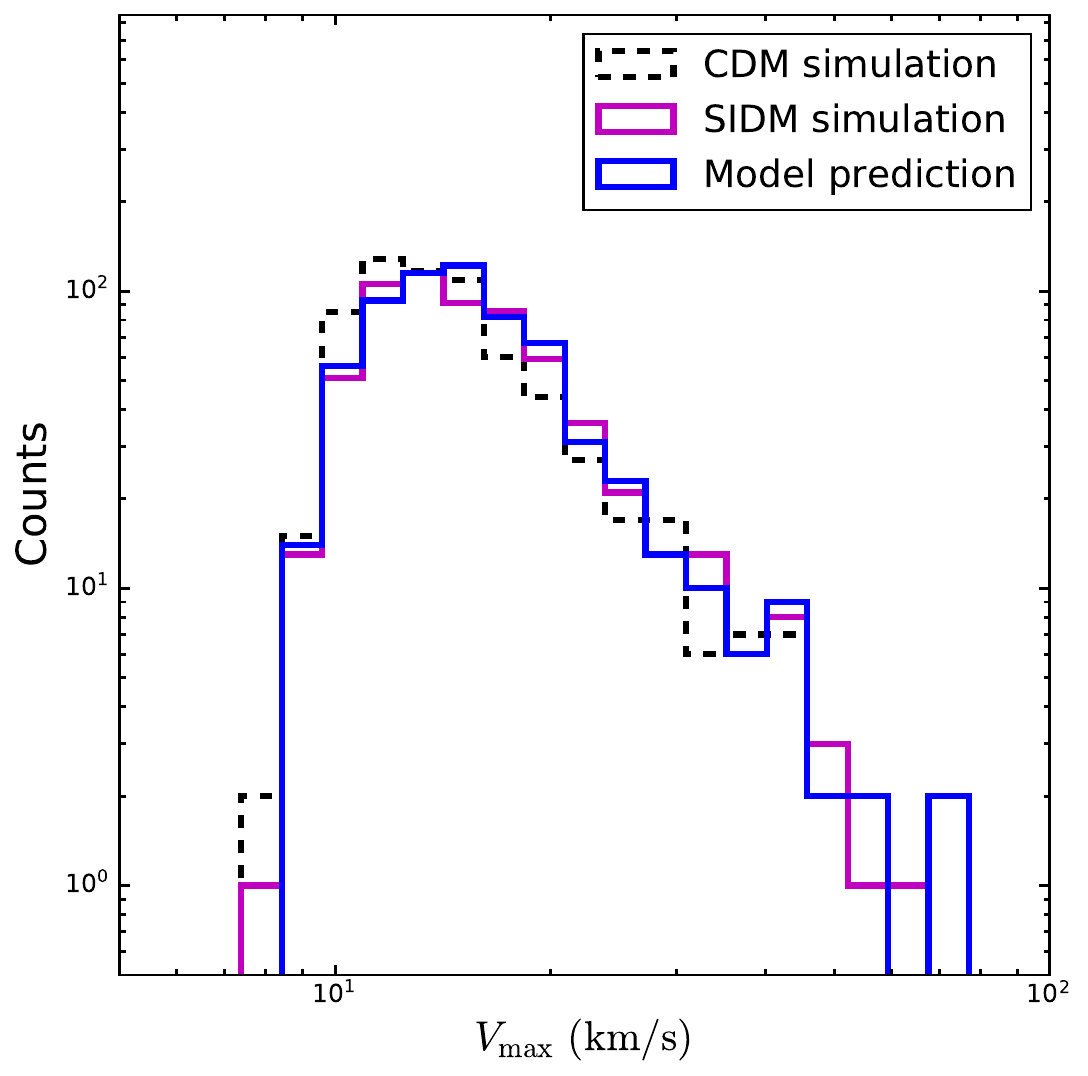}
  \includegraphics[height=4.8cm]{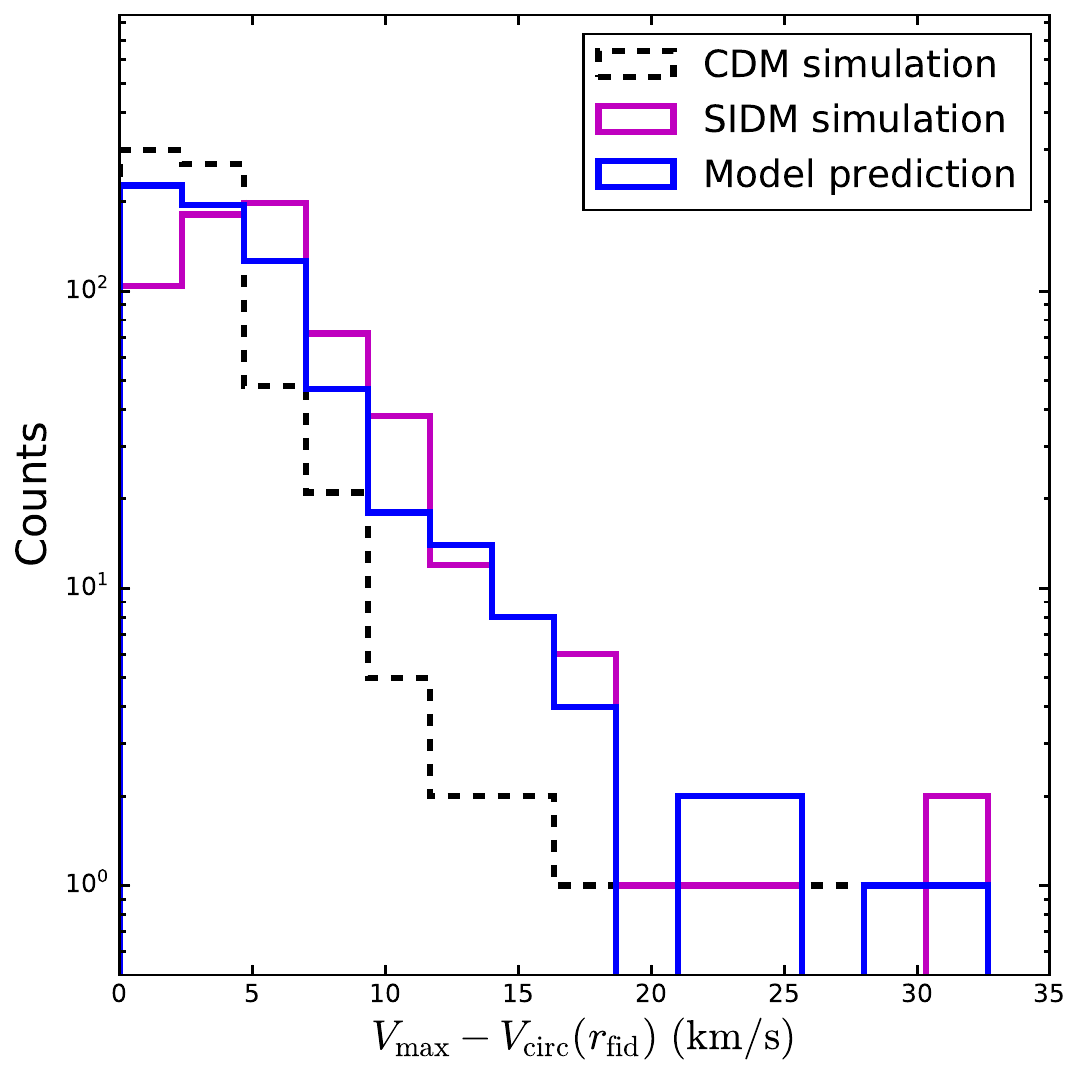}
  \caption{\label{fig:stats} The $\log_{10}({R_{\rm max}}/{R_{\rm max,median}})$ (left), $V_{\rm max}$ (middle), and $(V_{\rm max}-V_{\rm circ}(r_{\rm fid}))$ (right) distributions for the sample of isolated halos from the parametric model (solid-blue), SIDM (solid-magenta) and CDM (dashed-black) simulations~\cite{Yang:2022mxl}.}
\end{figure}

\begin{table}[bthp]
\begin{center}
\begin{tabular}{c|ccc}
\hline
\hline
Properties                                                           &  CDM simulation  & SIDM simulation  & Model prediction \\
\hline
$\mu\left(\log_{10}\frac{R_{\rm max}}{R_{\rm max,median}}\right)$    &  0.0144          & -0.0763          & -0.0918       \\
$\sigma\left(\log_{10}\frac{R_{\rm max}}{R_{\rm max,median}}\right)$ &  0.173           & 0.224            & 0.272         \\
\hline
$\mu\left(V_{\rm max}\right)$ $\rm (km/s)$                           &  15.9            & 16.9             & 16.7          \\
$\sigma\left(V_{\rm max}\right)$ $\rm (km/s)$                        &  7.40            & 7.75             & 7.46          \\
\hline
$\mu\left(V_{\rm max}-V_{\rm circ}(r_{\rm fid})\right)$ $\rm (km/s)$               &  3.00            & 5.40             & 4.16          \\ 
$\sigma\left(V_{\rm max}-V_{\rm circ}(r_{\rm fid})\right)$ $\rm (km/s)$            &  2.32            & 3.74             & 3.90          \\
\hline
\hline
\end{tabular}
\caption{\label{tab1} The mean and standard deviation of the $\log_{10}({R_{\rm max}}/{R_{\rm max,median}})$, $V_{\rm max}$, and $(V_{\rm max}-V_{\rm circ}(r_{\rm fid}))$ distributions.  }
\end{center}
\end{table}

Figure~\ref{fig:stats} further shows the distribution of $\log_{10}(R_{\rm max}/R_{\rm max,median})$ (left), $V_{\rm max}$ (middle), and $(V_{\rm max}-V_{\rm circ}(r_{\rm fid}))$ (right) for the sample, from the parametric model (solid-blue), as well as the N-body SIDM (solid-magenta) and CDM (dashed-black) simulations~\cite{Yang:2022mxl}. We see good agreement between parametric model and simulation. In addition, compared to CDM, SIDM predicts a higher population for $\log_{10}(R_{\rm max}/R_{\rm max,median})\lesssim-0.5$ due to core collapse, as well as a higher population for $(V_{\rm max}-V_{\rm circ}(r_{\rm fid}))>5~{\rm km/s}$ because of core formation. For a more quantitative comparison, we show the mean and standard deviation of the three distributions in table~\ref{tab1}.

\section{Modeling the evolution of SIDM subhalos}
\label{sec:subs}

Subhalos can be significantly affected by the host halo due to tidal interactions. As a result, their density profiles are different from those of isolated halos. To account for the tidal effects, we modify the density profile in eq.~(\ref{eq:cnfw}) as  
\begin{eqnarray}
\label{eq:tcnfw}
\rho_{\rm tSIDM}(r) = \frac{\rho_s}{\frac{(r^{\beta}+r_c^{\beta} )^{1/{\beta}}}{r_s} \left(1 + \frac{r}{r_s} \right)^2 \left(1 + \left(\frac{r}{r_t}\right)^{2-u} \right)^{1+3u}},  
\end{eqnarray}
where the additional parameters are the tidal radius $r_t$ and the numerical factor $u$ controlling the logarithmic slope of the truncated density profile in the outer region. We evaluate the tidal radius $r_t$ as~\cite{vandenBosch:2017ynq,Klypin:1997fb}
\begin{eqnarray}
r_t = d \left(\frac{M_{\rm sub}}{M_{\rm host}(r<d)} \right)^{1/3}, 
\end{eqnarray}
where $d$ is the distance of the subhalo from its host halo, $M_{\rm sub}$ is the subhalo virial mass and $M_{\rm host}(r<d)$ is the host halo mass enclosed within $d$. We further take into account the evolution of the tidal radius in our study. The smooth truncation is controlled by the term $1/(1+(r/r_t)^{2-u})^{1+3 u}$, which falls off at large radii as $r^{-2}$ at $u=0$ and as $r^{-4}$ at $u=1$, as in the case of a truncated NFW profile for CDM subhalos, see, e.g.,~\cite{Baltz:2007vq,Minor:2020hic}. In this work, we take $u(c_{\rm eff}) = {\rm Min}(1,0.0004 c_{\rm eff}^{2.2})$, where $c_{\rm eff}=R_{\rm vir}/(R_{\rm max}/2.1626)$ is the effective concentration~\cite{Yang:2022mxl}, and the corresponding truncated profile in eq.~(\ref{eq:tcnfw}) can fit both core-forming and -collapsing subhalos. We have also  checked that the SIDM evaporation effect~\cite{Nadler200108754} is negligible for the subhalos in the simulation~\cite{Yang:2022mxl}, although such an effect could be incorporated using a semi-analytical method~\cite{Shirasaki220509920}.

Since subhalos undergo tidal mass loss after falling into the host halo, the basic approach is insufficient for modeling their evolution. However, the integral approach based on eq.~(\ref{eq:int}) still works. As a demonstration, we select $6$ representative cases, ranging from core-forming subhalos to deeply collapsing ones, as summarized in table~\ref{tab3}. The evolution stage is quantified by $\tau_0 = \int_{t_f}^{t_0} dt /t_c(t)$, where $t_0=13.647~{\rm Gyr}$ and we take the formation time $t_f$ to be half of $13.647~{\rm Gyr} - t_{L}(z_f)$, see eq.~(\ref{eq:lookback}). In addition, the collapse time $t_c$ changes along with tidal mass loss, and we use eq.~(\ref{eq:tc0}) to evaluate it with $\rho_{\rm eff}=(V_{\rm max, CDM}/1.648r_{\rm eff})^2/G$ and $r_{\rm eff}=(R_{\rm max, CDM}/2.1626)$ for a given moment.

\begin{table}[bthp]
\begin{center}
\begin{footnotesize}
\begin{tabular}{l|cccc|cc}
\hline
\hline
Cosmo-ID & $M_{\rm vir,0} $  & $\tau_0$ & $V_{\rm max,SIDM,0}$ & $R_{\rm max,SIDM,0}$
                                  & $V_{\rm max,Model,0}$         & $R_{\rm max,Model,0}$ \\
Subhalos                &   $\rm (10^8\ M_{\odot})$    & - & $\rm (km/s)$ & $\rm (kpc)$  & $\rm (km/s)$ & $\rm (kpc)$ \\
\hline
71    &  57.1  & 0.092 & 44.6 & 4.93  & 43.0 & 5.60 \\
186   &  20.7  & 0.27  & 34.4 & 2.46  & 31.4 & 2.00 \\ 
567   &  6.04  & 0.85  & 31.7 & 0.58  & 30.3 & 0.82 \\ 
665   &  4.61  & 0.12  & 13.7 & 3.72  & 13.6 & 3.43 \\
945   &  3.92  & 1.10  & 38.8 & 0.24  & 38.3 & 0.045 \\
1357  &  2.38  & 1.18  & 26.0 & 0.37  & 24.6 & 0.42 \\
\hline
\hline
\end{tabular}
\caption{\label{tab3} Characteristic properties of subhalos at $z=0$, selected from the cosmological zoom-in SIDM simulation~\cite{Yang:2022mxl} for testing the parametric model.
The virial mass, $\tau_0 = \int_{t_f}^{t_0} dt /t_c(t)$, $V_{\rm max}$ and $R_{\rm max}$ are reported for the subhalos.
For comparison, their $V_{\rm max}$ and $R_{\rm max}$ values predicted in the parametric model with the integral approach are also listed in the last two columns; see the relevant discussion after eq.~(\ref{eq:int}).
}
\end{footnotesize}
\end{center}
\end{table}

\begin{figure*}[htbp]
  \centering
  \includegraphics[height=4.8cm]{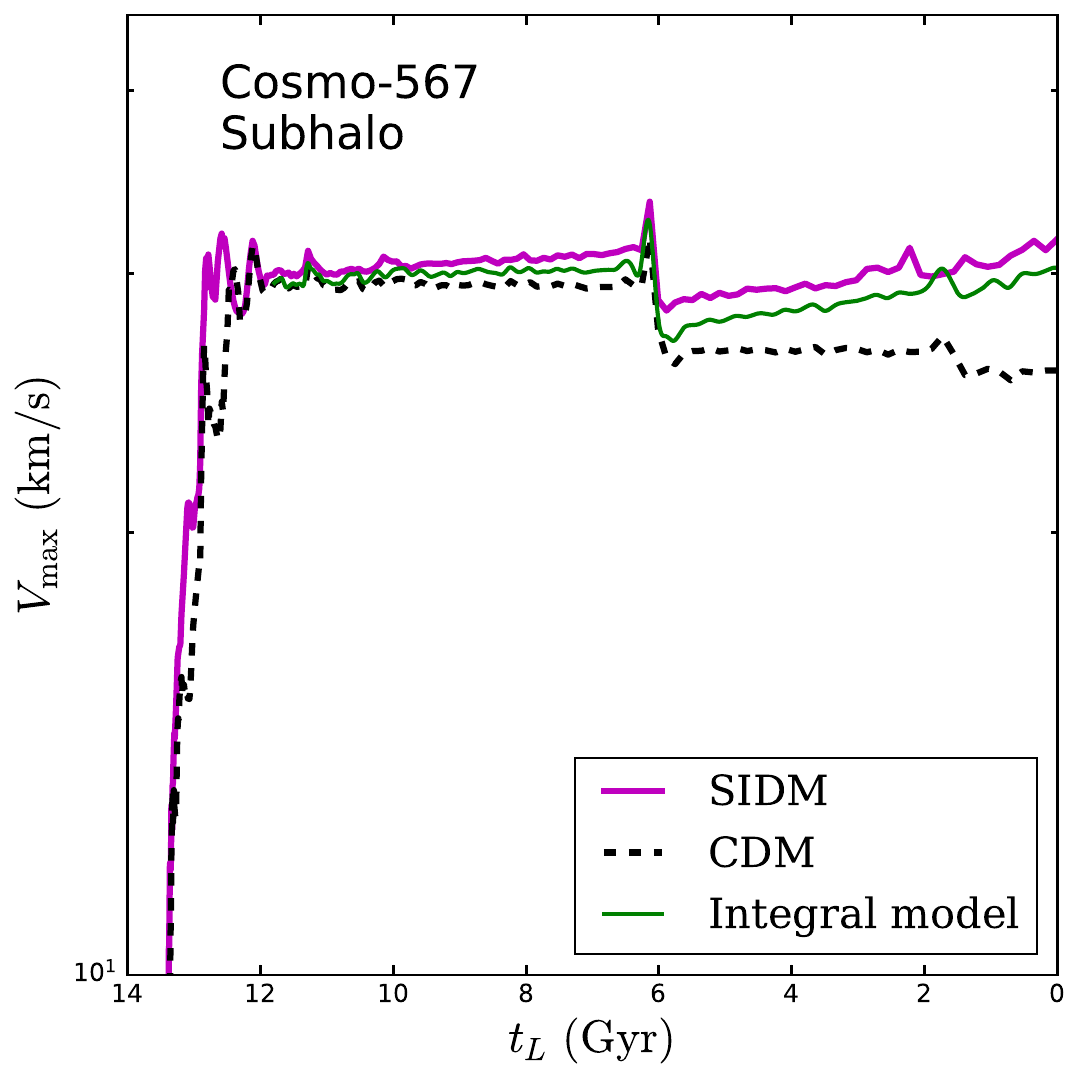}
  \includegraphics[height=4.8cm]{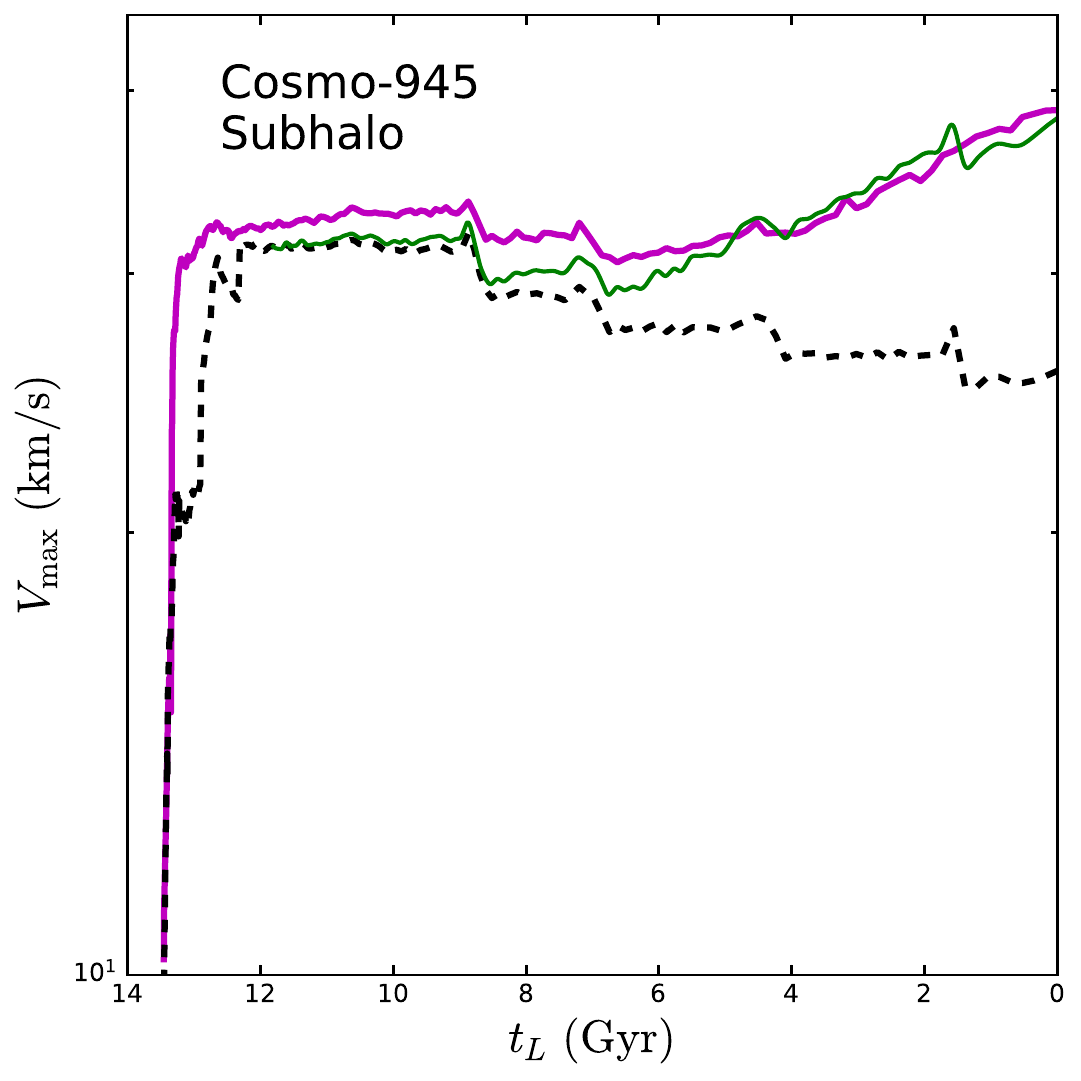}
  \includegraphics[height=4.8cm]{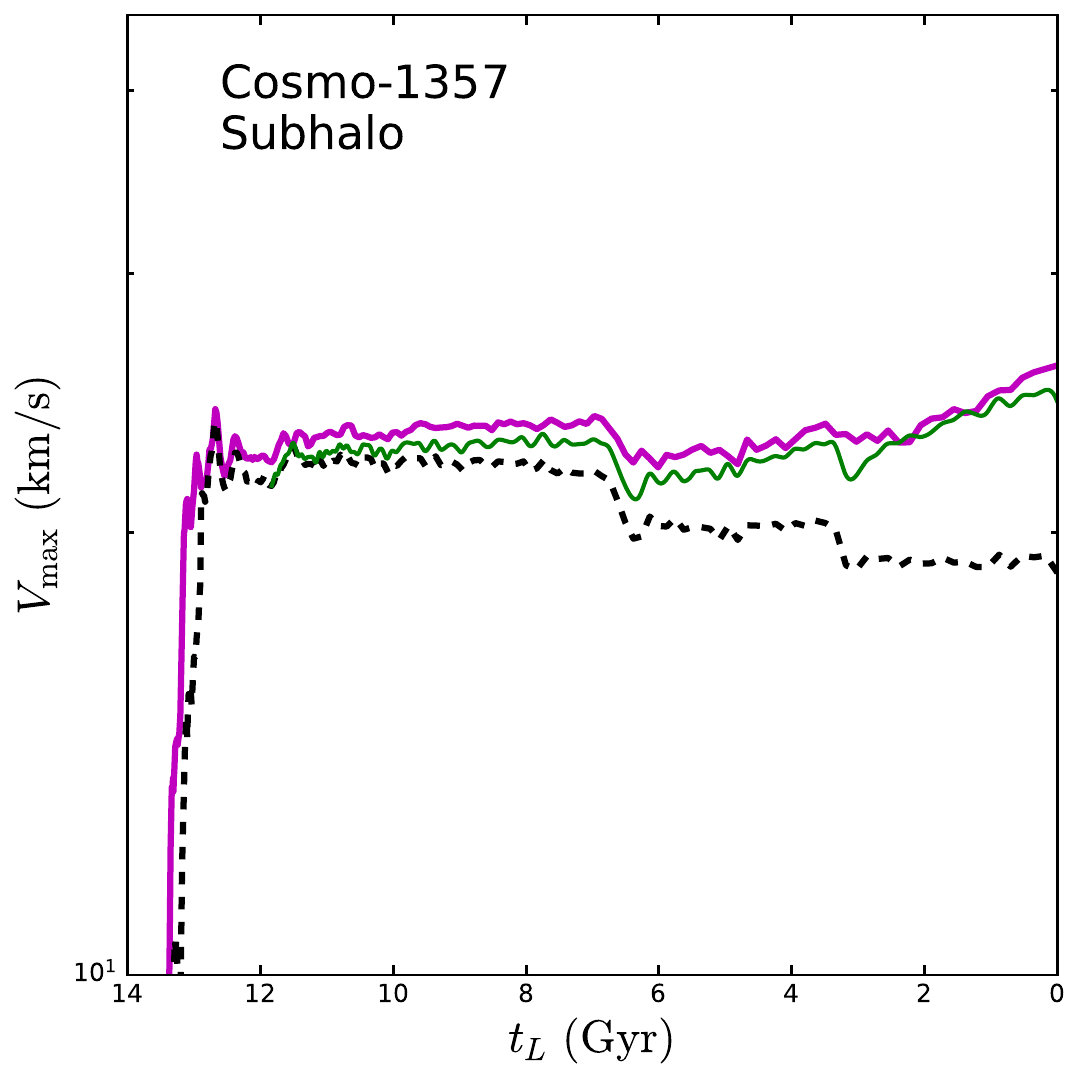}
  \includegraphics[height=4.8cm]{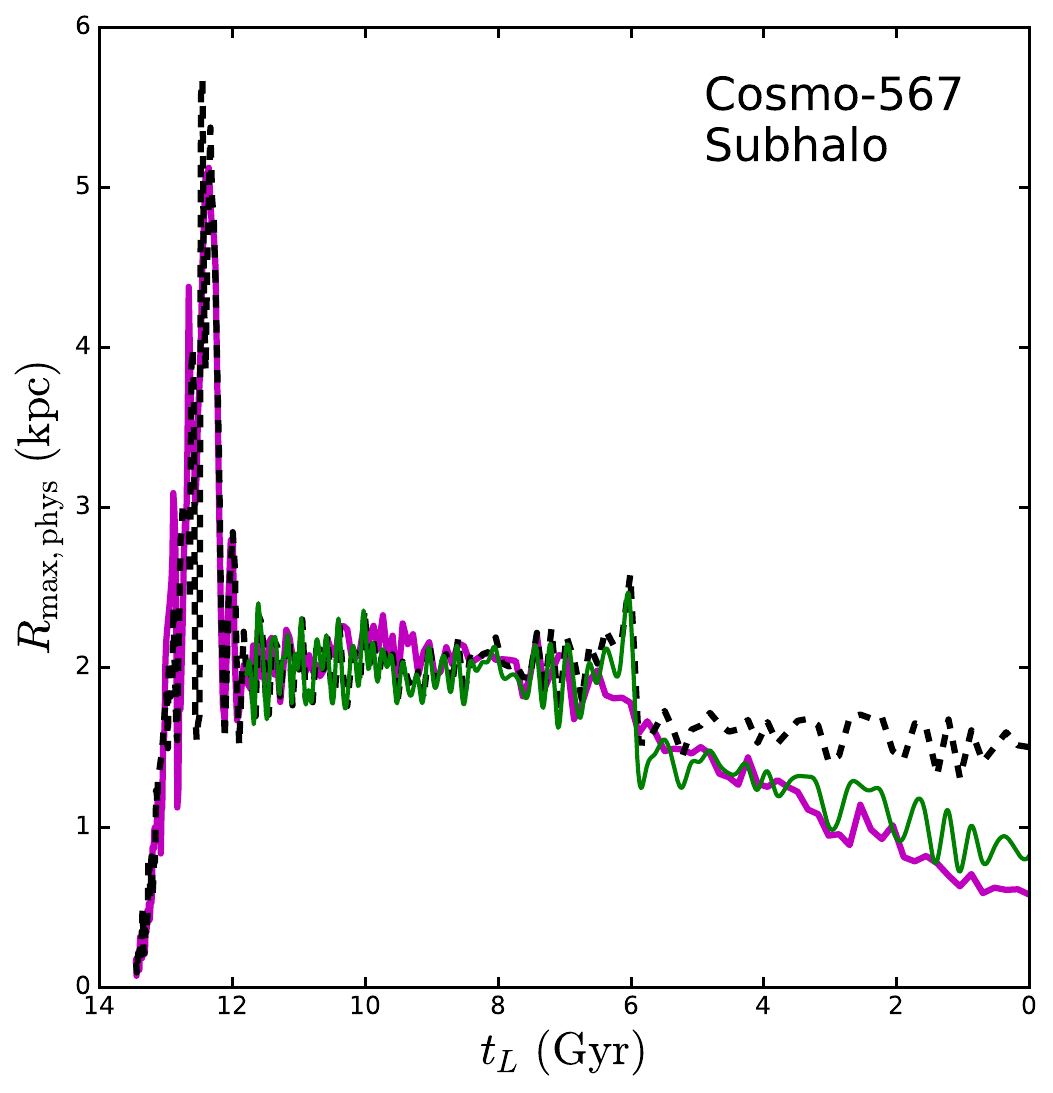}
  \includegraphics[height=4.8cm]{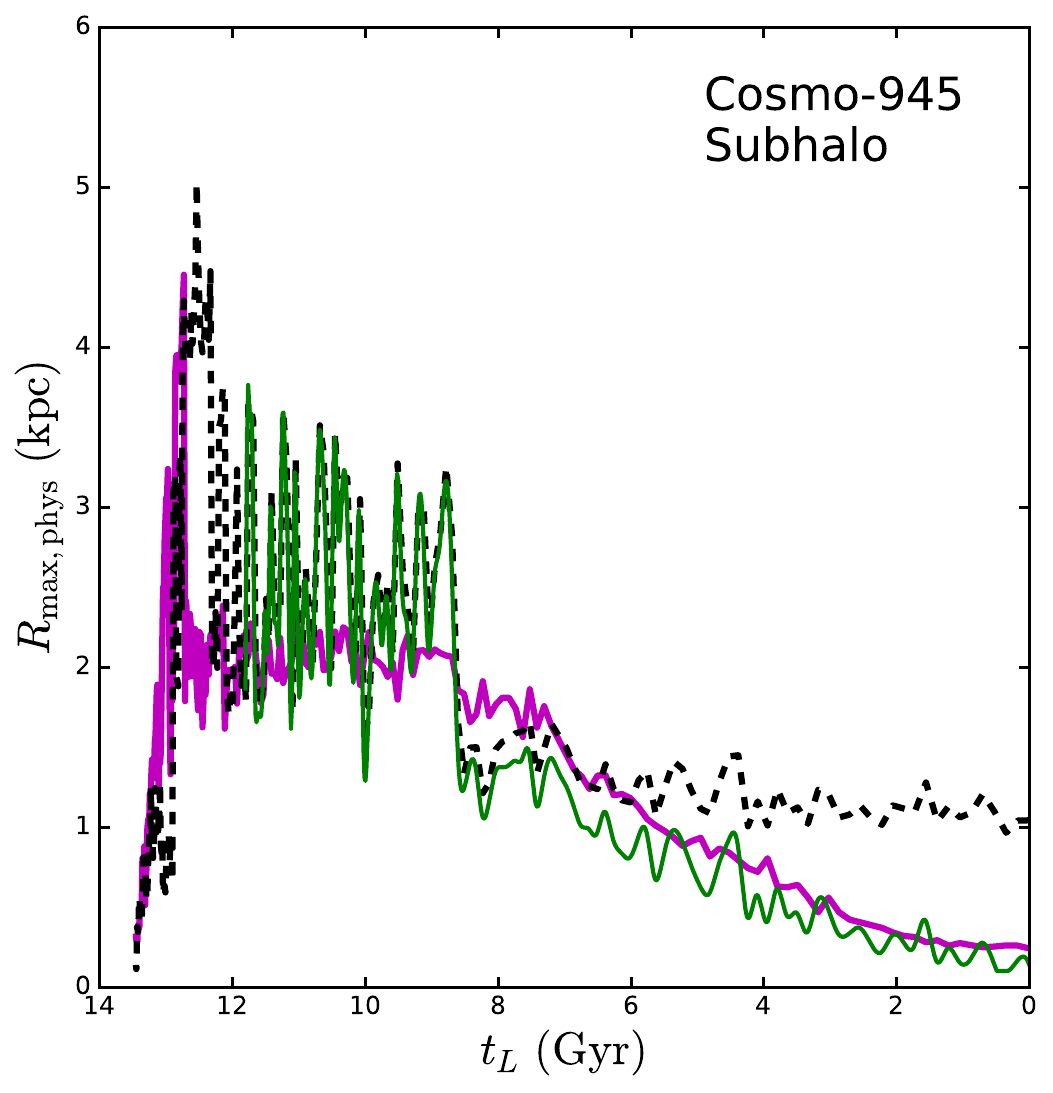}
  \includegraphics[height=4.8cm]{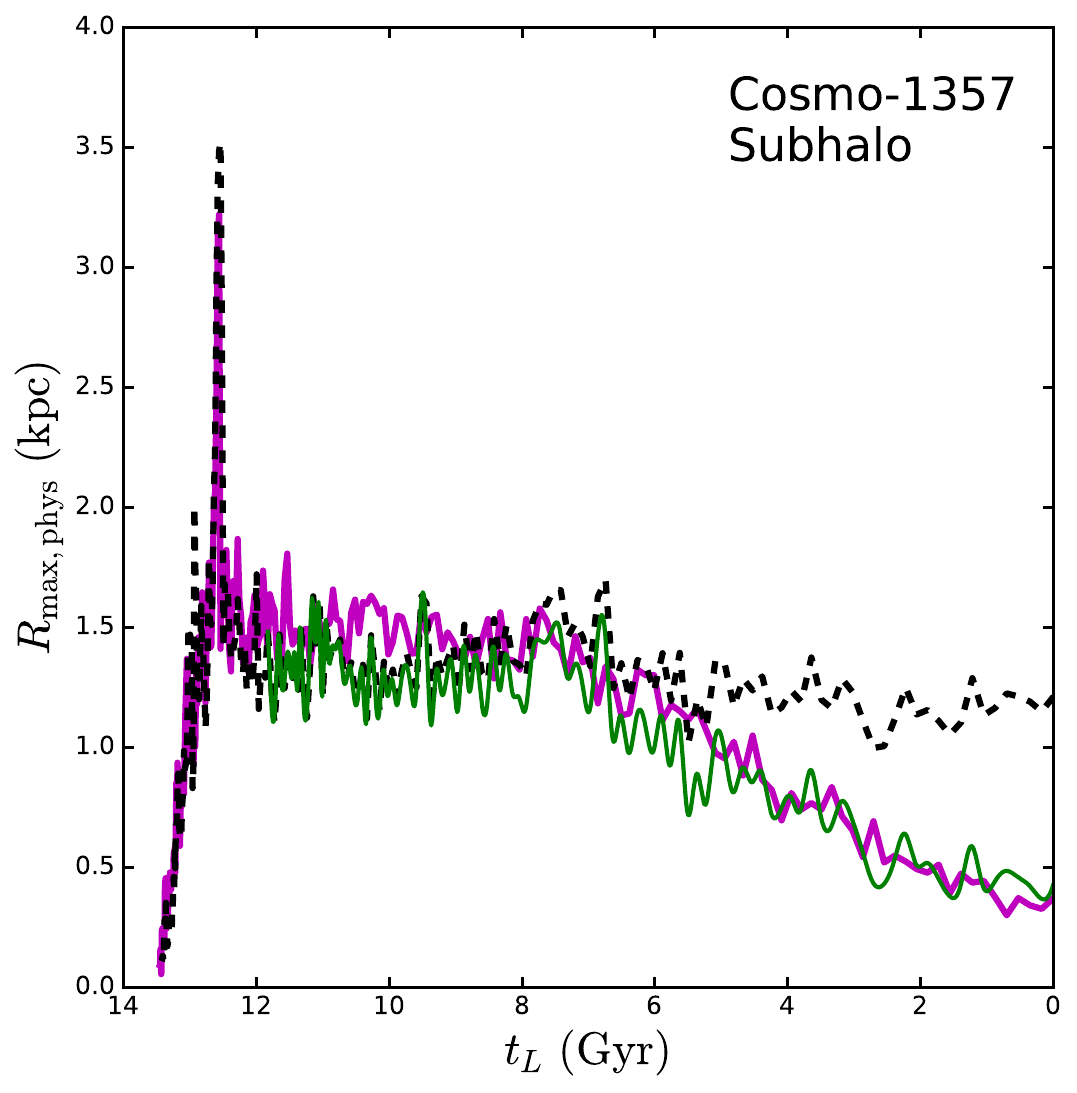} 
  \caption{\label{fig:subBM1} The $V_{\rm max}$ (top) and $R_{\rm max}$ (bottom) evolution of three SIDM subhalos that are in the deeply collapse phase at $z = 0$ from the parametric model with the integral approach (solid-green) and the SIDM simulation~\cite{Yang:2022mxl} (solid-magenta), as well as their CDM counterparts (dashed-black). }
\end{figure*}

\begin{figure*}[htbp]
  \centering
  \includegraphics[height=4.8cm]{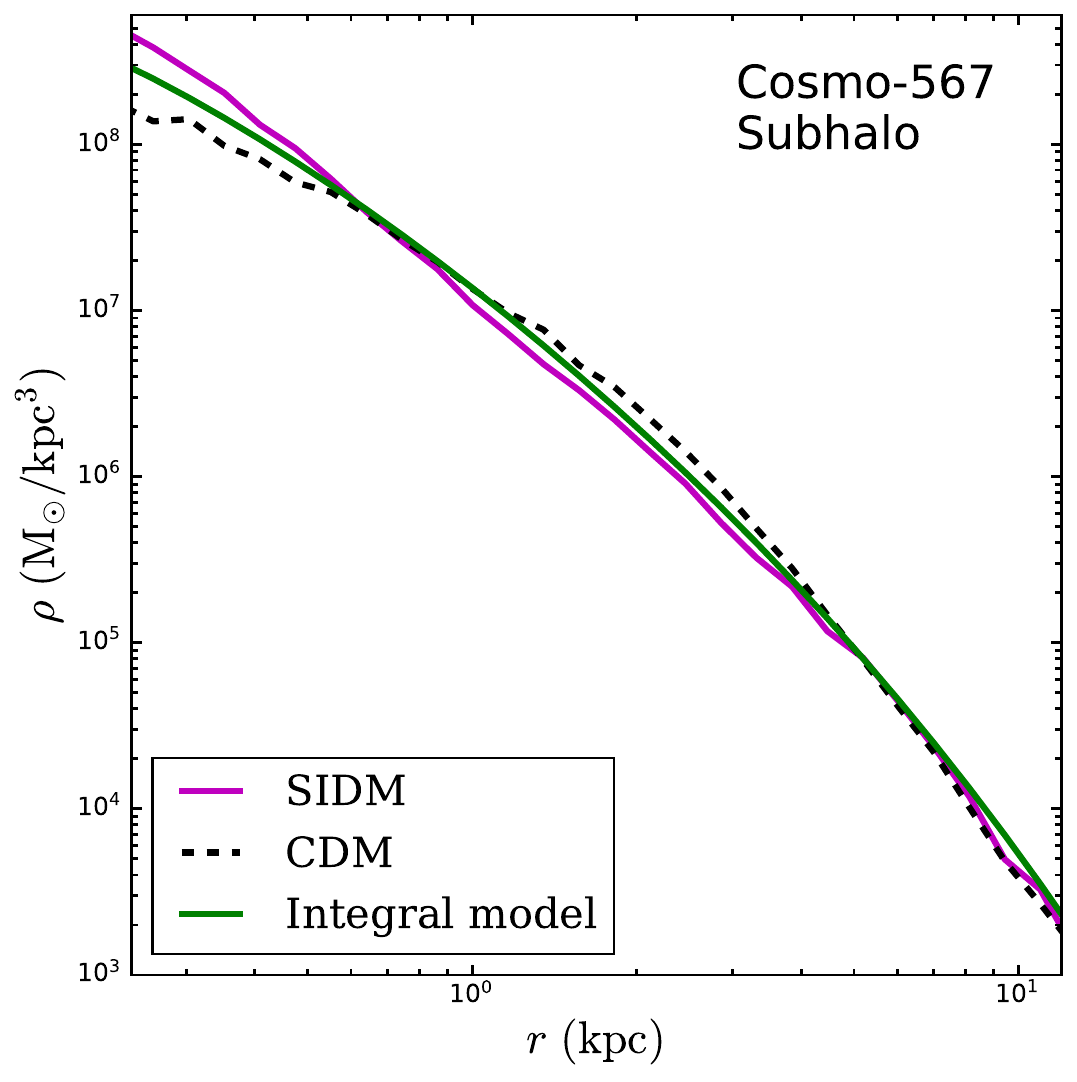}
  \includegraphics[height=4.8cm]{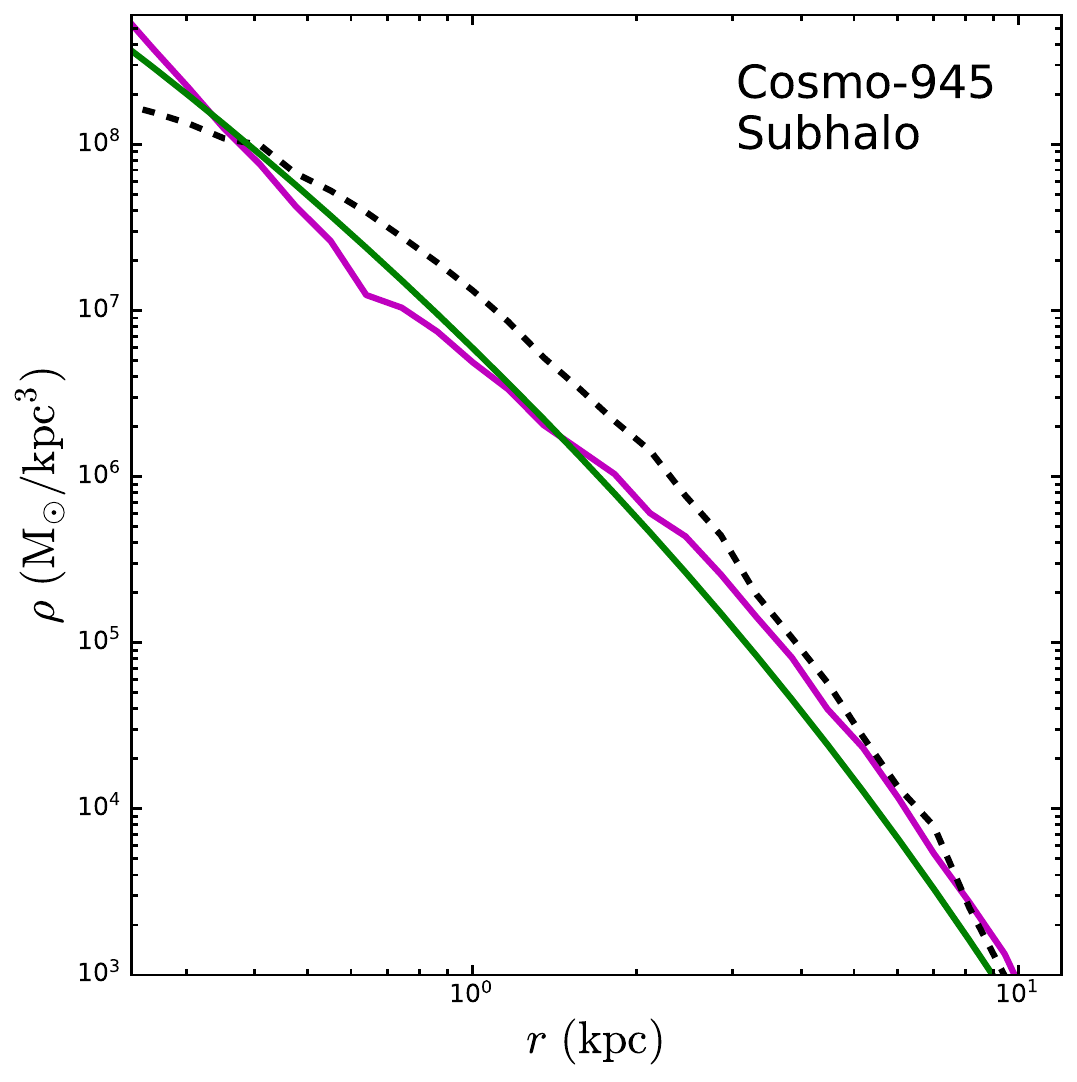}
  \includegraphics[height=4.8cm]{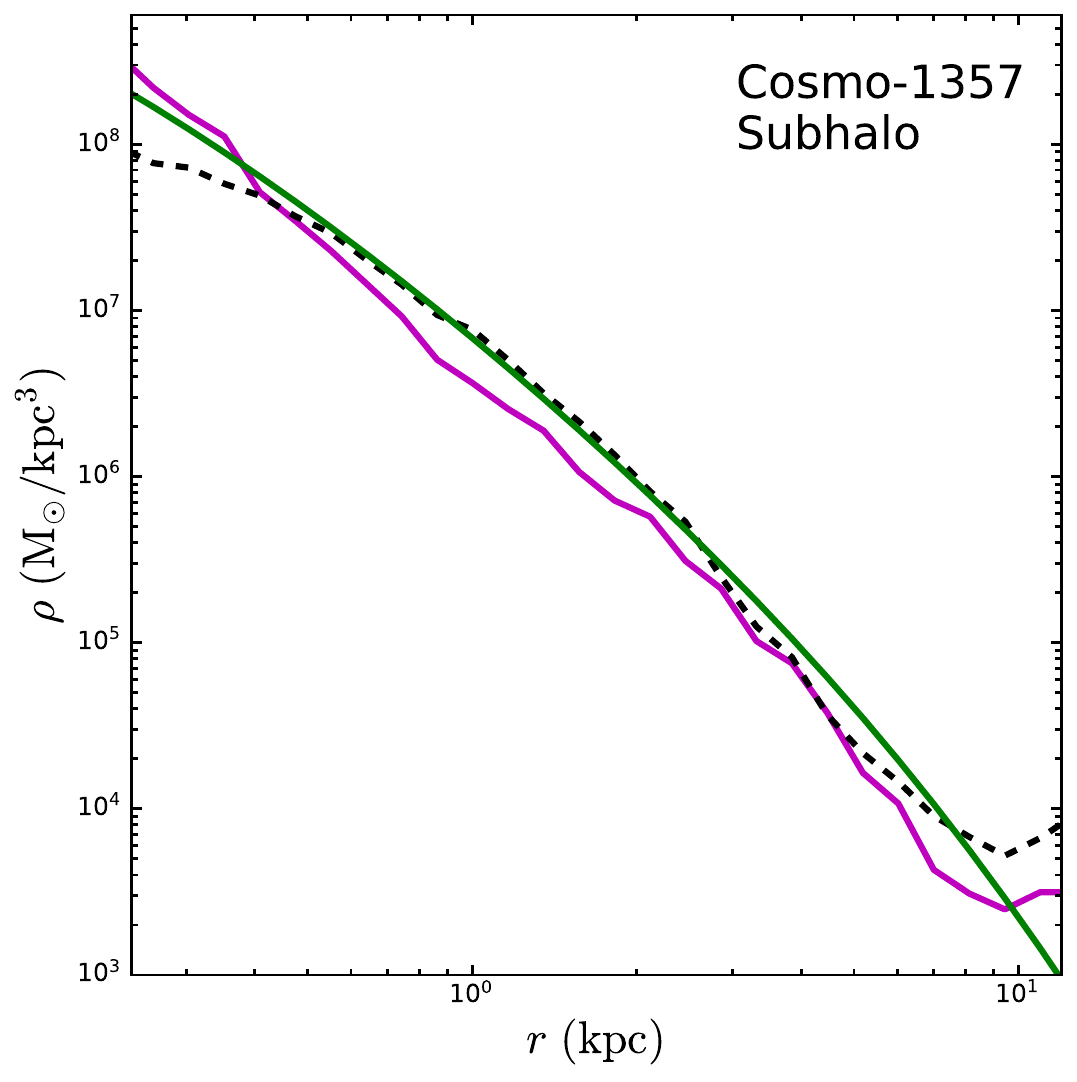}\\
  \includegraphics[height=4.8cm]{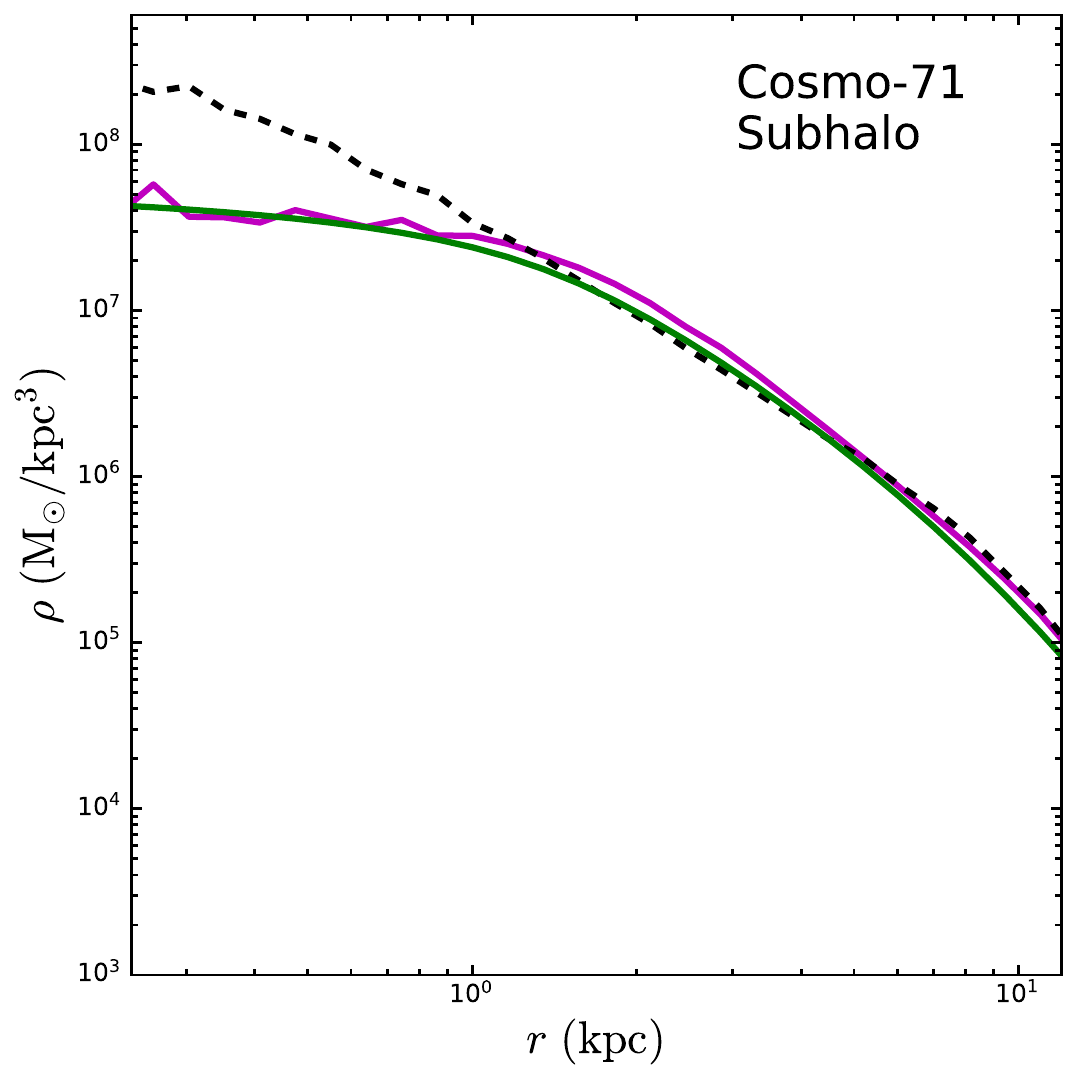} 
  \includegraphics[height=4.8cm]{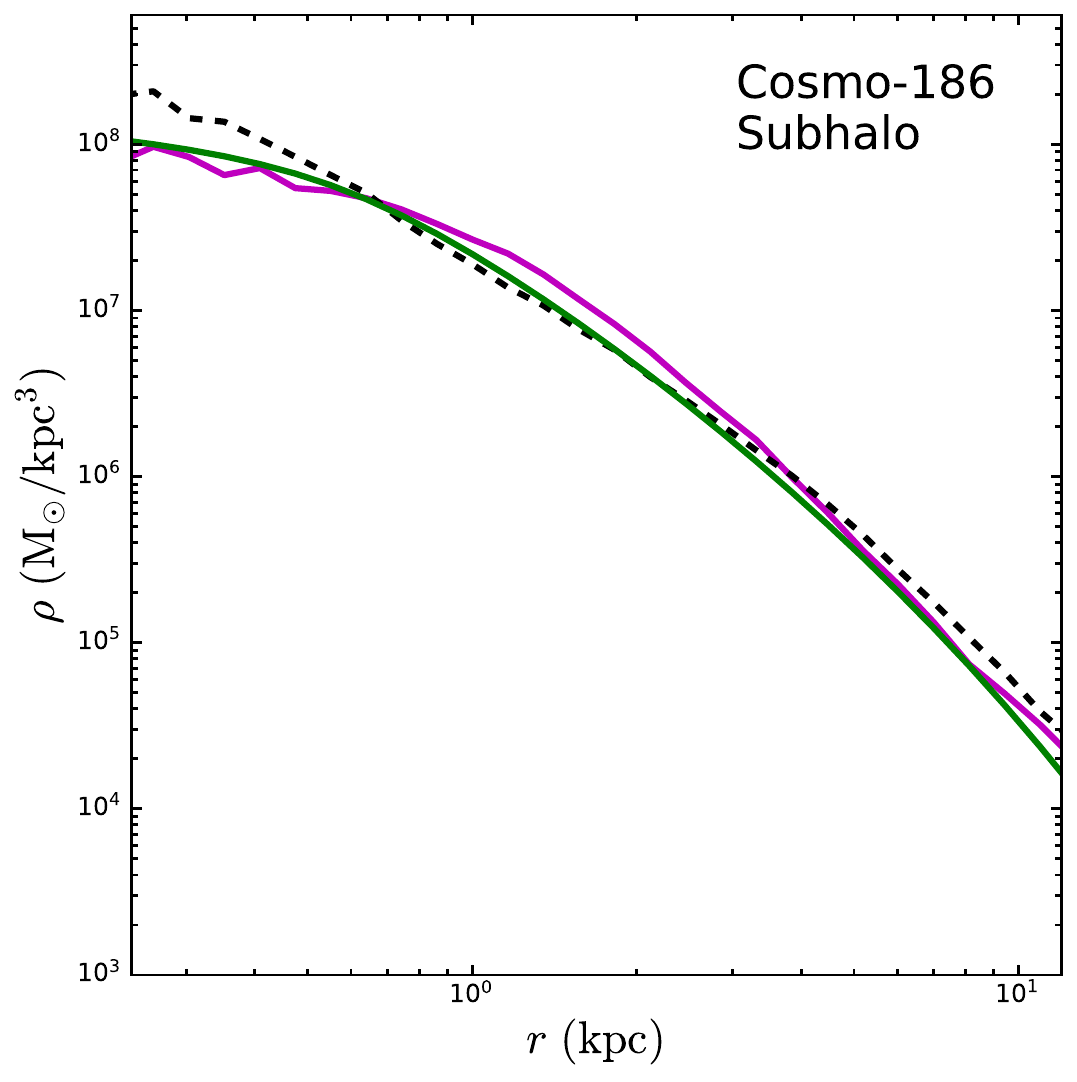}
  \includegraphics[height=4.8cm]{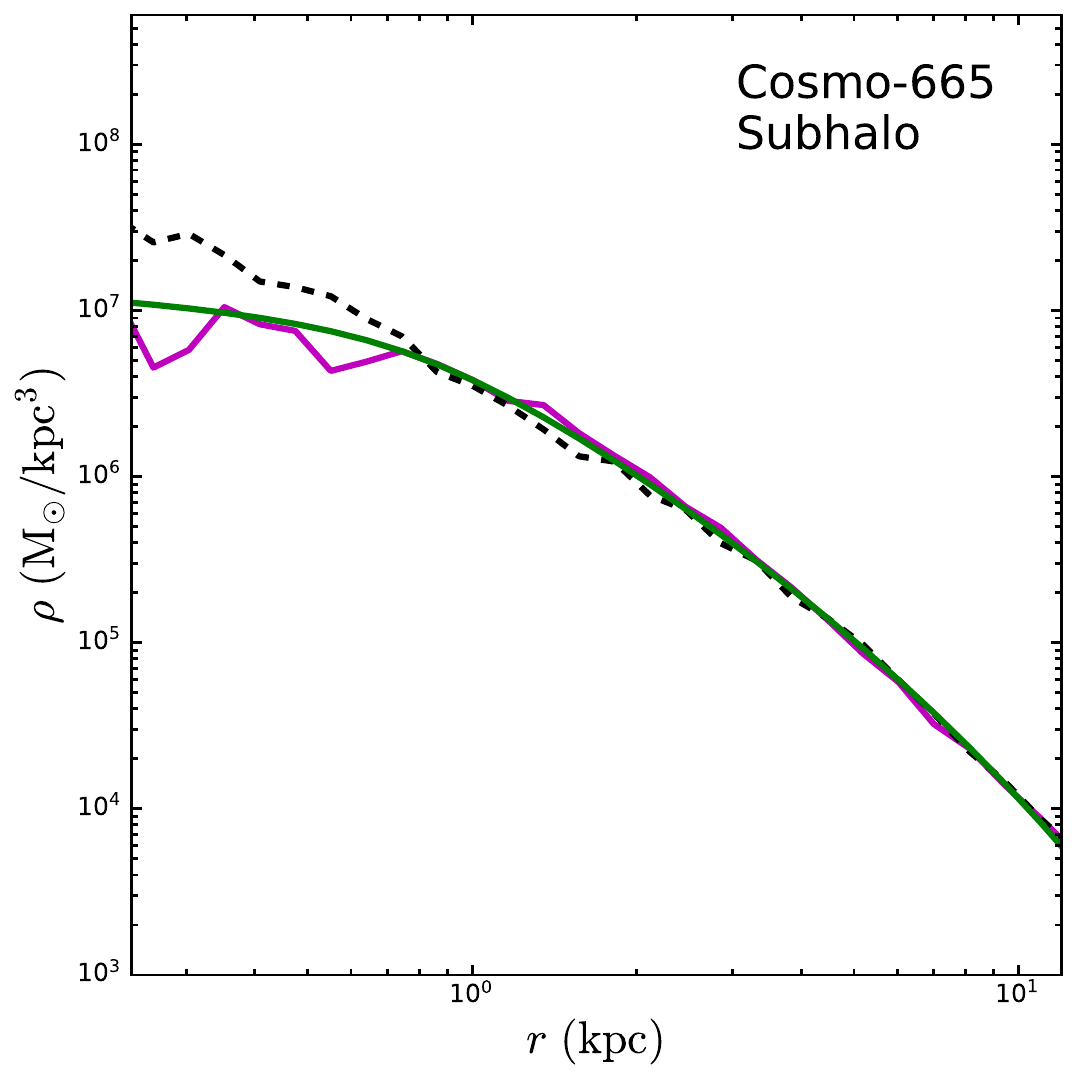}
  \caption{\label{fig:subBM2} Density profiles of core-collapsing (top) and -forming (bottom) SIDM subhalos at $z=0$ from the parametric model with the integral approach (solid-green) and the SIDM simulation~\cite{Yang:2022mxl} (solid-magenta), as well as their CDM counterparts~\cite{Yang:2022mxl} (dashed-black).
}
\end{figure*}

Figure~\ref{fig:subBM1} shows the $V_{\rm max}$ and $R_{\rm max}$ evolution of three SIDM subhalos that are in the collapse phase at $z=0$ from the parametric model (solid-green) and the N-body simulation~\cite{Yang:2022mxl} (solid-magenta), as well as their CDM counterparts (dashed-black). We see that the model well reproduces the evolution history of $V_{\rm max}$ and $R_{\rm max}$ of the simulated SIDM subhalos, and the agreement is within $10\%$. For the SIDM subhalos, $V_{\rm max}$ increases, while $R_{\rm max}$ decreases continuously for $t_L\gtrsim6~{\rm Gyr}$, a significant deviation from CDM. In figure~\ref{fig:subBM2}, we further show the density profiles for all subhalos listed in table~\ref{tab3} at $z=0$. We again see the agreement between parametric model (solid-green) and simulation (solid-magenta) predictions for both core-forming and -collapsing SIDM subhalos. Their density profiles different from the corresponding CDM ones (dashed-black).

\begin{figure*}[htbp]
  \centering
  \includegraphics[height=4.8cm]{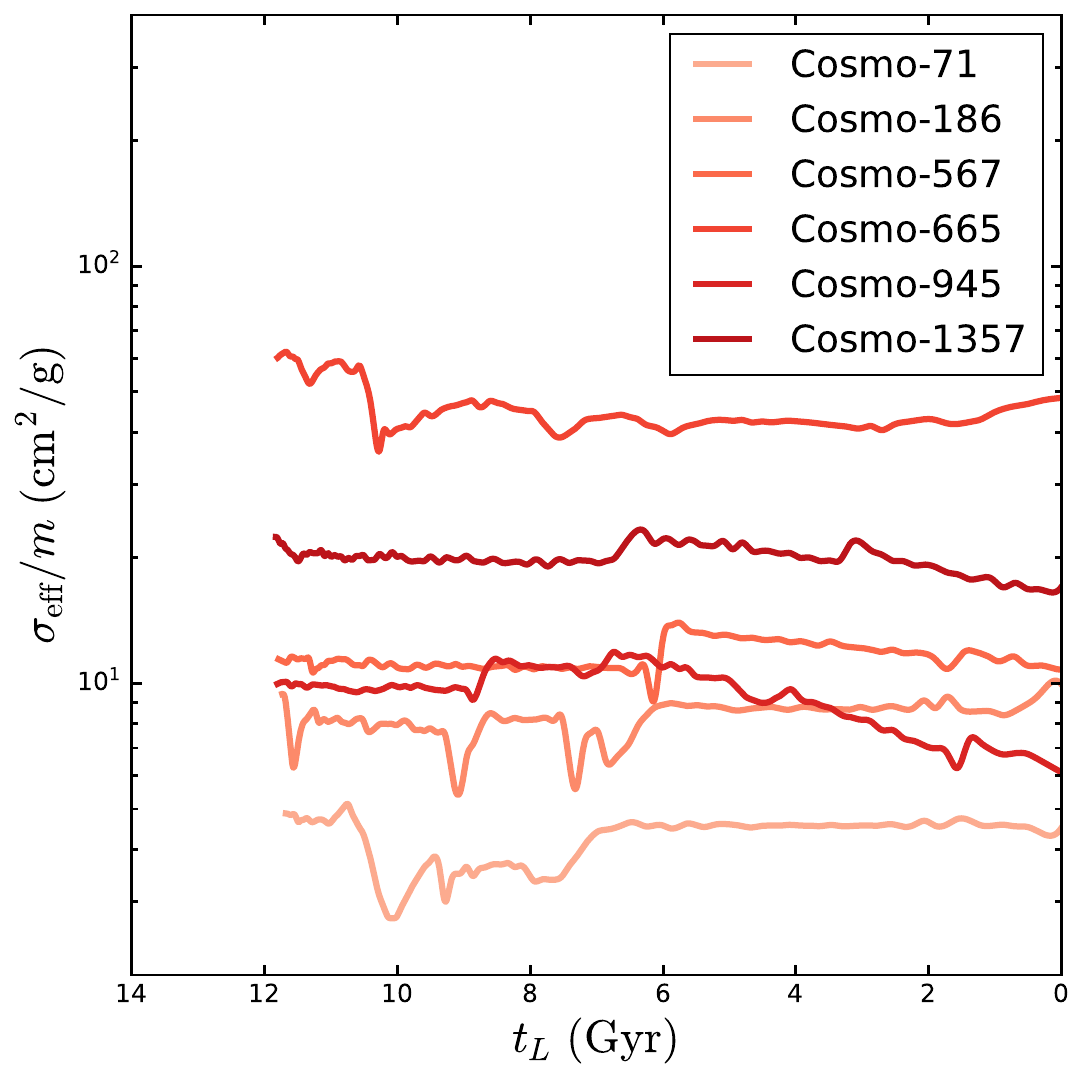}
  \includegraphics[height=4.8cm]{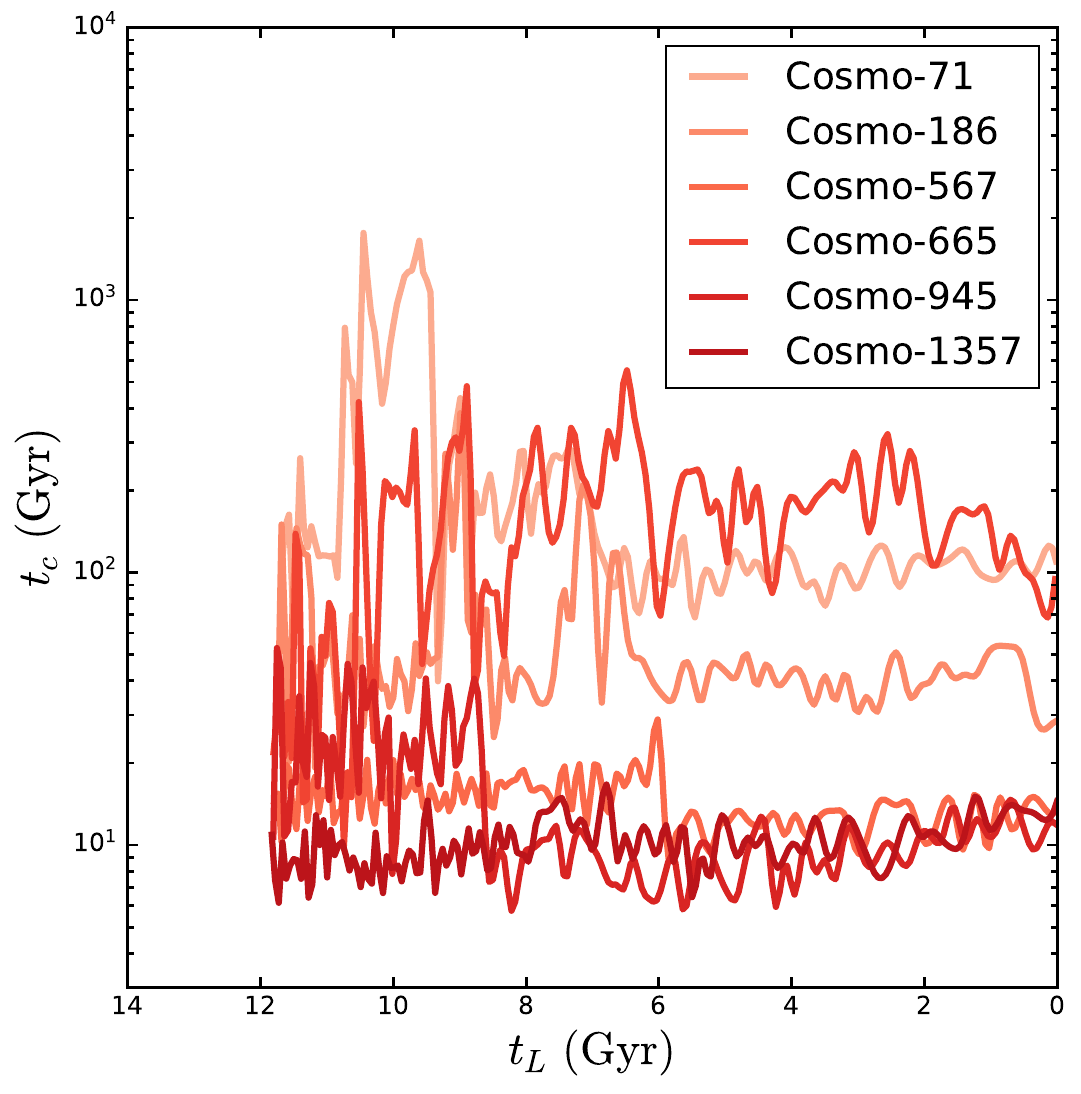}
  \includegraphics[height=4.8cm]{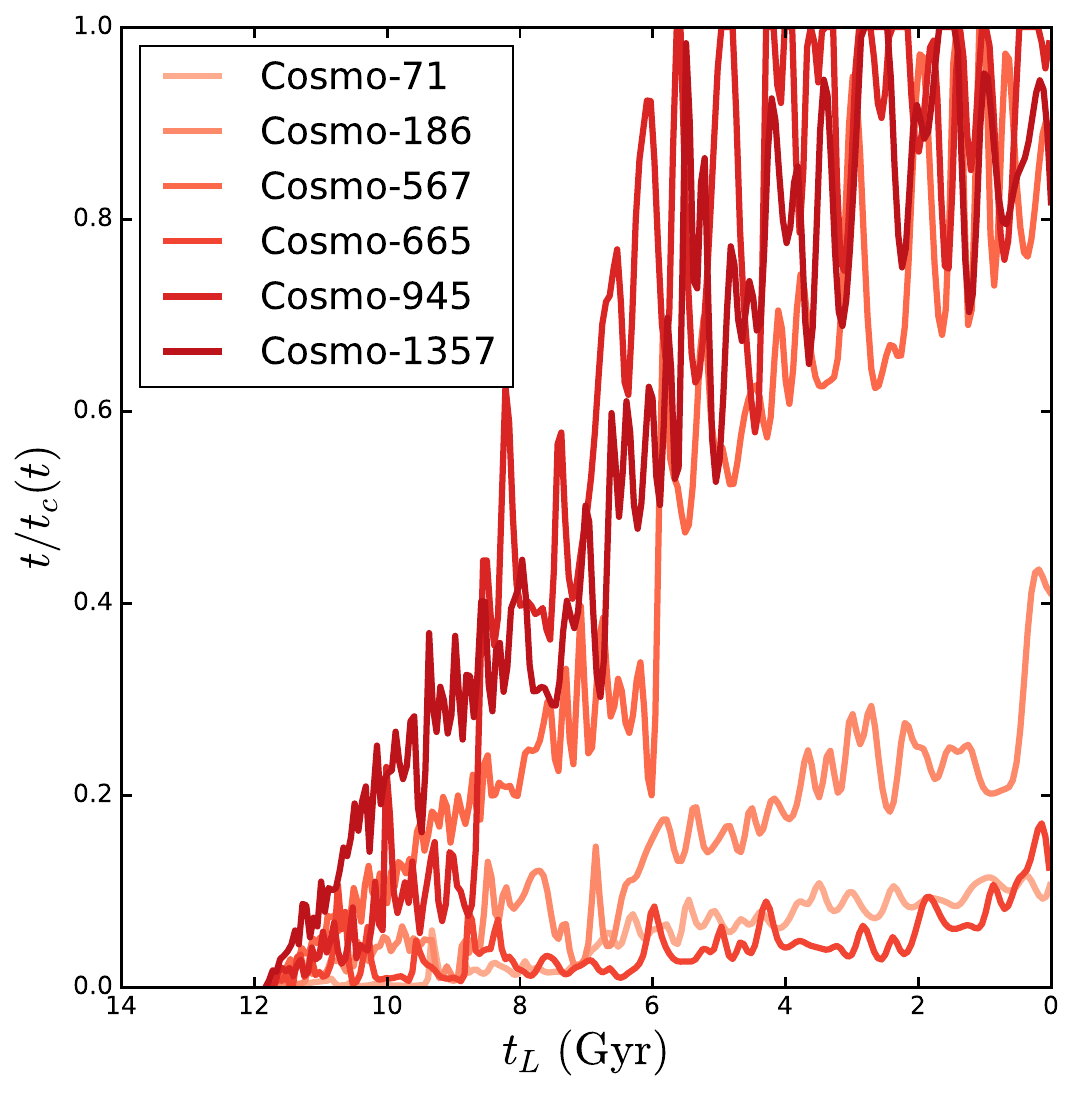}
  \caption{\label{fig:subTraj} The $\sigma_{\rm eff}/m$, $t_c(t)$, and $t/t_c(t)$ evolution of the SIDM subhalos in table~\ref{tab3}, where $t$ is the time duration the halo has evolved since its formation.}
\end{figure*}

As discussed in Sec.~\ref{sec:integral}, $\sigma_{\rm eff}/m$ and $t_c$ evolve with the mass change, which is captured in the integral approach.  
Figure~\ref{fig:subTraj} shows the $\sigma_{\rm eff}/m$, $t_c$ and $t/t_c$ evolution of the $6$ subhalos listed in table~\ref{tab3}. 
Apart from fluctuations, both $\sigma_{\rm eff}/m$ and $t_c(t)$ changed moderately on average, within an order of magnitude. 
The ratio $t/t_c$ has large fluctuations but gradually increases since the halo formation.

\begin{figure*}[htbp]
  \centering
    \includegraphics[width=4.8cm]{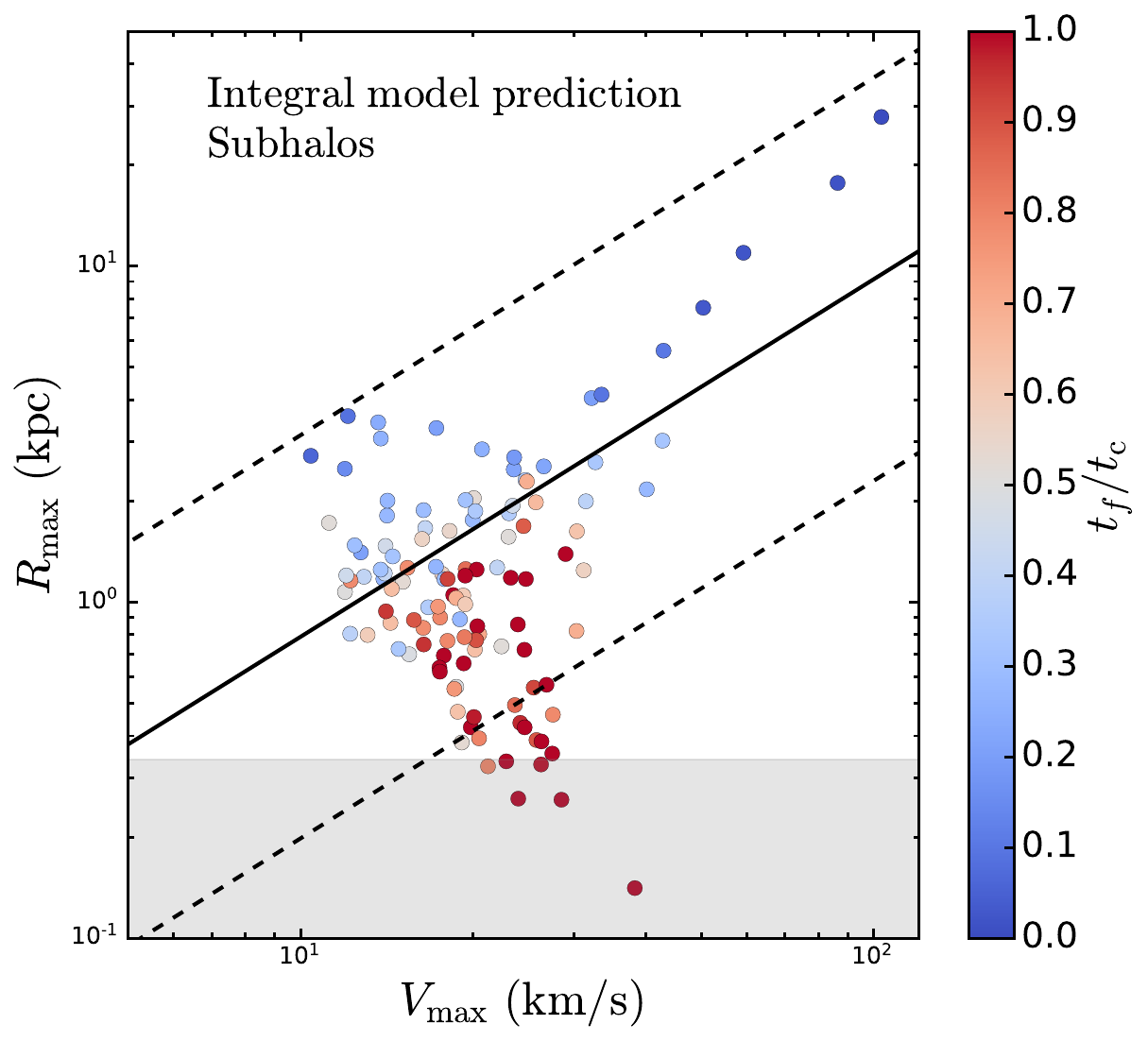}
  \includegraphics[width=4.8cm]{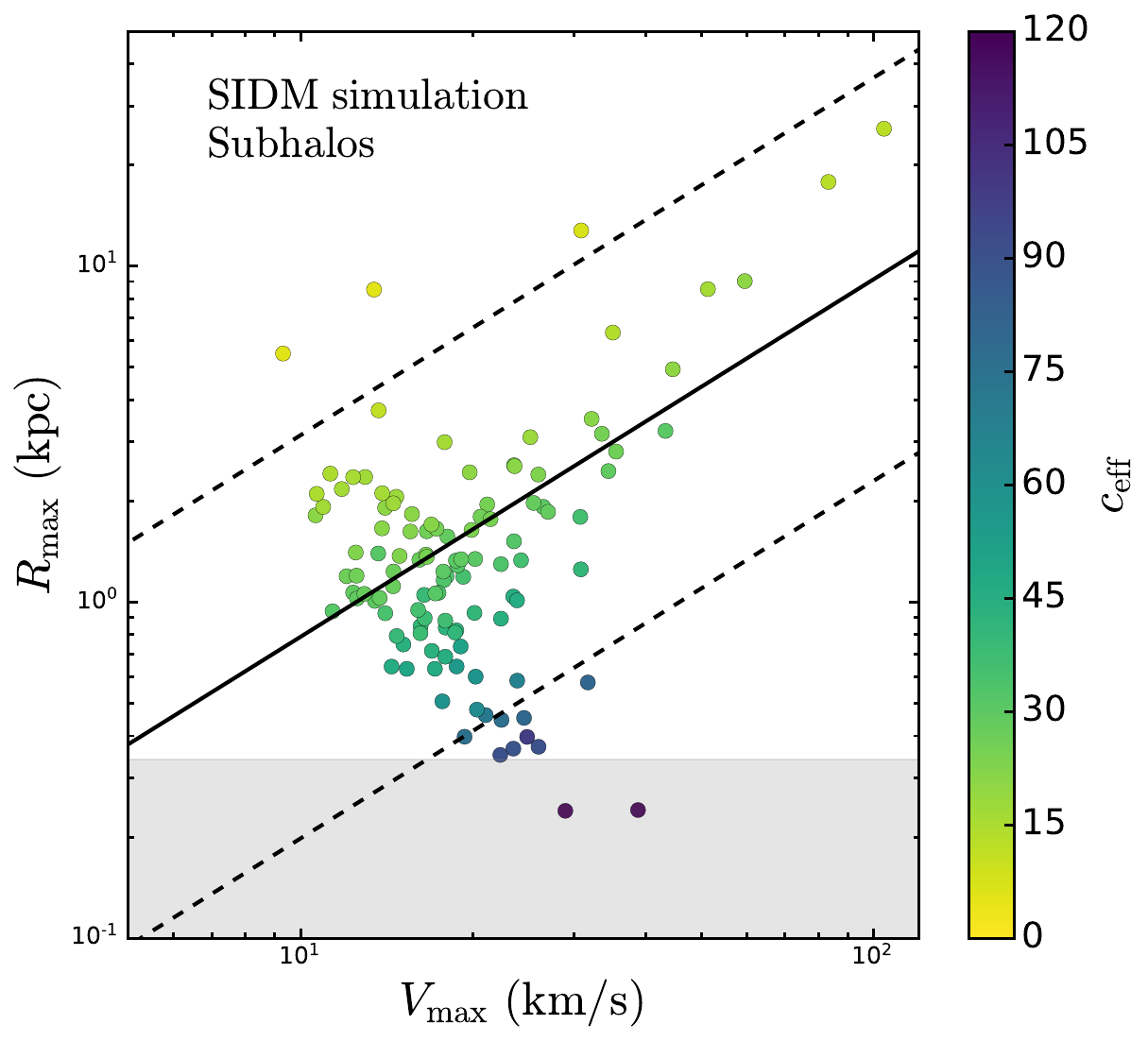}
 \includegraphics[width=4.8cm]{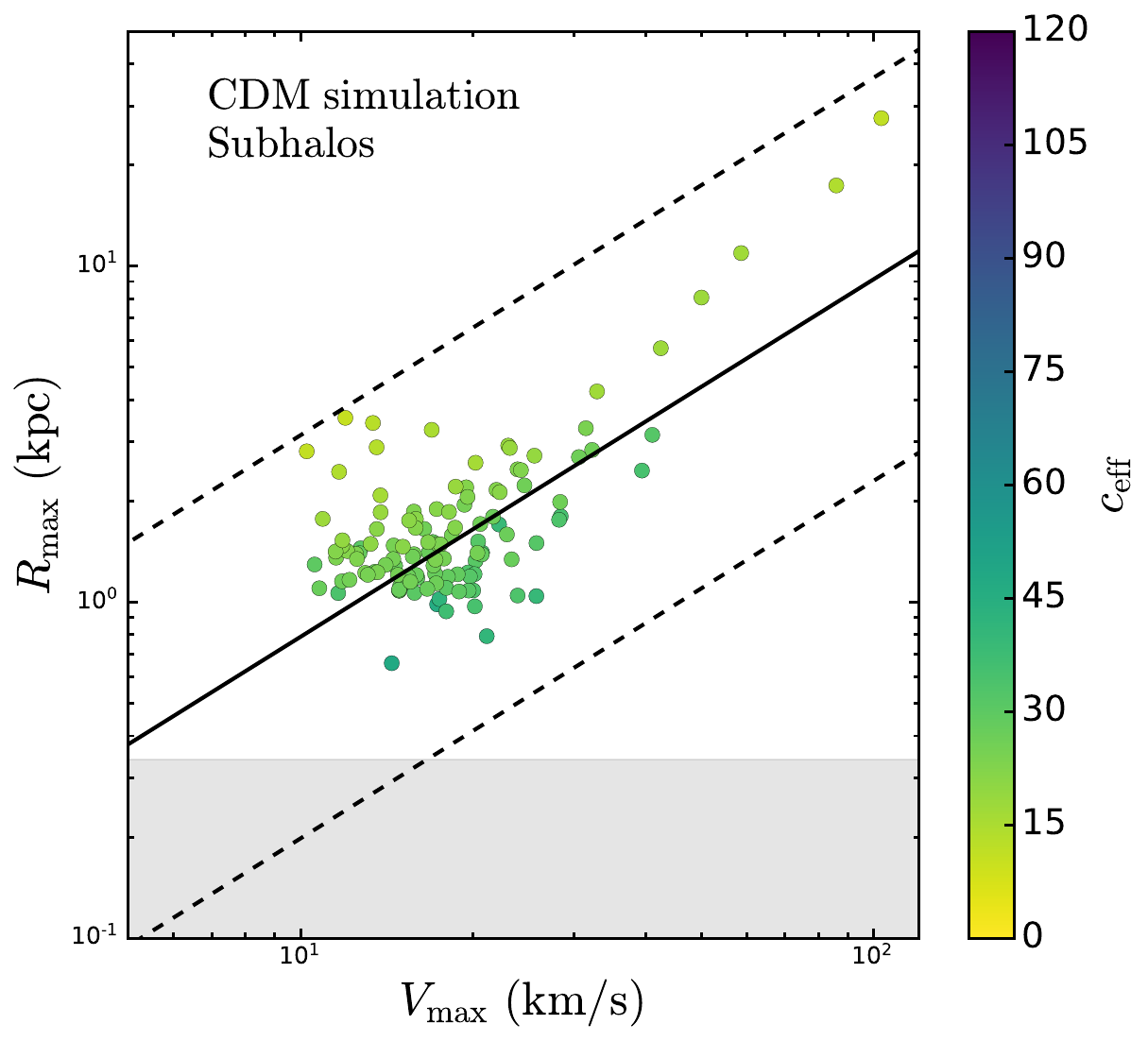}  
 \caption{\label{fig:subhalos} The $V_{\rm max}\textup{--}R_{\rm max}$ distribution of subhalos predicted using the parametric model with the integral approach (left), where each subhalo is colored according to its $t_L(z_f)/t_c$ value. For comparison, the distributions from the SIDM (middle) and CDM (right) simulations are shown; taken from ref.~\cite{Yang:2022mxl}. As in~\cite{Yang:2022mxl}, the median CDM subhalo $V_{\rm max}\textup{--}R_{\rm max}$ relation (solid-black) together with a $\pm0.6$ dex band (dashed-black), and resolution limit (gray-shaded) are shown. 
}
\end{figure*}

We have applied the parametric model to all $125$ resolved CDM subhalos in the cosmological simulation of a Milky Way analog in ref.~\cite{Yang:2022mxl} and obtain their SIDM predictions in the $R_{\rm max}\textup{--}V_{\rm max} $ plane. Since some of the low-mass simulated subhalos that undergo strong stripping may suffer from numerical artifacts, we set $V_{\rm max}=2~{\rm km/s}$ and $R_{\rm max}=0.1~{\rm kpc}$, if their values drop below the corresponding ones when performing the discrete summation in eq. (\ref{eq:int}). With this condition, we avoid producing unphysical results while preserving the numerical error in the same order as the numerical noise induced by the resolution limit.

Figure~\ref{fig:subhalos} shows the $R_{\rm max}\textup{--}V_{\rm max}$ distributions of the SIDM subhalos predicted using the parametric model (left) and the N-body simulation~\cite{Yang:2022mxl} (middle), compared to the simulated CDM subhalos~\cite{Yang:2022mxl} (right). We see that our model successfully reproduces the main trends of the simulated SIDM subhalos, especially the spread towards the lower-right region in the $R_{\rm max}\textup{--}V_{\rm max}$ plane, corresponding to a population of collapsed subhalos. In the upper-left region, the model prediction misses a few subhalos with large $R_{\rm max}$ seen in the SIDM simulation. For these subhalos, the tidal mass loss is amplified due to SIDM core formation and they become more diffuse, compared to CDM. Our current model does not capture this effect, as we have assumed that the tidal mass loss rate of an SIDM subhalo is similar to that of its CDM counterpart. The assumption could be violated in some extreme cases, see, e.g.,~\cite{Sameie:2019zfo,Yang:2020iya,2022NatAs...6..496M,Yang:2022mxl,Carleton180506896}. A dedicated study on this aspect is beyond the scope of this work, and we leave it for future work.

\section{Exploring SIDM parameter space using the parametric model}
\label{sec:scan}

In Sec.~\ref{sec:isolated}, we applied the parametric model to the isolated CDM halos in ref.~\cite{Yang:2022mxl} and obtained their SIDM counterparts, whose properties well agree with those from the cosmological N-body simulation. The simulation in ref.~\cite{Yang:2022mxl} assumes $\sigma_0/m=147.1~{\rm cm^2/g}$ and $w=24.33~{\rm km/s}$. Since our model is flexible, it becomes possible to efficiently sample large parameter space of $\sigma_0/m$ and $w$, which can be in turn mapped to fundamental particle physics parameters, such as mediator mass and coupling constant; see ref.~\cite{Yang220503392} for details. As a demonstration, we take the $647$ isolated CDM halos in~\cite{Yang:2022mxl} as input and apply the parametric model, while varying $\sigma_0/m$ and $w$ over a wide range.

\begin{figure*}[htbp]
  \centering
  \includegraphics[height=6.9cm]{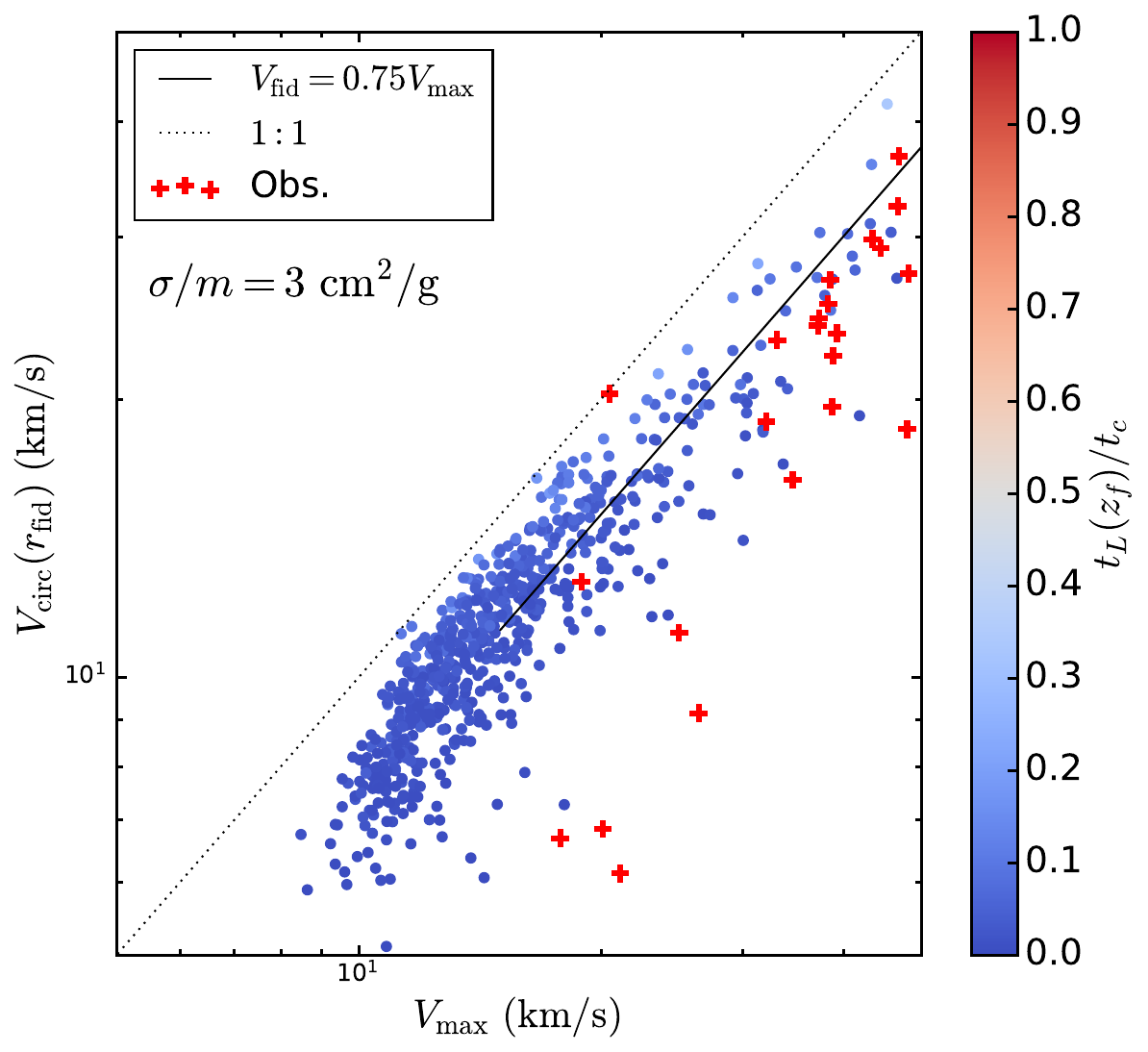} 
  \includegraphics[height=6.9cm]{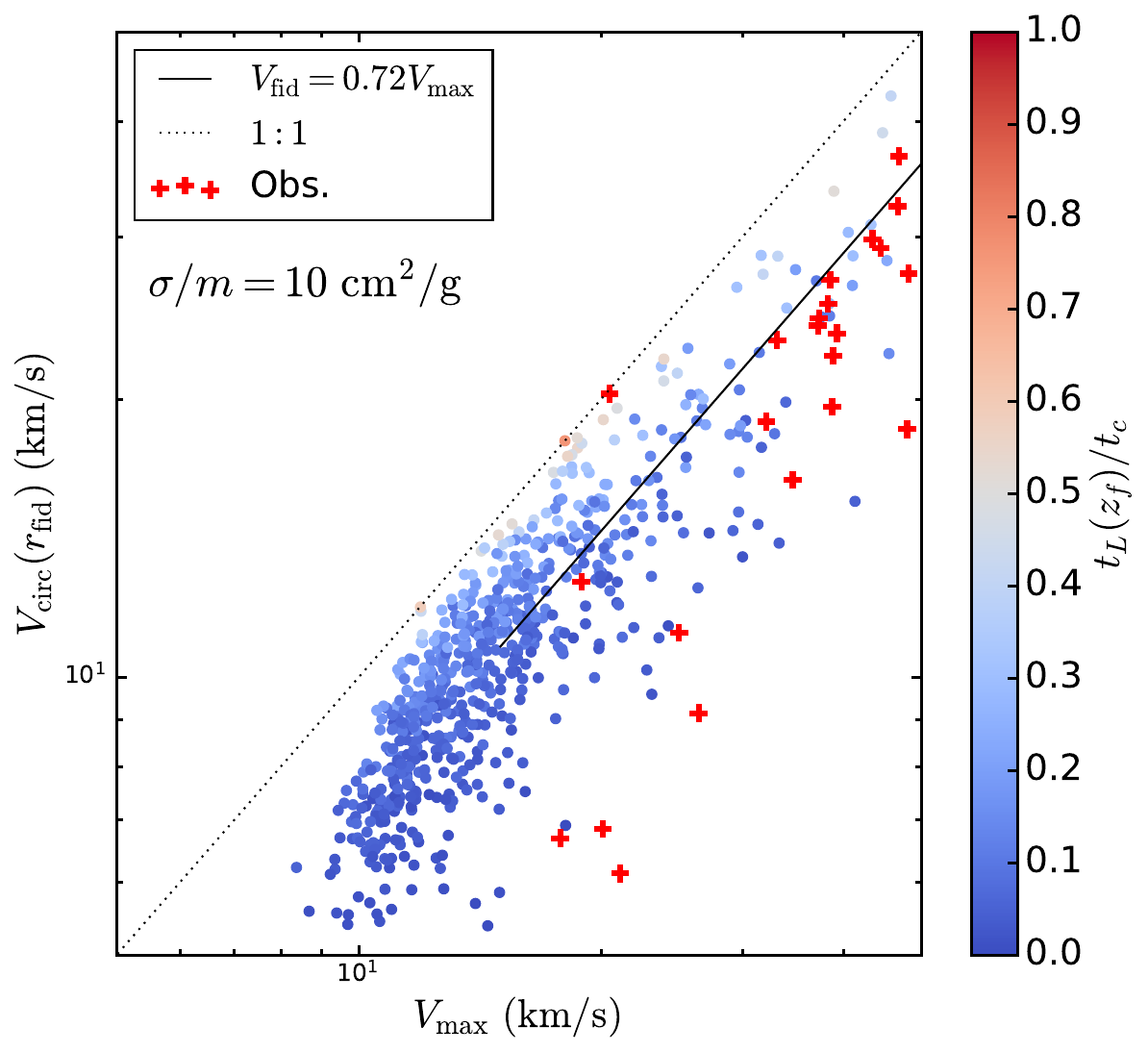} \\
  \includegraphics[height=6.9cm]{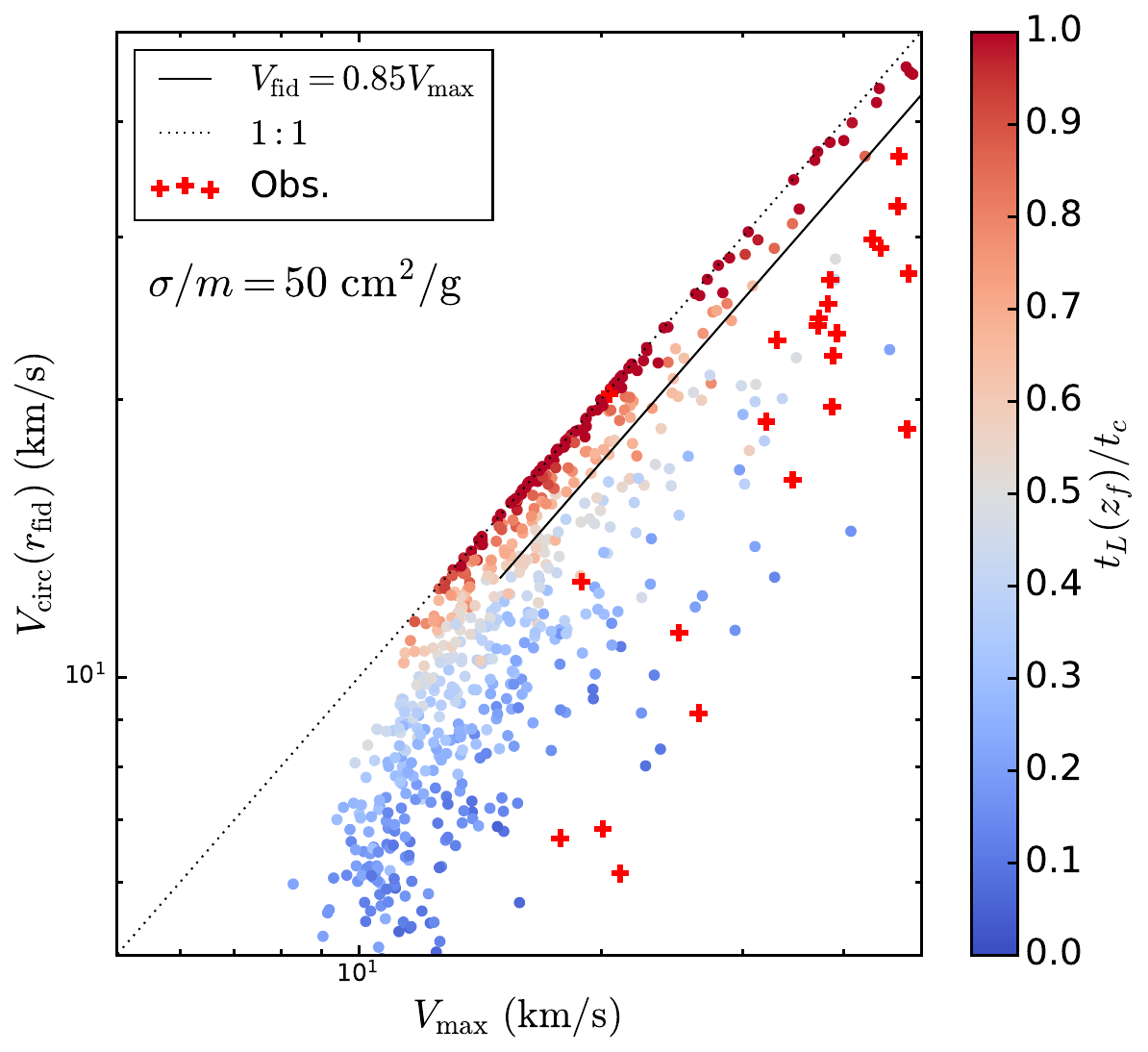} 
  \includegraphics[height=6.9cm]{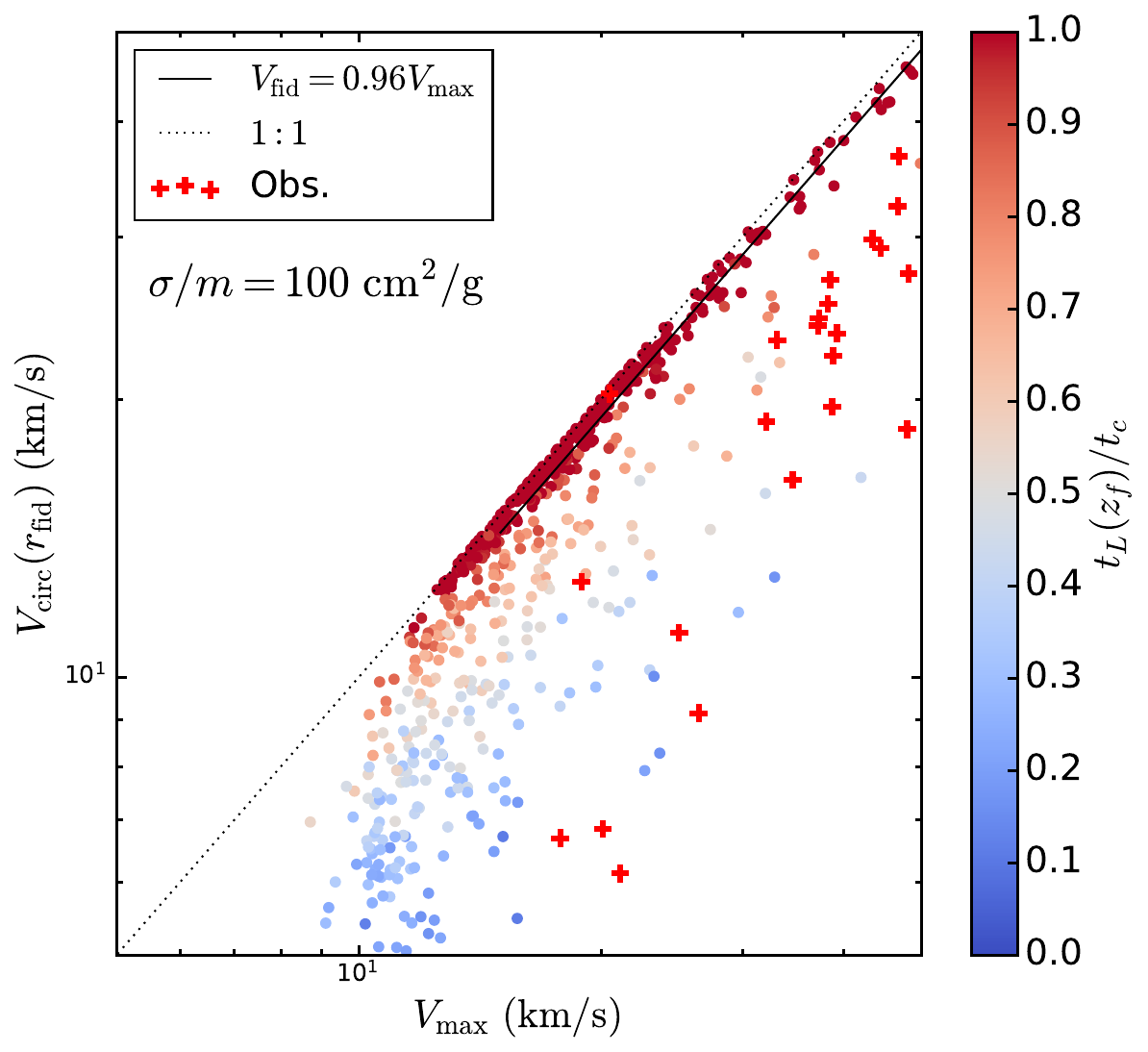} 
  \caption{\label{fig:eg3} The $V_{\rm circ}(r_{\rm fid})\textup{--}V_{\rm max} $ distribution of isolated halos predicted using the parametric model with the basic approach for the constant cross sections $\sigma/m=3~{\rm cm^2/g}$, $10~{\rm cm^2/g}$, $50~{\rm cm^2/g}$ and $100~{\rm cm^2/g}$ (circle). Each halo is colored according to $t_L(z_f)/t_c$. The $V_{\rm circ}(r_{\rm fid})\textup{--}V_{\rm max}$ relation from fitting to the halos in the range $15~{\rm km/s}<V_{\rm max}<50~\rm km/s$ (solid-black), the $1\textup{:}1$ relation (dotted-black), and the data points from observed galaxies compiled in~\cite{2020MNRAS.495...58S} (red cross) are shown for comparison.
}
\end{figure*}

We first consider four representative constant cross sections to gain insights: $\sigma/m=3~\rm cm^2/g$, $10~\rm cm^2/g$, $50~\rm cm^2/g$, and $100~\rm cm^2/g$. In figure~\ref{fig:eg3}, we show the $V_{\rm circ}(r_{\rm fid})\textup{--}V_{\rm max}$ distribution predicted using the parametric model with the basic approach (circle), where each halo is colored according to its $t_L(z_f)/t_c$ value. For comparison, we also include observed galaxies with data compiled in~\cite{2020MNRAS.495...58S} (cross). For $\sigma/m=3~\rm cm^2/g$, all halos are in the core-forming phase, $V_{\rm circ}(r_{\rm fid})$ is systematically lower and its spread for fixed $V_{\rm max}$ becomes larger, compared to CDM~\cite{Yang:2022mxl}. As the cross section increases, the number of core-collapsing halos increases accordingly. When $\sigma/m=50\textup{--}100~{\rm cm^2/g}$, a significant number of halos are in the collapse phase and they have a high inner density, close to the $1\textup{:}1$ line in the $V_{\rm circ}(r_{\rm fid})\textup{--}V_{\rm max}$ plane. Meanwhile, the most diffused halos are still in the core-forming phase and remain far off the $1\textup{:}1$ line, even for $\sigma/m=100~\rm cm^2/g$. These halos could potentially host ultra-diffuse galaxies in the field~\cite{Kong220405981}.

The deviation of the $V_{\rm circ}(r_{\rm fid})\textup{--}V_{\rm max}$ distribution from the $1\textup{:}1$ line provides a useful measure to characterize the SIDM effects. We fit the halos with a relation of $V_{\rm circ}(r_{\rm fid}) = b~ V_{\rm max}$ for the range of $15~{\rm km/s}<V_{\rm max} < 50~{\rm km/s}$, consistent with that for observed galaxies shown in figure~\ref{fig:eg3}, and minimize the absolute mean distance from the $1\textup{:}1$ line to determine the parameter $b$; see figure~\ref{fig:eg3} (solid-black line). Our fit tends to split the selected halos into two equal halves on two sides of the fitted line. For the CDM counterparts, $b=0.82$. As the cross section increases, the $b$ value decreases first as the core size increases for most halos, but it increases again as more and more halos are collapsed. For $\sigma/m=100~{\rm cm^2/g}$, $b=0.96$. As a reference, $b=0.60$ for the observed galaxies shown in figure~\ref{fig:eg3}. We expect that there is significant coverage bias for the sample of observed galaxies, and hence it is challenging to make a quantitative comparison between observations and SIDM predictions based on the $b$ value. Nevertheless, it is a useful indicator for the overall significance of dark matter self-interactions over a population of halos.

\begin{figure*}[htbp]
  \centering
  \includegraphics[height=8cm]{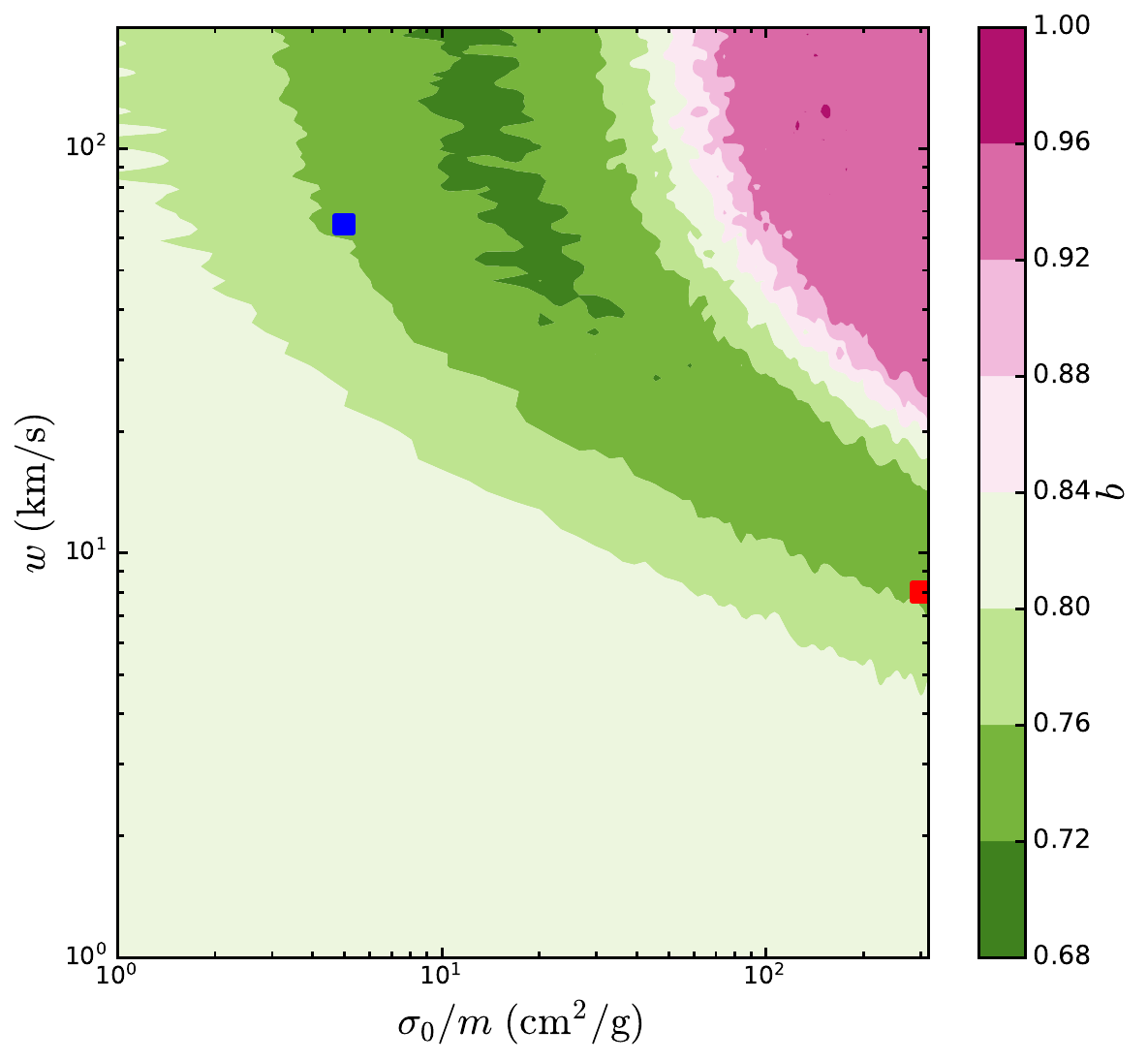}   
  \caption{\label{fig:scan} 
Contours of the fitted $b$ value in the $w\textup{--}\sigma_0/m$ plane, where $\sigma_0/m$ is the overall normalization factor of the self-scattering cross section and $w$ controls the velocity dependence; see eq.~(\ref{eq:xsr}). The squares denote the two SIDM models on the same contour with $b=0.75$: $(\sigma_0/m,w)=(300~\rm cm^2/g,8~\rm km/s)$ (red) and $(5~\rm cm^2/g, 65~\rm km/s)$ (blue), which will be studied further.}
\end{figure*}

In figure~\ref{fig:scan}, we show contours of the fitted $b$ value based on a scan with a grid of $100\times 100$ points in the $w\textup{--}\sigma_0/m$ plane, where $\sigma_0/m$ is the overall normalization factor of the cross section and $w$ controls the velocity dependence. We see the distribution of $b$ has nontrivial features, including a valley (darker green) and a hill (darker pink). In the lower-left region, the SIDM effect is small and the $b$ value reduces to the CDM case around $0.82$. For $w\sim{\cal O}(10)~\rm  km/s$, the increase of $\sigma_0/m$ first decreases $b$, enlarging the spread of halos in the $V_{\rm circ}(r_{\rm fid})\textup{--}V_{\rm max}$ from the $1\textup{:}1$ line. When $\sigma_0/m$ further increases, $b$ increases as well because more and more halos enter the collapse phase. These nontrivial features could help us distinguish different particle physics models of SIDM.

\begin{figure*}[htbp]
  \centering
  \includegraphics[height=6.7cm]{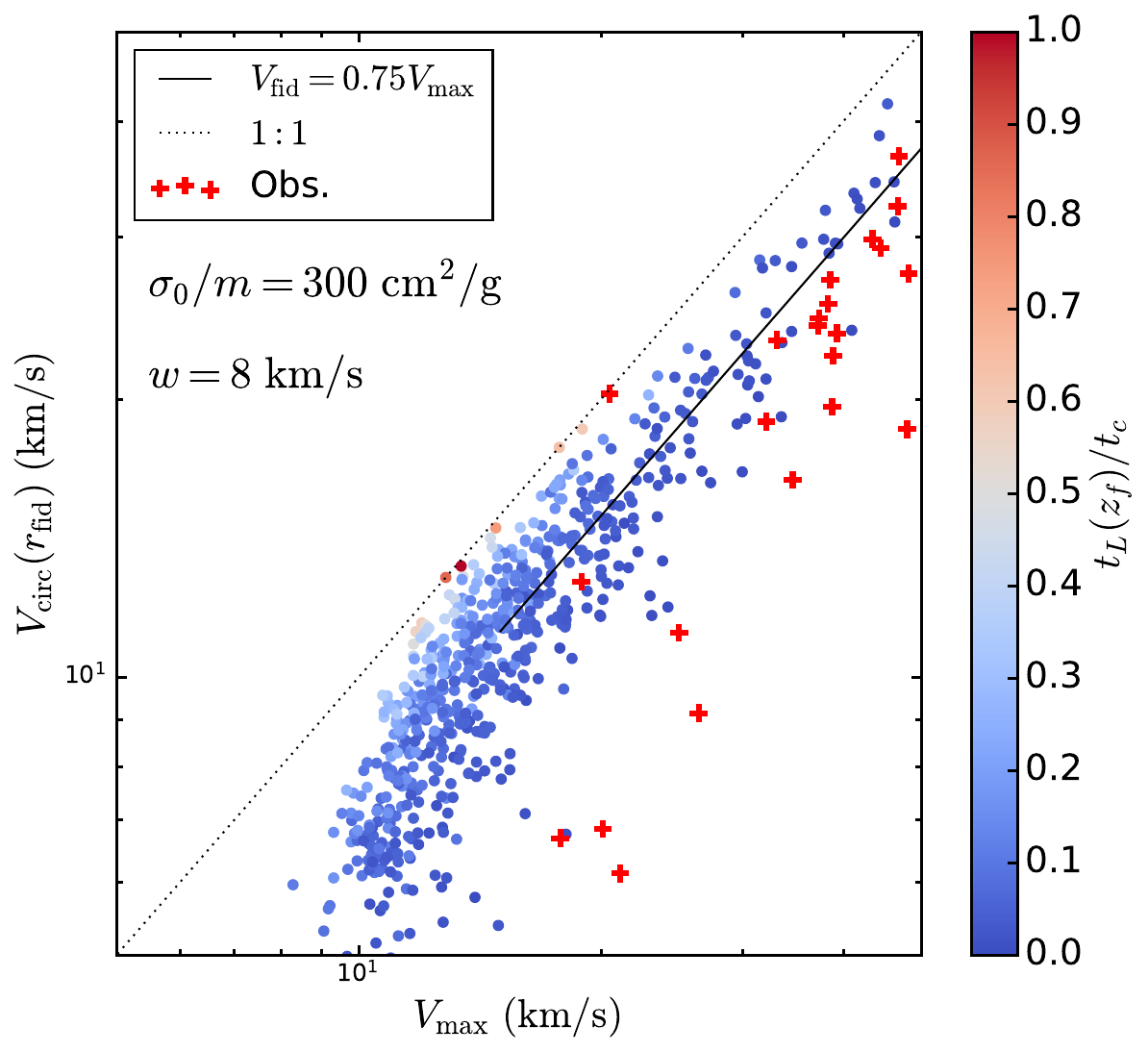}
  \includegraphics[height=6.7cm]{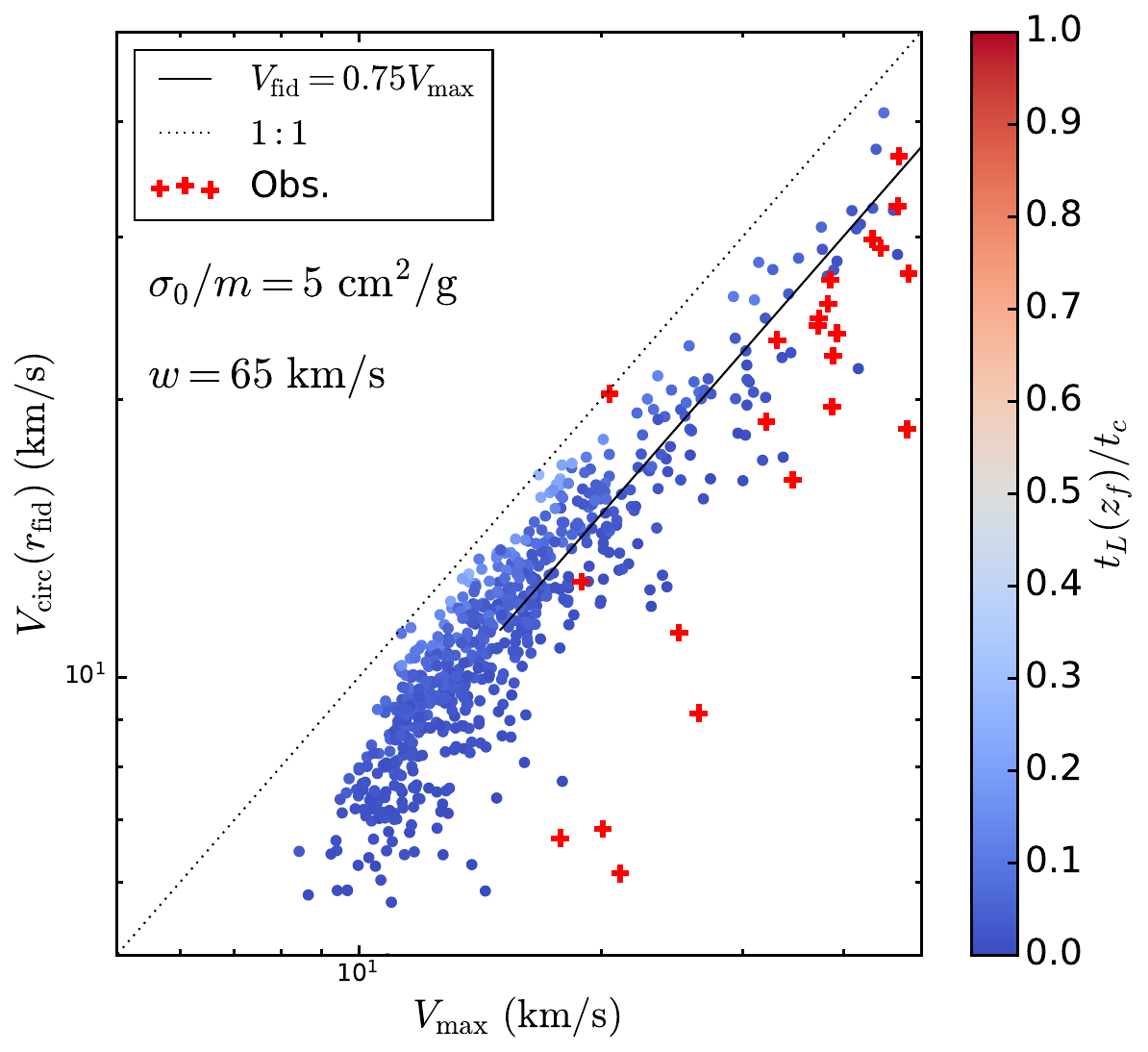}
  \caption{\label{fig:x1} The $V_{\rm circ}(r_{\rm fid})\textup{--}V_{\rm max} $ distribution of isolated halos predicted using the parametric model with the basic approach for two SIDM models that have the same $b$ value ($b=0.75$): $(\sigma_0/m,w)=(300~\rm cm^2/g,8~\rm km/s)$ (left) and $(5~\rm cm^2/g, 65~\rm km/s)$ (right). Each halo is colored according to $t_L(z_f)/t_c$. The $V_{\rm circ}(r_{\rm fid})\textup{--}V_{\rm max}$ relation from fitting to the halos in the range $15~{\rm km/s}<V_{\rm max}<50~\rm km/s$ (solid-black), the $1\textup{:}1$ relation (dotted-black), and the data points from observed galaxies compiled in~\cite{2020MNRAS.495...58S} (red cross) are shown for comparison. 
}
\end{figure*}

\begin{figure*}[htbp]
  \centering
  \includegraphics[width=7.2cm]{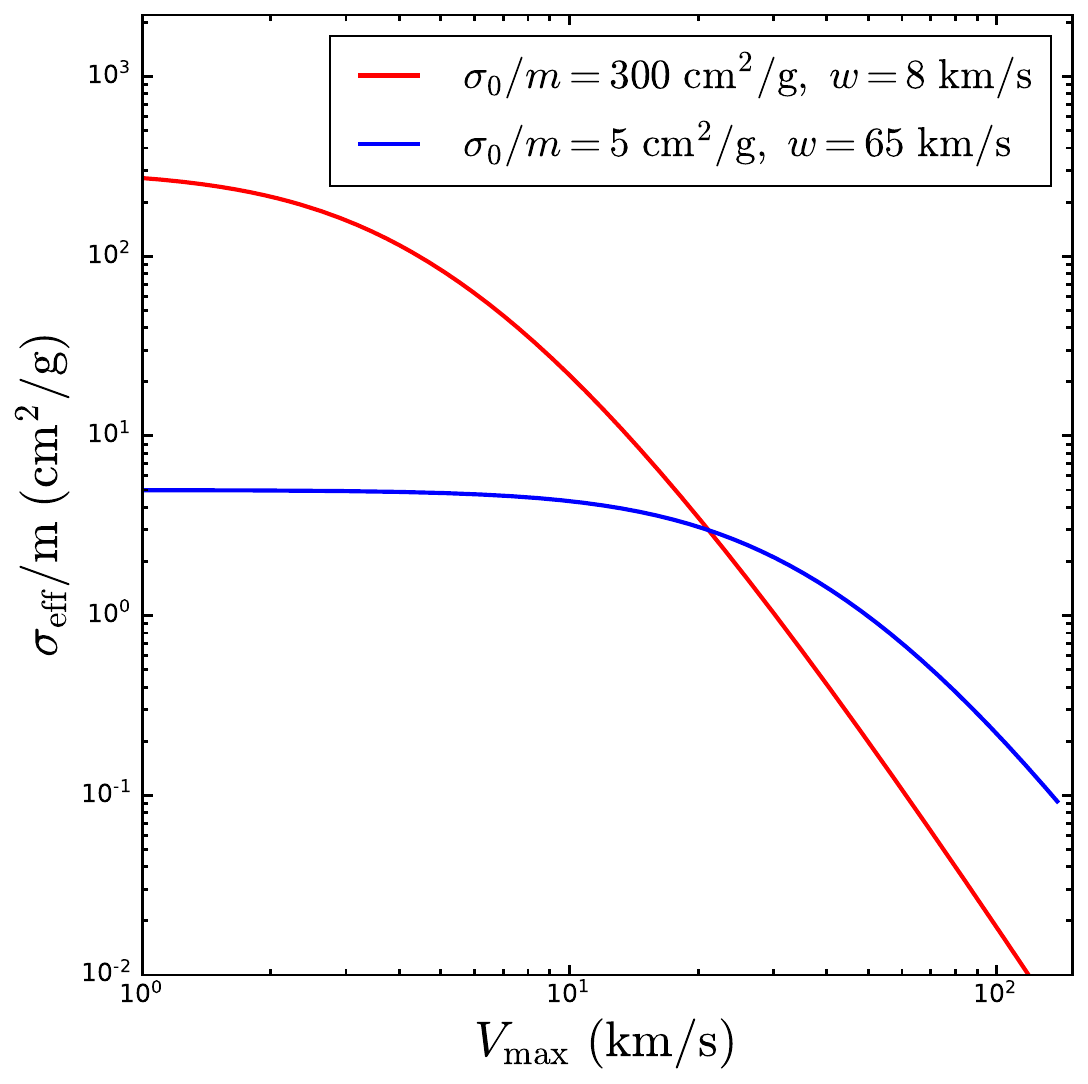}
  \includegraphics[width=7.8cm]{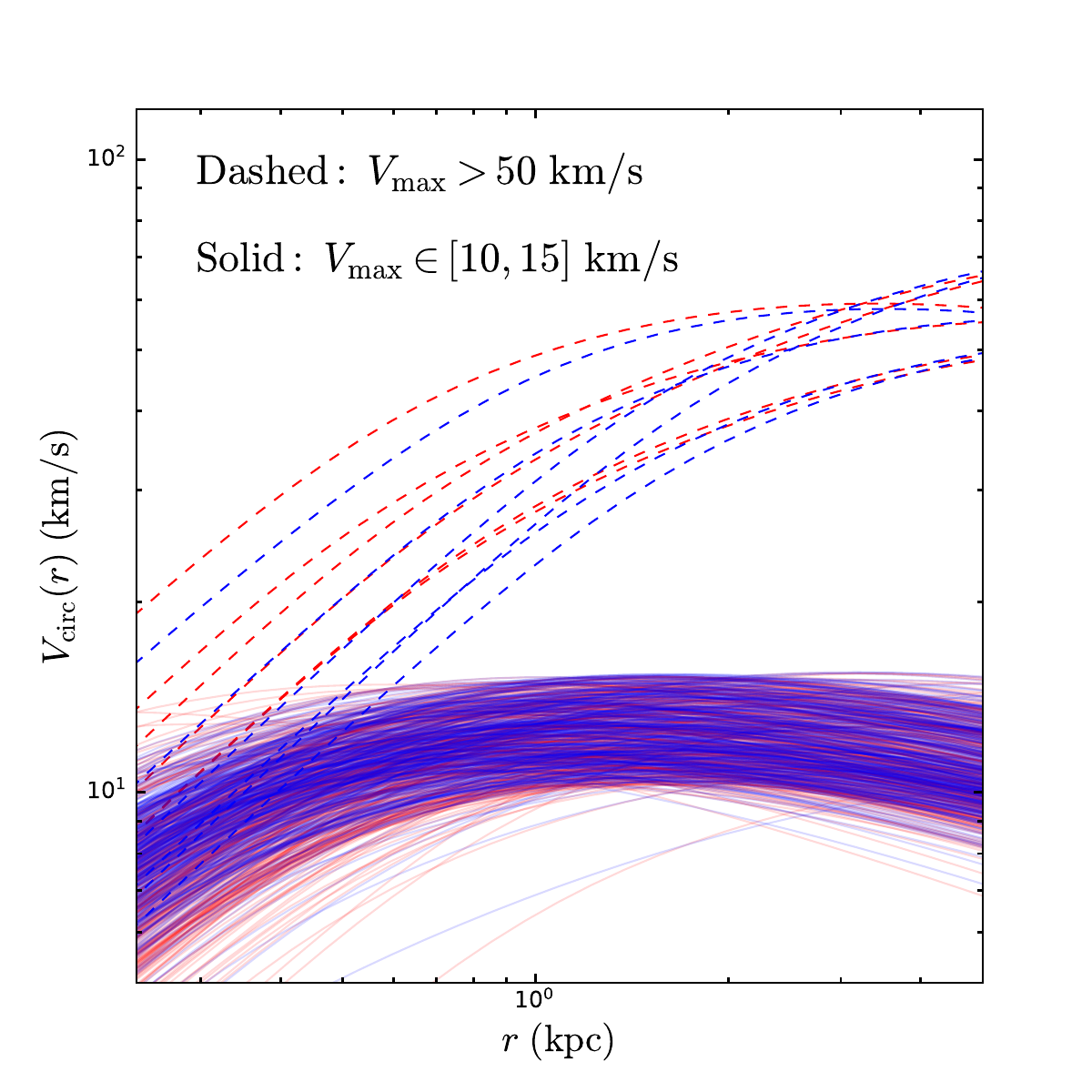}
  \caption{\label{fig:x3}The effective cross section per mass as a function of $V_{\rm max}$ in CDM for the velocity-dependent SIDM models (left): $(\sigma_0/m,w)=(300~\rm cm^2/g,8~\rm km/s)$ (red) and $(5~\rm cm^2/g, 65~\rm km/s)$ (blue); their corresponding rotation curves (right) for isolate halos with $10~{\rm km/s}<V_{\rm max}<15~{\rm km/s}$ (solid) and $V_{\rm max}>50~\rm km/s$ (dashed). 
}
\end{figure*}

To further disentangle the degeneracy effect associated with the parameter $b$, we select two example velocity-dependent SIDM models that have the same $b=0.75$ for a detailed comparison: $(\sigma_0/m,w)=(300~\rm cm^2/g,8~\rm km/s)$ and $(5~\rm cm^2/g, 65~\rm km/s)$, denoted by the red and blue squares in figure~\ref{fig:scan}, respectively. Figure~\ref{fig:x1} shows the $V_{\rm circ}(r_{\rm fid})\textup{--}V_{\rm max}$ distribution for the two models. For $15~{\rm km/s}<V_{\rm max} <50~\rm km/s$, the distribution is similar for both models as expected. Nevertheless, for the halos with $V_{\rm max} \lesssim15~\rm km/s$, the model with $(\sigma_0/m,w)=(300~\rm cm^2/g,8~\rm km/s)$ predicts much larger spread. The former has a much larger effective cross section in the regime $V_{\rm max} \lesssim15~\rm km/s$, as illustrated in figure~\ref{fig:x3} (left), and hence halos are more cored or collapsed, resulting in a larger spread.

On the other hand, for more massive halos with $V_{\rm max}\gg15~{\rm km/s}$, the effective cross section of the $(300~\rm cm^2/g,8~\rm km/s)$ model drops below that of the $(5~\rm cm^2/g, 65~\rm km/s)$ model, while approaching the CDM limit for $V_{\rm max}\gtrsim100~{\rm km/s}$. Thus the former predicts smaller cores and higher inner circular velocities than the latter model, as demonstrated in figure~\ref{fig:x3} (right), where we show the halo rotation curves in the two regimes $10~{\rm km/s}<V_{\rm max}<15~{\rm km/s}$ (solid) and $V_{\rm max}>50~{\rm km/s}$ (dashed) for the models. 

We have demonstrated it is possible to further break the degeneracy and distinguish different SIDM models with similar $b$ values by applying the parametric model to isolated halos over a wide range of masses (and thus velocities). We can further extend the analysis to satellite halos. For example, in the model with $(\sigma_0/m,w)=(300~\rm cm^2/g,8~\rm km/s)$, we expect that most of the ultra-faint dwarf galaxies of the Milky Way would be in the collapse phase, resulting in a dense core. In practice, it is important to take into account observational uncertainties, baryonic contributions to the velocity of tracers, and selection bias of galaxy samples. We will leave the application of the parametric model to observed galaxies for future work.

\section{Conclusions}
\label{sec:discussion}

In this work, we have proposed a parametric model to transfer CDM (sub)halos into their SIDM counterparts. The model is based on an analytical universal density profile, which can accurately describe the dark matter distribution of halos over the course of the evolution. We calibrated the density profile using controlled N-body simulations of an isolated halo and obtained a set of equations capturing the time evolution of the parameters of the density profile. We then introduced two ways of applying the model. The basic approach uses the scale radius and density of an isolated CDM halo ($z=0$) as input and reconstructs the halo's evolution history in the presence of dark matter self-interactions. The integral approach further extends the basic one to incorporate the mass change, and integrates the SIDM effects along the evolution history. Thus it can be used to model gravothermal evolution of subhalos, which suffer from tidal mass loss. 

We have tested and validated the parametric model using zoom-in cosmological SIDM simulations of a Milky Way analog for both isolated halos and subhalos. We further applied it to map out the relevant parameter space for a particle physics model of SIDM. In the future, we could apply the parametric model to state-of-the-art cosmological CDM simulations, e.g., FIRE2~\cite{2018MNRAS.480..800H,Wetzel:2022man} and IllustrisTNG~\cite{Pillepich:2017jle,Nelson:2018uso}, and transform existing CDM (sub)halo properties into SIDM predictions. We could integrate the parametric model together with existing semi-analytical models with halo merger trees~\cite{Behroozi11104370,Somerville:2008bx,Benson:2010kx} and subhalo evolution trajectories~\cite{Ogiya:2019del}. The parametric model is flexible and it provides an efficient tool for identifying favored SIDM parameter regions in light of latest observations of galactic systems, such as satellite galaxies of the Milky Way and ultra-diffuse galaxies.

We could further improve the model. While this work focuses on the dark matter-only case, it is crucial to include baryonic effects on gravothermal evolution in order to make better connections with observations of galaxies whose baryon concentration is high. For the integral approach, extra terms could be included to incorporate SIDM effects that couple to mass changes. In addition, we use the collapse timescale as a normalization factor to derive the universal density profile. It would be interesting to explore if there is a more accurate way to estimate the timescale for halos in cosmological environments. We will leave these interesting topics for future investigation. 

We provide example scripts for applying the parametric model at: \url{https://github.com/DanengYang/parametricSIDM}

\acknowledgments
We thank Manoj Kaplinghat and Daniel Gilman for helpful discussion, and the anonymous referee for suggesting the Read profile. D.Y. and H.-B. Y were supported by the John Templeton Foundation under grant ID \#61884 and the U.S. Department of Energy under grant No.\ de-sc0008541. Y.Z. was supported by the Kavli Institute for Cosmological Physics at the University of Chicago through an endowment from the Kavli Foundation and its founder Fred Kavli. The opinions expressed in this publication are those of the authors and do not necessarily reflect the views of the John Templeton Foundation. 

\clearpage

\appendix

\section{The parametric model based on the Read profile}
\label{sec:appA}

\begin{figure*}[htbp]
  \centering
  \includegraphics[height=4.8cm]{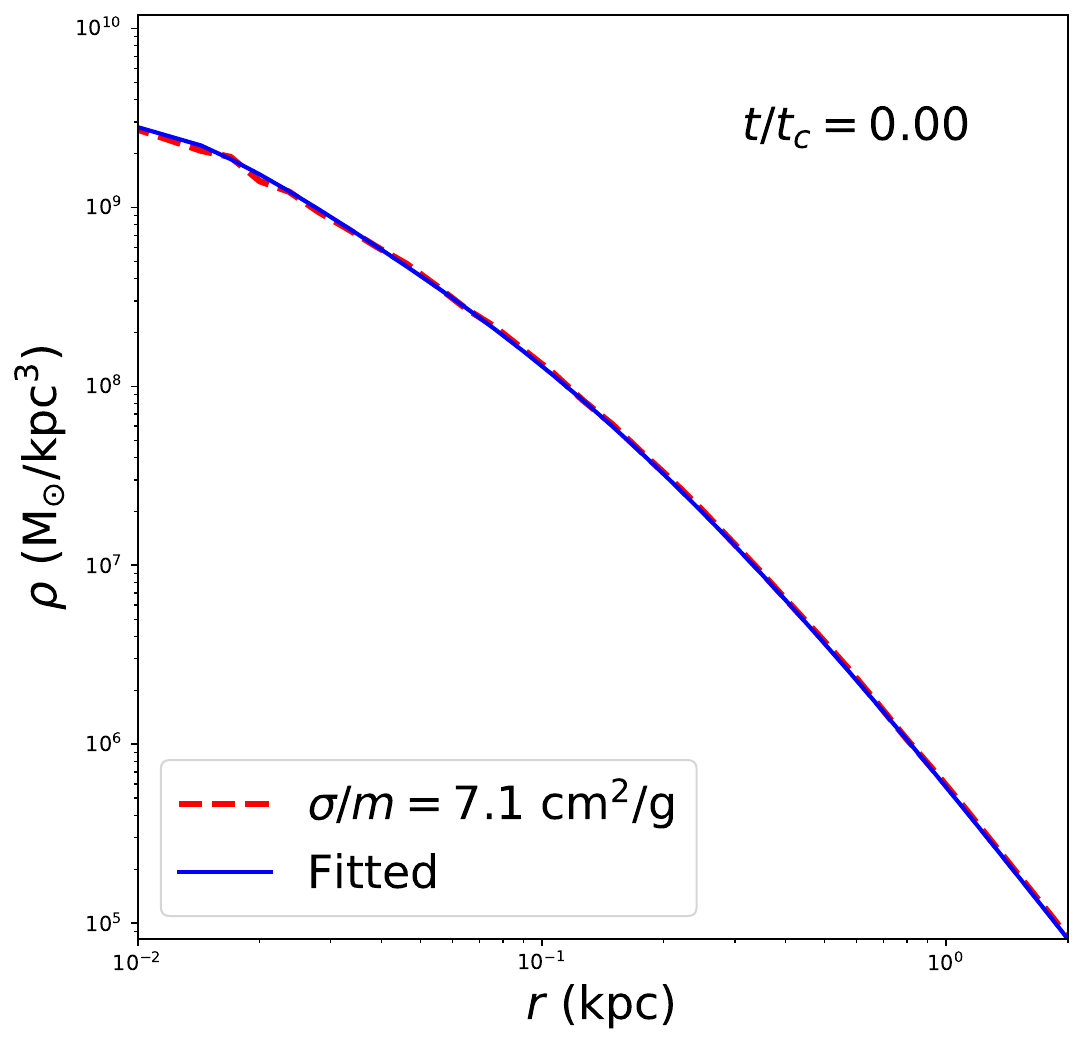}
  \includegraphics[height=4.8cm]{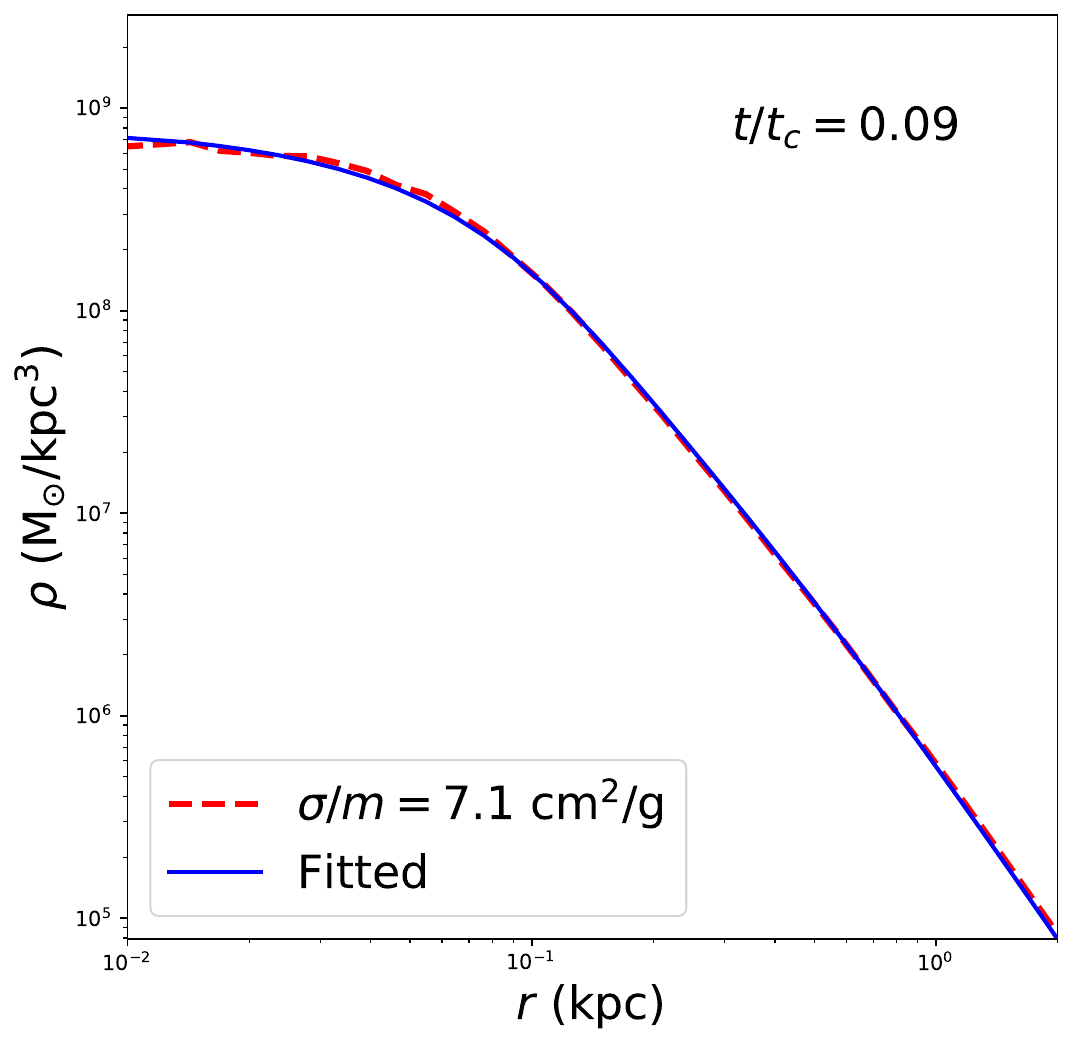}
  \includegraphics[height=4.8cm]{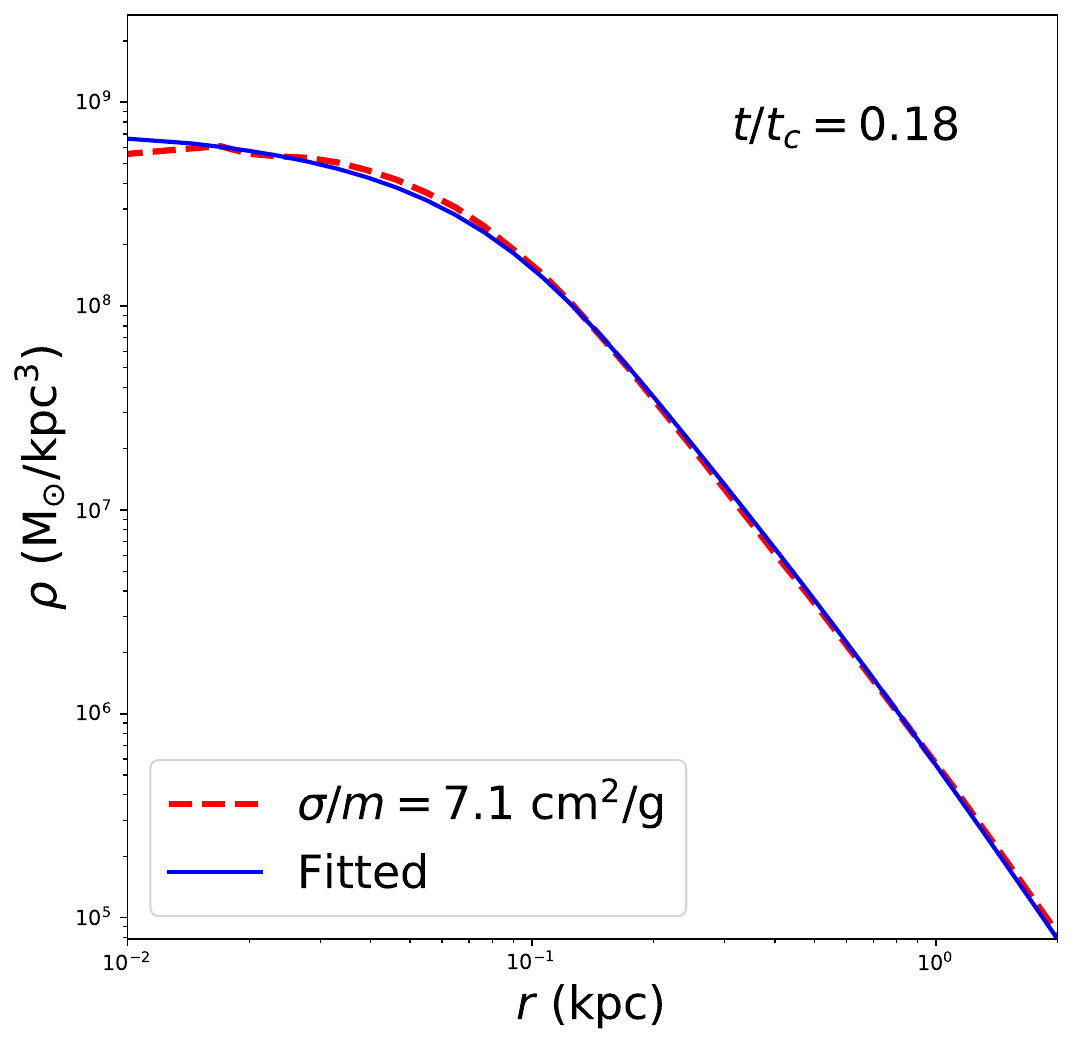}\\
  \includegraphics[height=4.8cm]{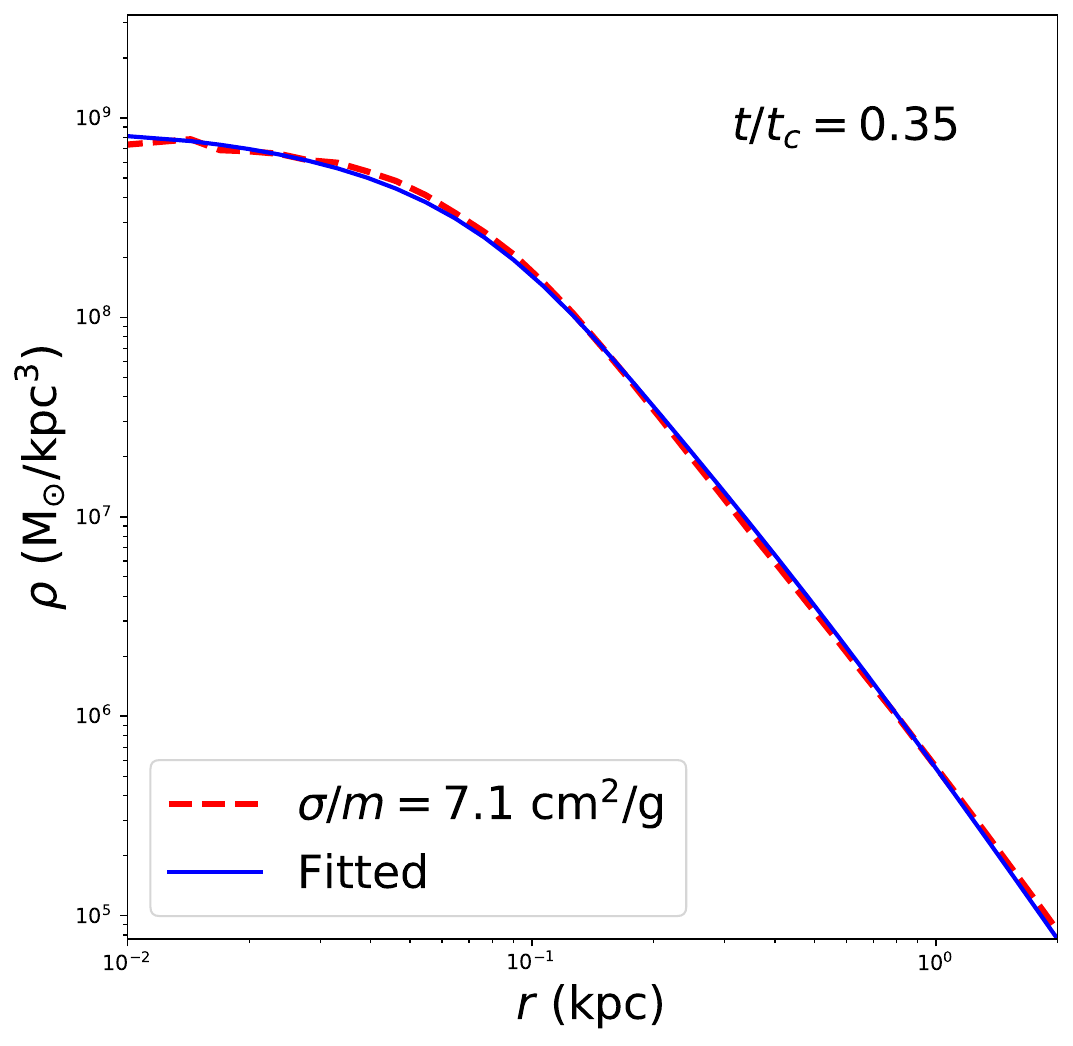}
  \includegraphics[height=4.8cm]{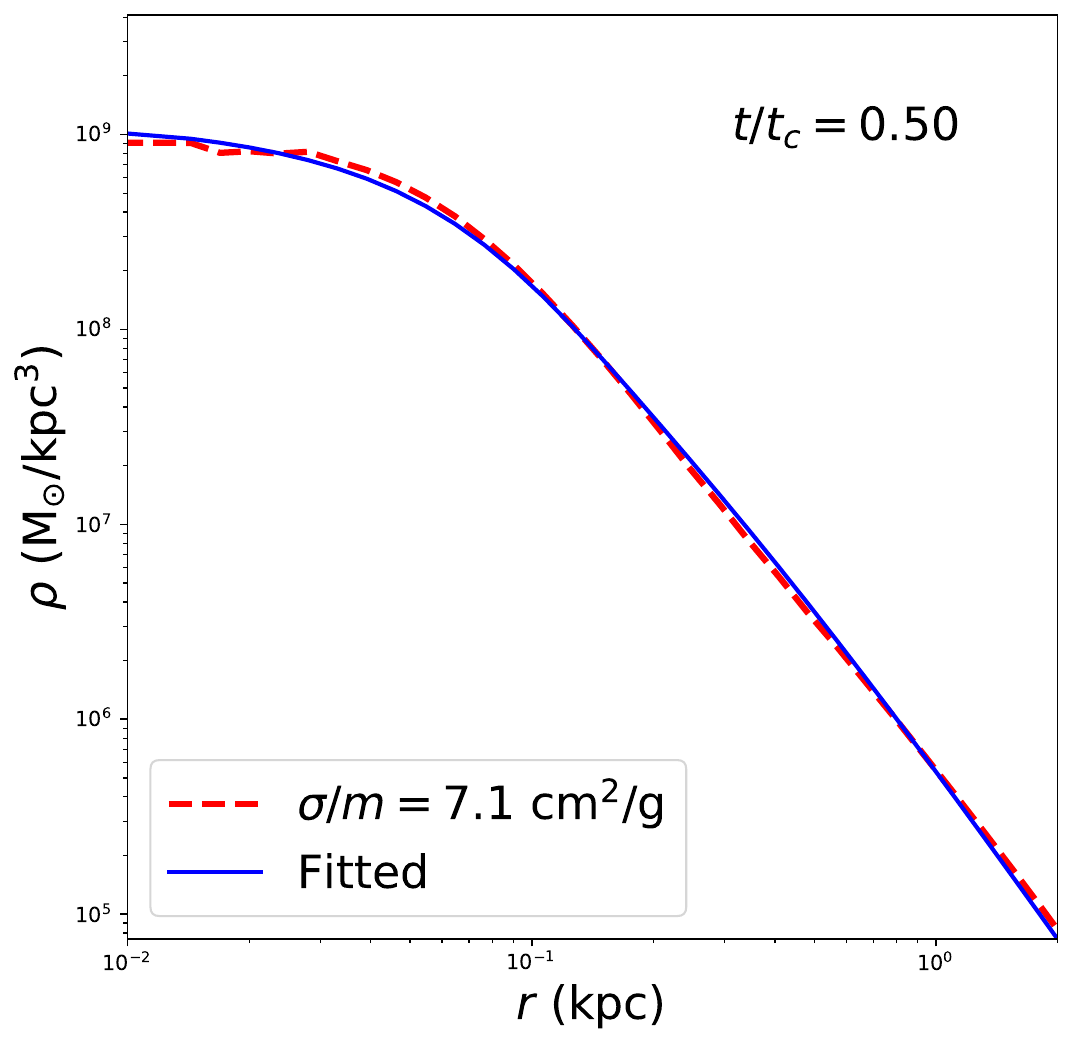}
  \includegraphics[height=4.8cm]{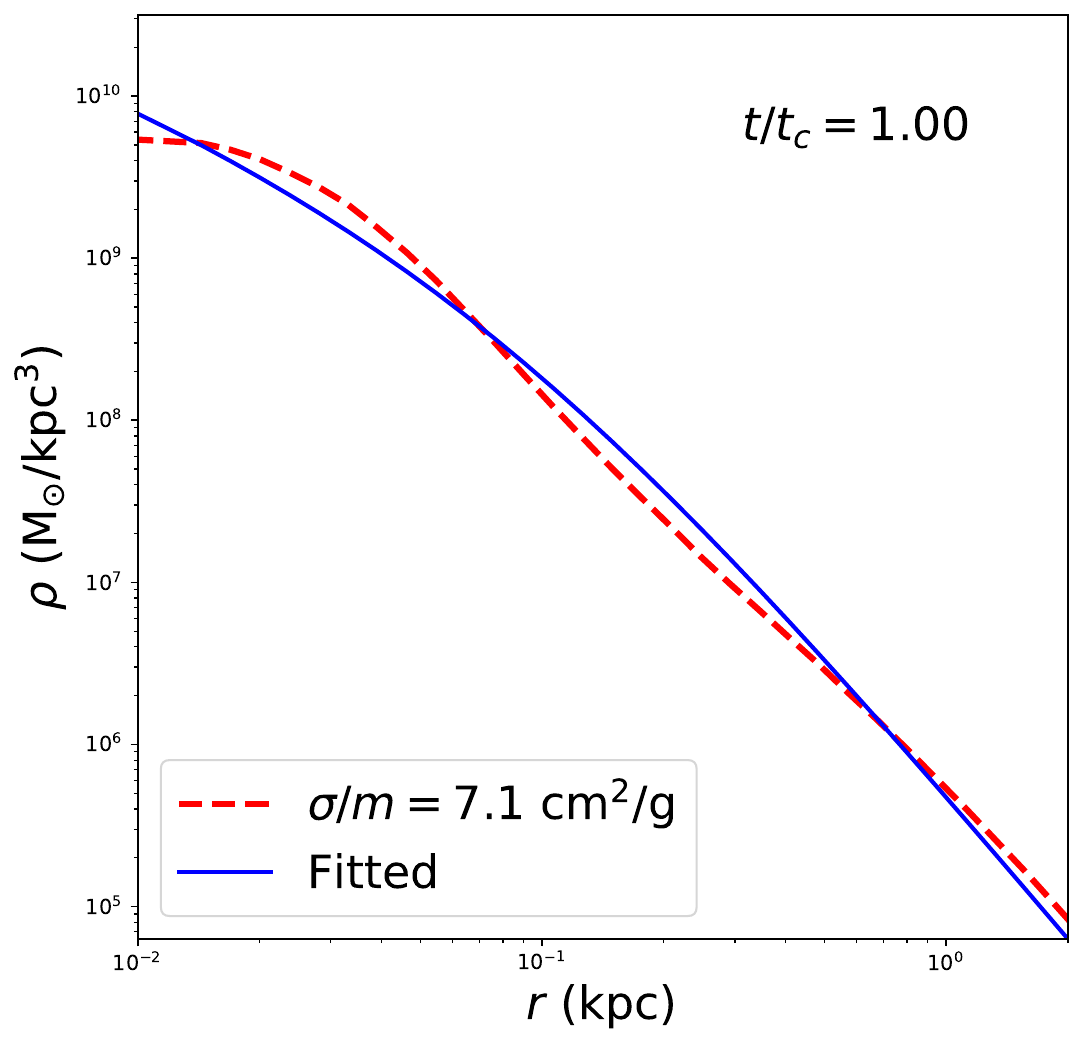}
  \caption{\label{fig:fitrhoRead} Density profiles of the simulated BM2 halo at different times for an effective constant cross section of $\sigma_{\rm eff}/m=7.1~{\rm cm^2/g}$ (dashed-red); data from ref.~\cite{Yang220503392}. Each simulated density profile is fitted using the Read profile eq.~(\ref{eq:read}) from refs.~\cite{2016MNRAS.459.2573R,2016MNRAS.462.3628R} (solid-blue). }
\end{figure*}

In this section, we show that the cored density (``Read'') profile proposed in refs.~\cite{2016MNRAS.462.3628R,2016MNRAS.459.2573R} can also be used to parametrize the evolution of the SIDM halo. The Read profile is formulated based on the NFW profile,
\begin{eqnarray}
\label{eq:read}
\rho_{\rm Read}(r) = f^n \rho_{\rm NFW} + \frac{nf^{n-1}(1-f^2)}{4\pi r^2 r_c } M_{\rm NFW},
\end{eqnarray}
where $r_c$ is the core radius, $f(r) = \tanh(r/r_c)$, and $n$ is a parameter in the range $0<n\leq1$. The NFW density and mass profiles are 
\begin{eqnarray}
\rho_{\rm NFW}(r) = \frac{\rho_s}{\frac{r}{r_s} \left(1 +\frac{r}{r_s} \right)^2},~
M_{\rm NFW}(r) = 4\pi \rho_s r_s^3 \left[ \ln\left(1+\frac{r}{r_s}\right) -\frac{r}{r+r_s} \right]. 
\end{eqnarray}
The enclosed mass follows the relation $M_{\rm Read} = f^n(r) M_{\rm NFW}(r)$. For $r\gg r_c$, $f^{n}(r)=1$, and $M_{\rm Read} = M_{\rm NFW}(r)$. In the limit of $r_c\rightarrow0$, $f\rightarrow1$ and $\rho_{\rm Read}(r)\rightarrow\rho(r)_{\rm NFW}$. In this work, we fix $n=1$, fit the profile in eq.~(\ref{eq:read}) to the simulated BM2 halo, and determine the parameters $\rho_s$, $r_s$, and $r_c$, accordingly.   

Figure~\ref{fig:fitrhoRead} shows the Read profile (solid-blue) fitted to the simulated density profile (dashed-red) of the BM2 halo at $t/t_{c} =0,\,0.09,\,0.18,\,0.25$, $0.50$ and $1$. We have checked that the overall fit quality of the Read profile is similar to that of our cored profile in eq.~(\ref{eq:cnfw}); see figure~\ref{fig:fitrho}. 

\begin{figure*}[tp]
  \centering
  \includegraphics[height=4.8cm]{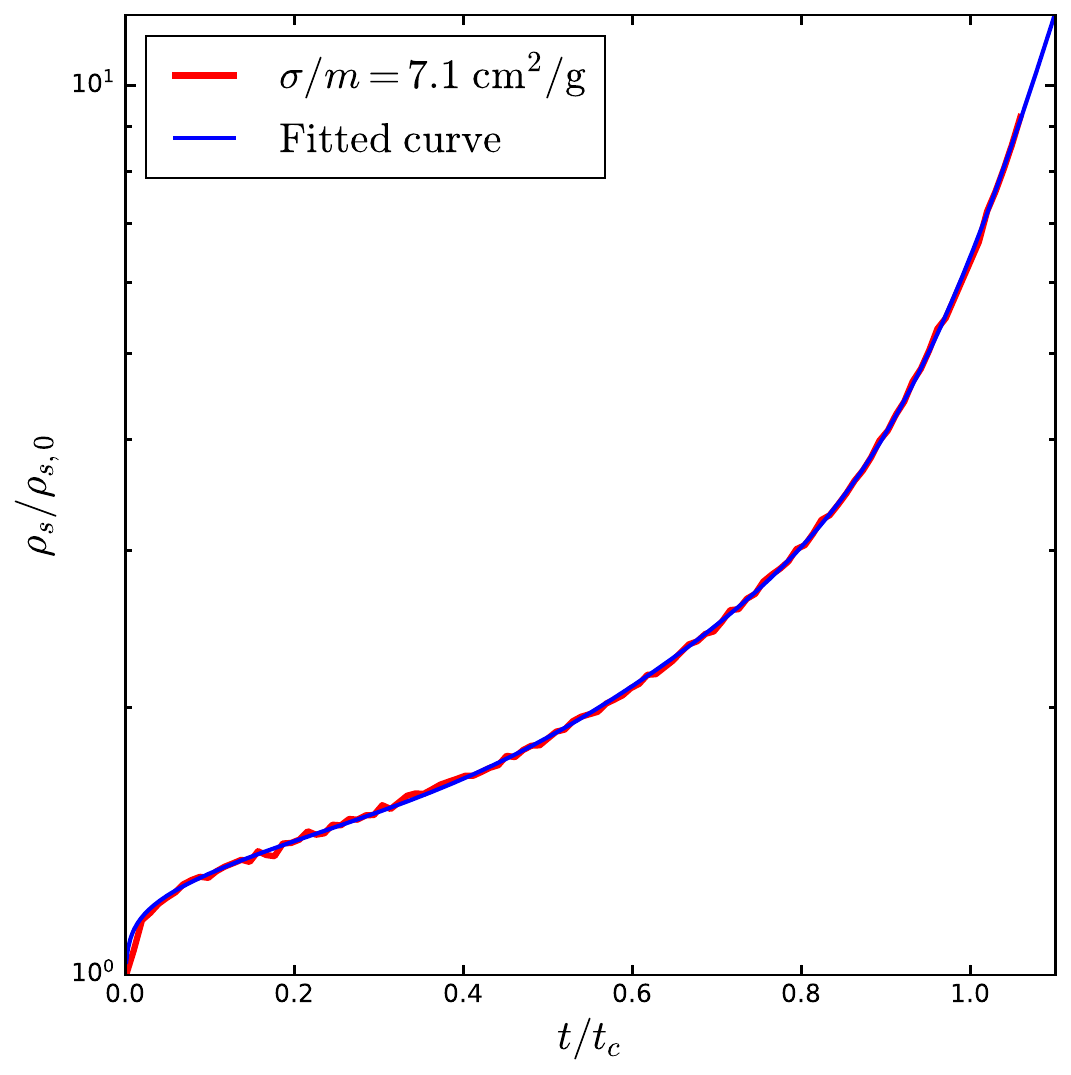}
  \includegraphics[height=4.8cm]{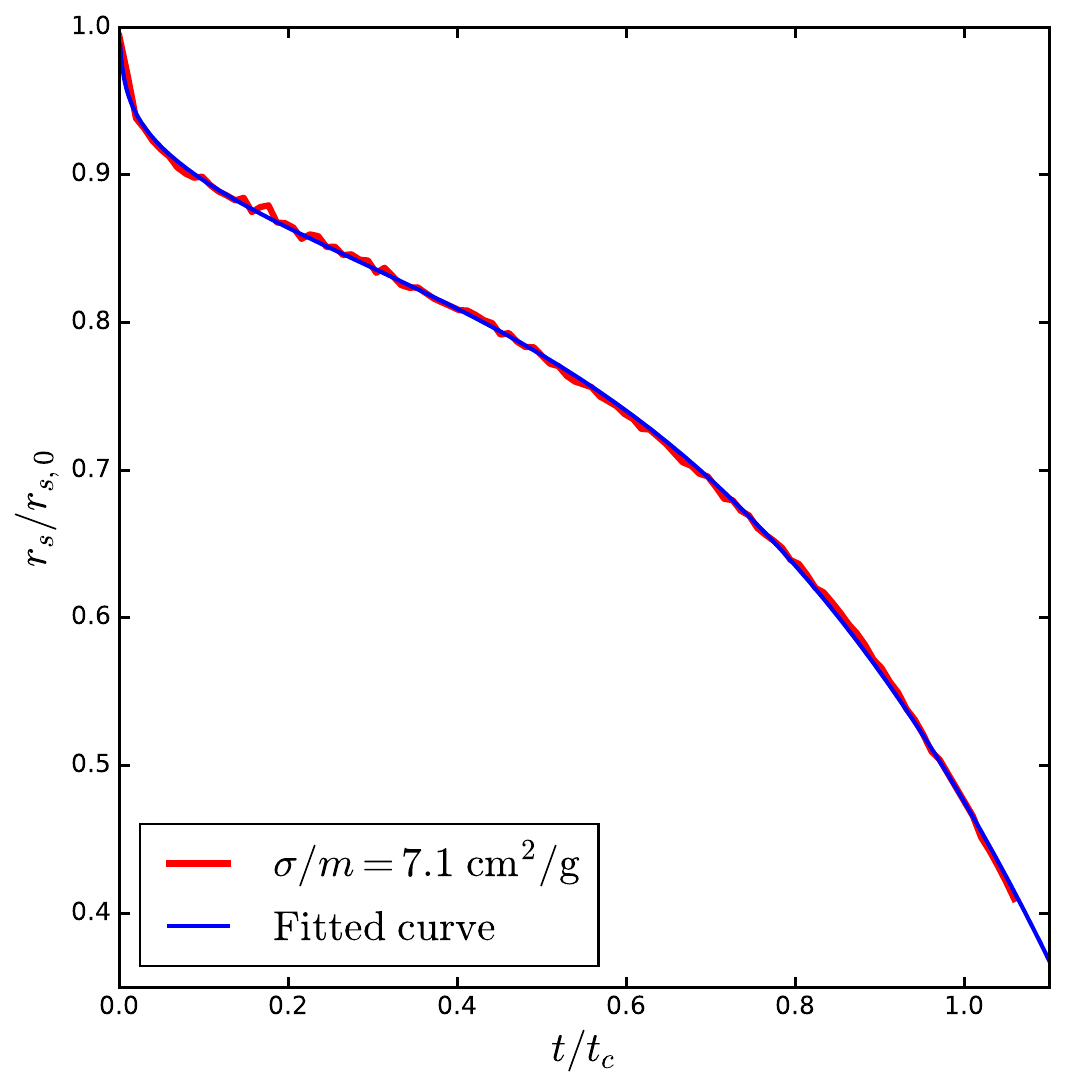}
  \includegraphics[height=4.8cm]{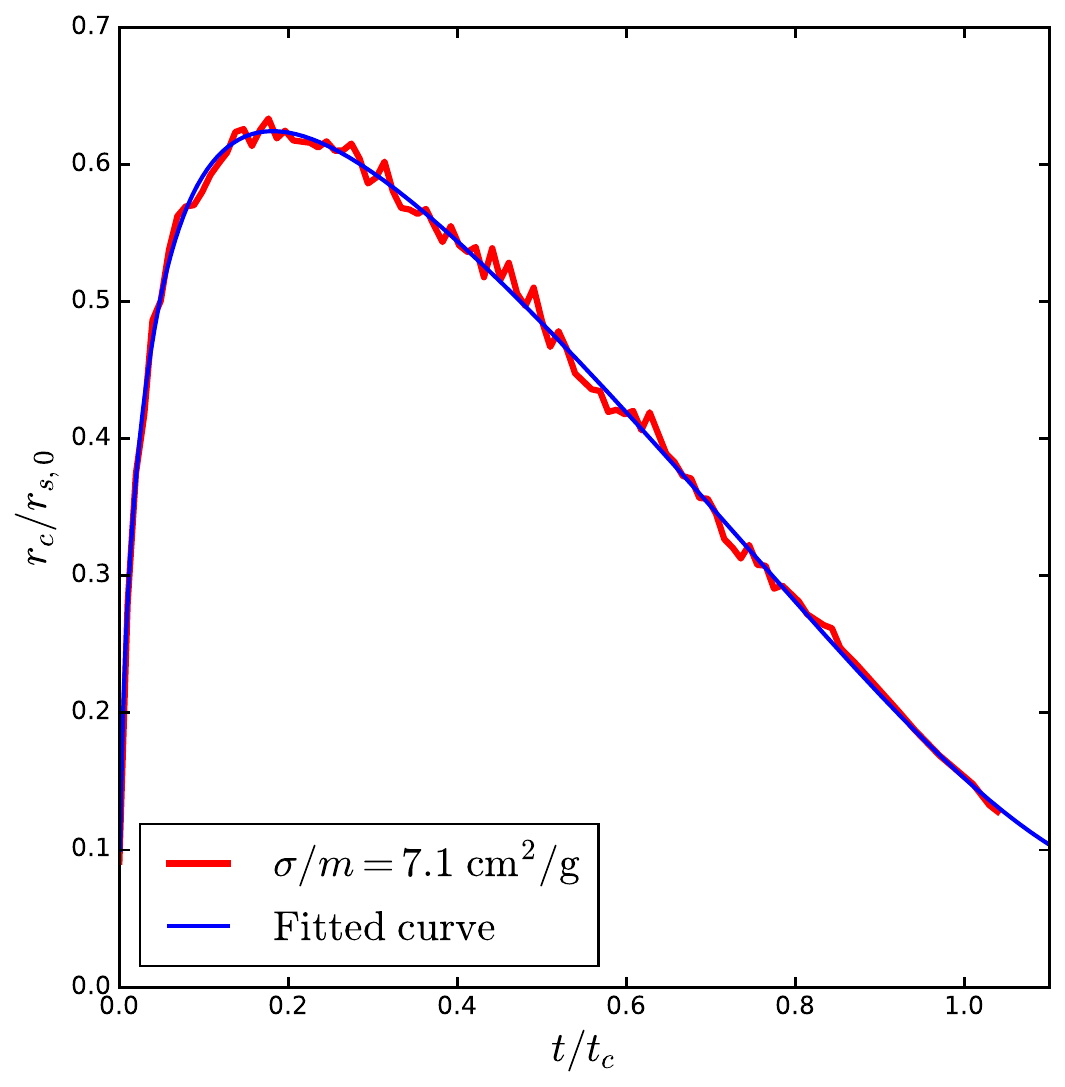}
  \caption{\label{fig:bm2read} Evolution of normalized density parameters $\rho_s/\rho_{s,0}$ (left), $r_s/r_{s,0}$ (middle), and $r_c/r_{s,0}$ (right) from the calibrated functions in eq.~(\ref{eq:m0read}) for the Read profile (solid-blue) and the N-body SIDM simulation with the effective constant cross section $\sigma/m=7.1~{\rm cm^2/g}$~\cite{Yang220503392} (solid-red). 
 }
\end{figure*}

As was done for the profile of eq.~(\ref{eq:cnfw}), we fit the Read profile to the simulated density one of the BM2 halo at a successive time interval of $0.2~{\rm Gyr}$, and obtain the evolution trajectories for the parameters $\rho_s$, $r_s$, and $r_c$,
\begin{eqnarray}
\label{eq:m0read}
\frac{\rho_s}{\rho_{s,0}} &=& 1.335 + 0.7746 \tau + 8.042 \tau^5 -13.89 \tau^7  + 10.18 \tau^9 + (1-1.335) (\ln 0.001)^{-1} \ln \left( \tau + 0.001 \right), \nonumber \\
\frac{r_s}{r_{s,0}} &=& 0.8771 - 0.2372 \tau +  0.2216 \tau^2 -0.3868 \tau^3 + (1-0.8771) (\ln 0.001)^{-1} \ln \left( \tau + 0.001 \right), \nonumber \\
\frac{r_c}{r_{s,0}} &=& 3.324 \sqrt{\tau} -4.897 \tau + 3.367 \tau^2 -2.512 \tau^3 + 0.8699 \tau^4,  
\end{eqnarray}
where the subscript $``0"$ denotes the corresponding value of the initial NFW profile. 
We found the same functional forms of eq.~(\ref{eq:m0}) work well for the Read profile, with the adjustment of the coefficients; see figure~\ref{fig:bm2read}. It is important to note that the fitted values of $\rho_s$, $r_s$, and $r_c$ of the Read profile are different from those of our cored profile in eq.~(\ref{eq:cnfw}).  We have also numerically checked $V_{\rm max}$ and $R_{\rm max}$ using the two fitted density profiles and found they agree within $10\%$.

\bibliographystyle{apsrev}

\bibliography{reference}

\end{document}